\documentclass[usenatbib]{mn2e}
\usepackage{times}
\usepackage{graphicx}
\usepackage{natbib}
\usepackage{subfigure}
\usepackage{dcolumn}
\usepackage{longtable}
\usepackage{lscape}
\usepackage[usenames,dvipsnames]{xcolor}
\usepackage{xspace}
\usepackage{aas_macros}

\newcommand{\galapagos}{{\scshape galapagos}\xspace}
\newcommand{\galfit}{{\scshape galfit}\xspace}
\newcommand{\galfitthree}{{\scshape galfit3}\xspace}
\newcommand{\galfitm}{{\scshape galfitm}\xspace}

\newcommand{\sersic}{S\'ersic\xspace}

\newcommand{\sex}{{\scshape SExtractor}\xspace}
\newcommand{\montage}{{\scshape Montage}\xspace}
\newcommand{\ferengi}{{\scshape ferengi}\xspace}
\newcommand{\kms}{km~s$^{-1}$\xspace}
\newcommand{\fita}{\texttt{S3}\xspace}
\newcommand{\fitb}{\texttt{SM}\xspace}
\newcommand{\fitc}{\texttt{MM}\xspace}

\newcolumntype{d}{D{.}{.}{-1}}

\title[MegaMorph II]{MegaMorph -- multi-wavelength measurement of galaxy structure: \sersic profile fits to galaxies near and far}
\author[Vika et al.]
{Marina~Vika,$^{1}$  Steven~P.~Bamford,$^{2}$ Boris~H{\"a}u{\ss}ler,$^{1,2}$ Alex~L.~Rojas,$^{1}$ Andrea~Borch,$^{1}$ \newauthor Robert~C.~Nichol$^3$
\smallskip\\
$^1$ Carnegie Mellon University in Qatar, Education City, PO Box 24866, Doha, Qatar \\
$^2$ School of Physics and Astronomy, The University of Nottingham, University Park, Nottingham, NG7 2RD, UK \\
$^3$ Institute of Cosmology and Gravitation (ICG), University of Portsmouth, Dennis Sciama Building, Burnaby Road Portsmouth, PO1 3FX, UK
}

\begin{document}

\date{Accepted ... Received ...; in original form ...}

\pagerange{\pageref{firstpage}--\pageref{lastpage}} \pubyear{2012}

\maketitle

\label{firstpage}


\begin{abstract}
We demonstrate a new multi-wavelength technique for two-dimensional parametric modelling of galaxy surface-brightness profiles, which we have incorporated into the widely used software \galfit. Our new method, named \galfitm, extends \galfitthree's current single-band fitting process by simultaneously using multiple images of the same galaxy to constrain a wavelength-dependent model. Each standard profile parameter may vary as a function of wavelength, with a user-definable degree of smoothness, from constant to fully free. The performance of \galfitm is evaluated by fitting elliptical \sersic profiles to $ugriz$ imaging data for 4026 galaxies, comprising the original SDSS imaging for 163 low redshift ($v \la 7000$~\kms) galaxies and 3863 artificially redshifted ($0.01 \la z \la 0.25$) images of the same galaxies.

Comparing results from single-band and multi-band techniques, we show that \galfitm significantly improves the extraction of information, particularly from bands with low signal-to-noise ratio (e.g., $u$ and $z$ SDSS bands) when combined with higher signal-to-noise images.  We also study systematic trends in the recovered parameters, particularly \sersic index, that appear when one performs measurements of the same galaxies at successively higher redshifts.  We argue that it is vital that studies investigating the evolution of galaxy structure are careful to avoid or correct for these biases.

The resulting multi-band photometric structural parameters for our sample of 163 galaxies are provided.  We demonstrate the importance of considering multi-band measurements by showing that the \sersic indices of spiral galaxies increase to redder wavelengths, as expected for composite bulge-disk systems. Finally, for the ellipticals in our sample, which should be well-represented by single-\sersic models, we compare our measured parameters to those from previous studies.
\end{abstract} 


\section{Introduction}
\label{sec:intro}

The radiation output of typical galaxies is dominated by starlight, with most energy emerging at optical--near-infrared wavelengths (e.g., \citealt{tex:DR12}).  From integrated photometry alone, although preferably combined with distance information, e.g., cosmological redshifts, we can learn a great deal about galaxy stellar populations (e.g., \citealt{tex:WG11}).  Measurements of luminosity at a number of wavelengths may be used to infer physical quantities, such as stellar mass, age and metallicity \citep[e.g.][]{tex:PM12}.  Including near-infrared (NIR) data is particularly useful for breaking the degeneracy between age and metallicity \citep{tex:W94}. However, the integrated stellar population is only one aspect of a galaxy's character.  By studying how the stellar population is spatially distributed, e.g. via surface photometry, we may gain further insights into a galaxy's history. In particular, the stellar distribution provides information on the dynamical state of a galaxy, without requiring observationally expensive resolved spectroscopy (e.g., \citealt{tex:KE11}).

The most basic spatially resolved measurement is simply the size of a galaxy.  Just measuring this can place strong constraints on models of galaxy formation (e.g., \citealt*{tex:MM98}), although defining and estimating a galaxy's size is by no means trivial.  More sophisticated descriptions of a galaxy's surface brightness distribution include visual morphology  (e.g., \citealt{tex:VV91,tex:LS11}), concentration (e.g., \citealt{tex:SF01}), asymmetry and clumpyness (e.g., \citealt{tex:C03}), multi-Gaussian expansion \citep{tex:C02}, \sersic models \citep{tex:S68,tex:GD05} and their extensions \citep{tex:AM11}, bulge-disk models (e.g., \citealt{tex:SW02}), and more sophisticated multi-component models (e.g., \citealt{tex:PH10}).

Studying spatially resolved surface brightness distributions is obviously easier and more precise for relatively nearby galaxies.  Indeed, for some it is possible to determine such distributions from counts of individual stars (e.g., \citealt{tex:IM09}).  Even without going to this extreme, it is possible to discern a great level of detail: thin disks, thick disks, bulges, bars, rings, spiral arms, nuclei, cores and streams  (e.g., \citealt{tex:YD06}).  Each of these features may be associated with different stellar populations  (e.g., \citealt{tex:SJ98}).  Clearly, characterising the structure of galaxies to this level is a extremely complex task.  Even for very nearby galaxies, studies of early-type galaxy kinematics have shown that visual morphologies and photometric structural parameters can be misleading \citep{tex:EC11}.

 Once we move beyond our local neighbourhood fine structural distinctions become more difficult, and only broad features may be discerned.  Furthermore, for large samples there is a need for robust automated measurements, which favours simpler approaches. In this context, integrated properties, such as total luminosity and colour, provide an important starting point (e.g., \citealt{tex:BD01,tex:BG04}), but additional physical insight is afforded by even simple size and structure information (e.g., \citealt{tex:BH03,tex:KF03,tex:SM03}).  Such studies demonstrate a clear division of the galaxy population in terms of spheroid- and disk-dominated systems, which have different relationships with mass, colour and environment (e.g., \citealt{tex:GM09}).  The fundamentally different nature of spheroid and disk components highlights the need to separate these structures, and consider their individual properties, whenever possible  \citep{tex:MC03,tex:AD06,tex:CD09,tex:SM11}.  It is even becoming clear that there exist different types of spheroids, formed via distinct processes, which may be distinguished by structural measurements (e.g., \citealt{tex:KK04,tex:A05,tex:FD10, tex:FS12}).

At still higher redshifts, decomposing galaxy components becomes difficult.  However, single-component parameters, such as effective radius and \sersic index, can still be used to explore variations in the structures of galaxies.  One active area of study concerns the evolution of early-type galaxy sizes \citep{tex:HB10}.  Members of the massive early-type galaxy population at $z \sim 2$ appear to be substantially more compact than those today (e.g., \citealt{tex:BT08, tex:TC09,tex:CG10}).  This suggests, when combined with expectations from hierarchical merging, that some of these early-formed objects must increase their sizes dramatically, without significantly changing their stellar mass (although see \citealt{tex:MD10} and \citealt{tex:TC12}).  Structural measurements have also been used to quantify evolution in the morphological composition of the galaxy population, finding that massive galaxies were considerably more disky at $z \sim 2$ than they are in the local Universe \citep{tex:WR11,tex:BD12,tex:BT13}.

Studies of galaxy structure typically require robust estimates of stellar mass and recent star formation, in order to study trends and ensure that comparisons are made between similar populations of objects.  These quantities, in turn, require accurate measurements of galaxy colour, which are compared with representative stellar population models to deduce the desired physical estimates.  These models necessarily make simplifying assumptions regarding star-formation history, metallicity, and dust extinction \citep{tex:WG11}.  They are also usually constrained using global colours, discarding any information on spatial variations due to multiple components.  However, the fact that colour typically varies radially within a galaxy indicates the presence of multiple star formation histories that are, to some degree, localised.  Incorporating this knowledge into our estimates of stellar mass and star formation history -- for example, by allowing different structural components of a galaxy to have different colours -- can potentially make these estimates more robust and meaningful.

In addition to the variation of stellar populations between galaxy components, one should consider population gradients within individual components. This is especially true in the case of low-quality imaging, where decomposing each galaxy into more than one component becomes difficult, or for fundamentally single-component systems, e.g. elliptical galaxies \citep{tex:SJ10}. Approaches include extracting radial colour profiles using annular photometry (e.g., \citealt{tex:Bd00,tex:WK09}), or estimating physical parameters pixel-by-pixel from colour maps (e.g., \citealt{tex:WC08,tex:LC12}). These techniques have been used very effectively, although they are primarily suitable for data with reasonably high spatial resolution, as they do not account for the point spread function (PSF).  This means that galaxies with smaller angular sizes will be systematically biased to flatter gradients.  For some studies one can choose to consider only large and/or nearby galaxies, and thereby avoid resolution issues.  However, in order to obtain sufficiently large samples for environmental studies, or to ensure completeness as a function of size, one must necessarily push to the resolution limits of current surveys.

For galaxies which can be reasonably well described by smooth, axisymmetric models, assuming such a model allows one to account for the effect of the PSF.  Assuming an appropriate model also enables more accurate surface brightness profile measurements, particularly for low signal-to-noise data.  However, measuring colours using model-based methods typically assumes that each component (the entire galaxy in the case of single-component models) has a homogeneous colour.  Determining structural parameters independently in each band results in colours that are hard to interpret, as the fluxes can correspond to models with very different shapes in each band.  Usually, therefore, structural parameters are first determined on one image, and these are subsequently applied to images in other wavebands with total flux as the only free parameter (e.g., \citealt{tex:LG12}).  \citet{tex:Ld09} take a hybrid approach, performing independent profile fits to each waveband, and then performing elliptical annular photometry on the resulting models to determine PSF-corrected colour gradients (utilised to great effect in \citealt{tex:Ld10}).  While some techniques allow their models to be fit to two images simultaneously (e.g., \citealt{tex:SW02}), until very recently (Bamford et al., in prep.) no solutions permitted the use of an arbitrary number of images at different wavelengths.

Prompted by all of the above issues, we embarked upon a project, named `MegaMorph', to investigate ways of improving our ability to extract physically-meaningful structural information from galaxy images \citep{tex:BH11}.  In this paper we present results obtained by applying our new technique for measuring galaxy profiles in multi-wavelength imaging.  The development of this technique was motivated by a desire to preserve the concept of distinct structural components, while allowing for realistic colour gradients within them.  It was also driven by a determination to make more effective use of the wealth of spatially-resolved multi-wavelength imaging available from modern surveys.  To achieve our goals we have extended standard two-dimensional profile fitting techniques to include wavelength-dependent models for which the degree of variation with wavelength is controllable.  We chose to implement these techniques by modifying \galfitthree \citep{tex:PH02,tex:PH10}, a widely used and respected profile-fitting tool.  Our approach has the benefits of greater physical consistency and fit reliability, while making more complete use of available data.  It also possesses various convenient features, such as straightforward restframe corrections. Throughout this paper we refer to our modified, multi-band version of \galfit as \galfitm. 

This paper is one of a series that will explore the advantages of our multi-wavelength approach to measuring galaxy structural properties.  In Bamford et al. (in prep.; hereafter Paper I) we present this new tool in more detail, describing the new features and demonstrating its use through some specific examples.  In \citet{tex:HB12} (Paper III) we test our new method on large datasets, by automating both the preparation of the data and the fitting process itself. For this purpose we adapt \galapagos \citep{tex:BH12} to use \galfitm, and quantify its performance by applying it to data from the GAMA survey \citep{tex:DH11} and to simulations.  The resulting measurements of GAMA galaxies, in particular the variation of structural parameters with wavelength, will be studied further in Vulcani et al. (Paper IV; in prep.). In the present paper, as well as Papers III and IV, we perform and utilise the results of single-\sersic profile fitting.  Ultimately, however, we expect the greatest benefits of multi-band fitting when dealing with multi-component models.  A further set of papers will present the results of bulge-disk decompositions on the same data.

This paper investigates the performance of \galfitm for fitting single-\sersic profiles to galaxy images with a wide range of resolution and signal-to-noise.  This is achieved by analysing SDSS images of large, nearby galaxies, as well as versions of these images that have been convolved and resampled to simulate their appearance at a range of redshifts.  These results are used to study trends in statistical scatter and potential systematic biases in measured parameters as the galaxies are considered at increasing distance.  As a further test, the structural parameters of the elliptical galaxies in our sample, as obtained on the original images, are compared to those obtained by independent studies.

In Section 2 we present the criteria for selecting our parent galaxy sample and the methodology used to construct the artificially redshifted sample. In Section 3 we briefly introduce our new technique and utilise it to fit wavelength-dependent profiles to $u$, $g$, $r$, $i$, $z$ images for both original and artificially redshifted galaxy samples.  We present a comparison of the results from our multi-band fitting method to those obtained by fitting each band independently. In Section 4 we discuss these comparisons and their implications, and compare our multi-band structural parameters for a subsample of elliptical galaxies with parameters derived by previous studies. We conclude and discuss future work in Section 5. Throughout the paper, we adopt a standard cosmological model with $\Omega_{\Lambda}=0.7$, $\Omega_{\rm m}=0.3$ and $H_{0}=70$~km~s$^{-1}$~Mpc$^{-1}$ .
   
   

\section{Data}
\label{sec:data}

\subsection{Sample selection and standard imaging}

Our primary aim in this paper is to evaluate the benefits of modelling galaxy structure by simultaneously fitting imaging data at a variety of wavelengths. For this purpose we select galaxies that have parametric surface brightness profile measurements by previous studies (using single-band data), and which are located within the footprint of the Sloan Digital Sky Survey (SDSS; \citealt{tex:AA09}).  We compile a sample of 168 galaxies with images in $u$, $g$, $r$, $i$, $z$ passbands, which have profile fits presented in one of the following papers: \citet{tex:PT06}, \citet{tex:M04}, \citet{tex:MG98} and \citet{tex:CC93}.

The galaxies in our sample are typically larger than a single SDSS frame. The production of images for these galaxies is accomplished with the help of \montage \citep{tex:JK10}, which performs the transformations, rebinning and background adjustment necessary to combine the individual frames into a single mosaic. We ensure that each mosaic contains the entire galaxy as well as sufficient sky area for an accurate background estimation. In the right-hand panels of Figures \ref{fig:gal1}, \ref{fig:gal2} and \ref{fig:gal3} we show three example of this imaging data for galaxies NGC4570, NGC4274 and NGC4431.

The \montage SDSS tool scales each image to a constant photometric zeropoint ($26$~mag arcsec$^{-2}$ s$^{-1}$, corresponding to $23.682$~mag~s$^{-1}$ with an exposure time of $53.91$~s, or $28.011$~mag).  In order for \galfit to construct a reasonable sigma image (giving an estimate of the noise on each pixel, and hence affecting the relative weighting of pixels and final parameter uncertainties), the images produced by \montage were rescaled to approximate electron counts.  This was done by assuming values of the zeropoint and gain for each band that were obtained by averaging over the survey.

Five out of 168 galaxies are excluded from the further analysis. NGC0988 and UGC04684 are removed because of flux contamination due to nearby saturated stars which hinder accurate photometry. We also exclude the blue-shifted galaxies NGC3031, NGC4406 and NGC4419 as they cannot be appropriately redshifted by \ferengi (see Section~\ref{sec:ferengi}). The final sample consists of 163 low redshift  ($v \la 7000$~\kms) galaxies of mixed morphology (from ellipticals to late-type spirals). We do not attempt to construct a complete or representative sample of galaxies, but a wide variety of galaxies are present in the sample. The global properties of the sample are tabulated in Table \ref{table:first}.

\begin{figure}
\centering
\includegraphics[width=0.5\textwidth]{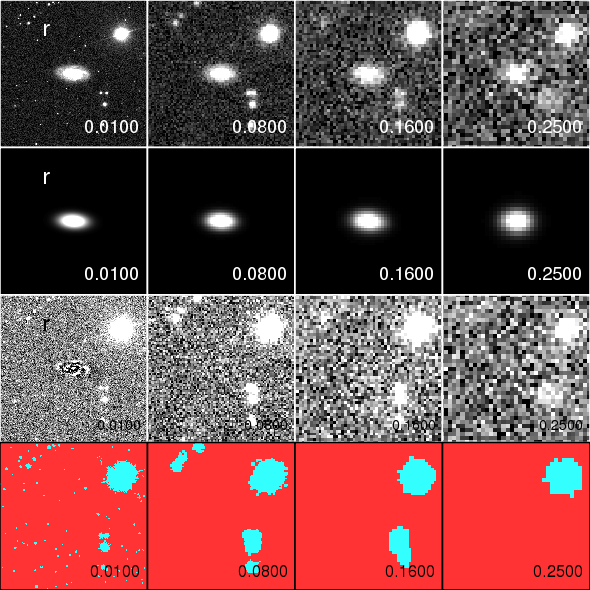}
\caption{Artificially redshifted $r$-band images of galaxy NGC2742 at redshifts $0.01$, $0.08$, $0.16$ and $0.25$, together with their masks and outputs of our multi-band fitting procedure. The redshifted input images are shown in the upper panels. The corresponding \galfitm best-fitting model and its residuals (image $-$ model) are shown in the middle panels. The lower panels display the masks that were used during the fit.  Note that the resolution and flux changes with redshift, images are displayed at constant physical size with an adaptive intensity scale.}
\label{fig:ferengi}
\end{figure}

\subsection{Artificial redshifting}
\label{sec:ferengi}

We wish to assess the accuracy and reliability of galaxy profile fitting over wide ranges of spatial resolution and signal-to-noise, and, in particular, compare the performance of single- and multi-band fitting techniques.
In order to do so, we require a sample of galaxies with known properties, or at least properties measured on very high quality imaging, which have been observed with a variety of resolutions and signal-to-noise levels.  Samples of real observations with such a wide variety of imaging qualities are very limited, and would make for a highly heterogeneous dataset.  In principle, completely simulated images could be used.  However, it is difficult to produce simulations which reproduce the detailed and varied appearance of real galaxies, with spiral arms, bars, star-formation regions and disturbances.  We utilise simulations based on smooth profiles in Paper III, but in this paper we wish to test our fitting technique on more realistic images.

We therefore choose to construct a sample of artificially redshifted images.  These are produced using \ferengi \citep{tex:BJ08}. For each of the 163 nearby galaxies we create a set of artificially redshifted $u$, $g$, $r$, $i$, $z$ images. The artificial redshifting applies cosmological changes in angular size, surface brightness and, optionally, spectral energy distribution, to simulate the observation of a given galaxy at a greater distance (for details see \citealt{tex:BJ08}).  \ferengi can also account for changes in instrumentation, such as pixel scale, PSF, filters and zeropoint.  However, in our case we use real SDSS imaging to simulate SDSS observations, so these details remain unchanged.

While \ferengi can account for the k-correction, i.e. the change in flux due to the change in the restframe wavelength sampled by a given observational filter for galaxies at different redshifts, we choose to disable this feature.  In this way the redshifted galaxies should retain a constant absolute magnitude in each band as a function of redshift. This is less physically realistic, but allows us to test the ability of \galfit to measure magnitudes as the noise increases and resolution decreases in a more stringent manner, without the uncertainties that would be introduced by bandpass shifting.

As galaxies become fainter and less resolved with increasing redshift, the fitting process becomes more challenging.  Faint components that are visible at low redshifts may no longer be apparent at greater distances.  We therefore anticipate greater uncertainties on the recovered structural parameters and potentially systematic biases.

The galaxies are artificially redshifted to values in the range $0.01<z<0.25$ with a step of $0.01$. However, if the true redshift of a galaxy is higher than $0.01$ the production of the artificial images starts at the next higher redshift, so that the artificial galaxies are always more distant than in the original imaging. In addition, we cease the redshifting process when the resulting image would be smaller than 21 pixels on a side, as at this point the galaxy is comparable in size to the PSF.  Smaller galaxies, therefore, do not have artificial images all the way out to $z = 0.25$.

The top panels of Fig.~\ref{fig:ferengi} illustrate 4 of the 25 artificially redshifted $r$-band images constructed for the galaxy NGC2742. Notice the reduction in resolution (a result of the decreasing angular size with respect to both the PSF and pixel scale) and diminishing signal-to-noise with increasing redshift.

One limitation of our method of artificially redshifting real galaxies is that in some cases the galaxies have neighbouring objects.  \ferengi treats all objects in the image in the same way, and so these neighbours are also artificially redshifted.  In the case of true galaxy groups this is realistic, and provides a way of assessing the impact of such systems on structural measurements.  However, in the case of stars, as well as foreground or background galaxies, the resulting image is not entirely realistic.  Furthermore, as our original images are large, they include a relatively high number of stars.  In this respect, the artificially-redshifted images therefore contain more stars than would be expected for real observations of higher-redshift galaxies.  On the other hand, the stars are faded by the same amount as the galaxy, which counteracts this effect.  In the end we expect our artificial images to present a reasonably realistic distribution of neighbouring objects.

\begin{figure}
\centering
\includegraphics[width=0.4\textwidth]{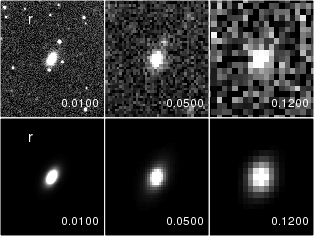}
\caption{ Artificially redshifted images of galaxy NGC4387 at redshifts $0.01$, $0.05$ and $0.12$, together with models fit using our multi-band technique. This Figure illustrates the artificial blending of nearby objects due to the artificial redshifting done with \ferengi.}
\label{fig:blend}
\end{figure}

At high redshifts we notice blending occurring: neighbouring galaxies or stars tend to be merged by the worsening resolution.  This process distorts the shape of the target object, resulting in changes to the structural parameter measurements. Such changes in shape are also accompanied, in some cases, by small changes of the magnitude (mostly increasing of the flux of the target galaxy). For instance, in Fig.~\ref{fig:blend} at $z = 0.01$ we can see, around galaxy NGC4387, three additional sources that are well separated from the central object of interest. At $z = 0.05$ two of the three sources have `blended' with the main galaxy and finally, at redshift $z = 0.12$, all four objects have blended into one extended source.

\subsection{Data preparation}
\label{sec:reduction}

\galfitthree, and hence \galfitm, can make use of mask images during the fitting process. These masks indicate the areas of each image that are contaminated with light from sources that are not intended to be modelled by \galfit. The masks are additional FITS files created by using \sex \citep{tex:BA96} segmentation maps.  Pixels flagged in the mask image are not included in the fitting process. Initially, we create separate masks for each of the five bands, and then take the union of these to produce a single merged mask.  We perform a detailed visual check on every mask to ensure that they do not include the main galaxy or parts thereof. When a mask was found to cover part of the galaxy flux the mask image was corrected. The final merged mask is used for all wavelengths in the \galfit fitting process.  This ensures that the fitted area is identical in each band and for fits with both \galfitthree and \galfitm.  Differences between bands and methods are therefore not attributable to variations in masking. The masks are generated independently for each redshift, in order to mimic the deblending issues faced by real observations. The lower panels of Fig.~\ref{fig:ferengi} show four masks at different redshifts for the galaxy NGC2742.


The estimation of the background level is important for a successful fit (e.g., \citealt{tex:HM07}). For all galaxies, the sky background value is measured before running \galfit. To estimate the sky for both the real images and artificial \ferengi images we use a similar method to that employed by \galapagos.  This method applies elliptical annuli centred on the galaxy to measure the surface brightness as a function of radius. The sky value is selected at the point where the surface brightness gradient is robustly and conservatively judged to be flat to the accuracy permitted by the sky noise. The procedure is described in more detail in \citet{tex:BH12}. In the case of the original images we have also employed a second approach for estimating the sky values. For this we simply calculate the resistant (sigma-clipped) mean of the large \montage mosaics. Each \montage image has at least 60 per cent of its area free of bright objects, which enables an accurate sky determination. The sky background values are kept fixed during the fitting process.

The galaxy light distribution, especially in the inner region, is distorted by the effect of atmospheric seeing, which needs to be corrected for (\citealt{tex:TA01a}). \galfit achieves this by convolving the \sersic function with the PSF in the fitting process. PSFs for our images are provided by SDSS, however for simplicity we have used a single typical PSF for each passband throughout this work.  For all our highly-resolved real images, small variations in the PSF will make minimal difference to the obtained parameters.  For our artificially-redshifted images, \ferengi produces output images which match the provided PSF.  The same PSF is also suppled to \galfit, so the two are entirely consistent.
	
\begin{figure*}
\centering
\includegraphics[height=13cm,width=15.0cm]{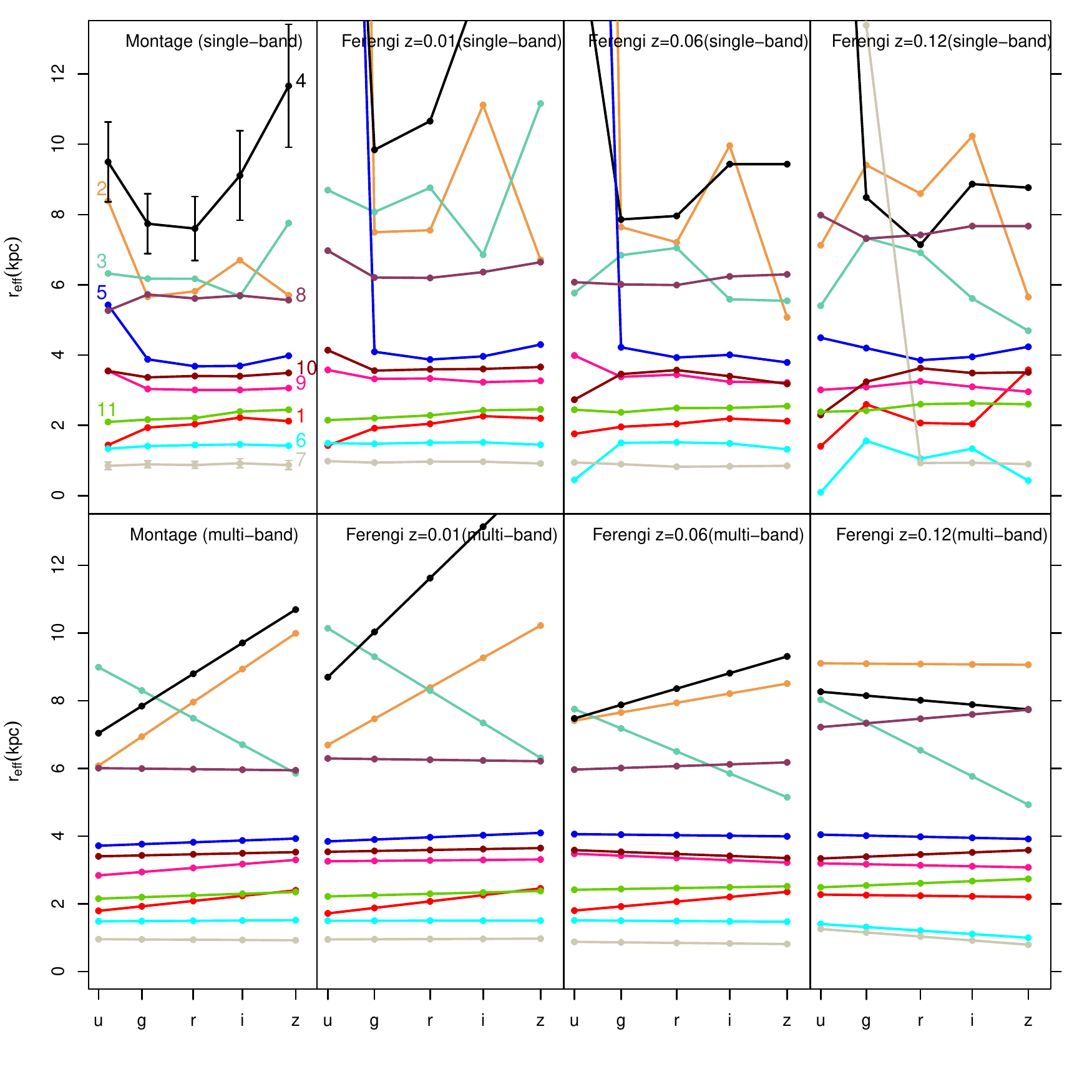}
\caption{ Example fitting results for eleven galaxies: 1-NGC853, 2-NGC2775, 3-NGC3521, 4-NGC3992, 5-NGC4274, 6-NGC4431, 7-NGC4434, 8-NGC4496A,9-NGC4570, 10-NGC4623, 11-NGC4900. We show recovered effective radius for both single-band (top row) and multi-band (bottom row) fitting methods. The leftmost column shows results from fitting the original images, while the remaining columns show results from fitting images artificially redshifted to redshifts 0.01, 0.06 and 0.12. Multi-band fitting results have a smooth dependence on wavelength as, by design, $r_{\rm e}$ is only permitted to vary linearly with wavelength. A typical one sigma error is given for the largest and smallest galaxy for the \montage images. The errors have estimated in Section~\ref{sec:uncert}. The eleven galaxies were chosen to show a variety of parameters, while ensuring some large galaxies were included. Note that in the panels with redshifted images there are a few cases where the effective radius, mostly in $u$-band, takes values above the upper limit of the plot, e.g., for NGC3992 at $z=0.01$ the $r_{{\rm e},u}=35$~kpc. }
\label{fig:reffband}
\end{figure*}

\begin{figure*}
\centering
\includegraphics[height=13cm,width=15.0cm]{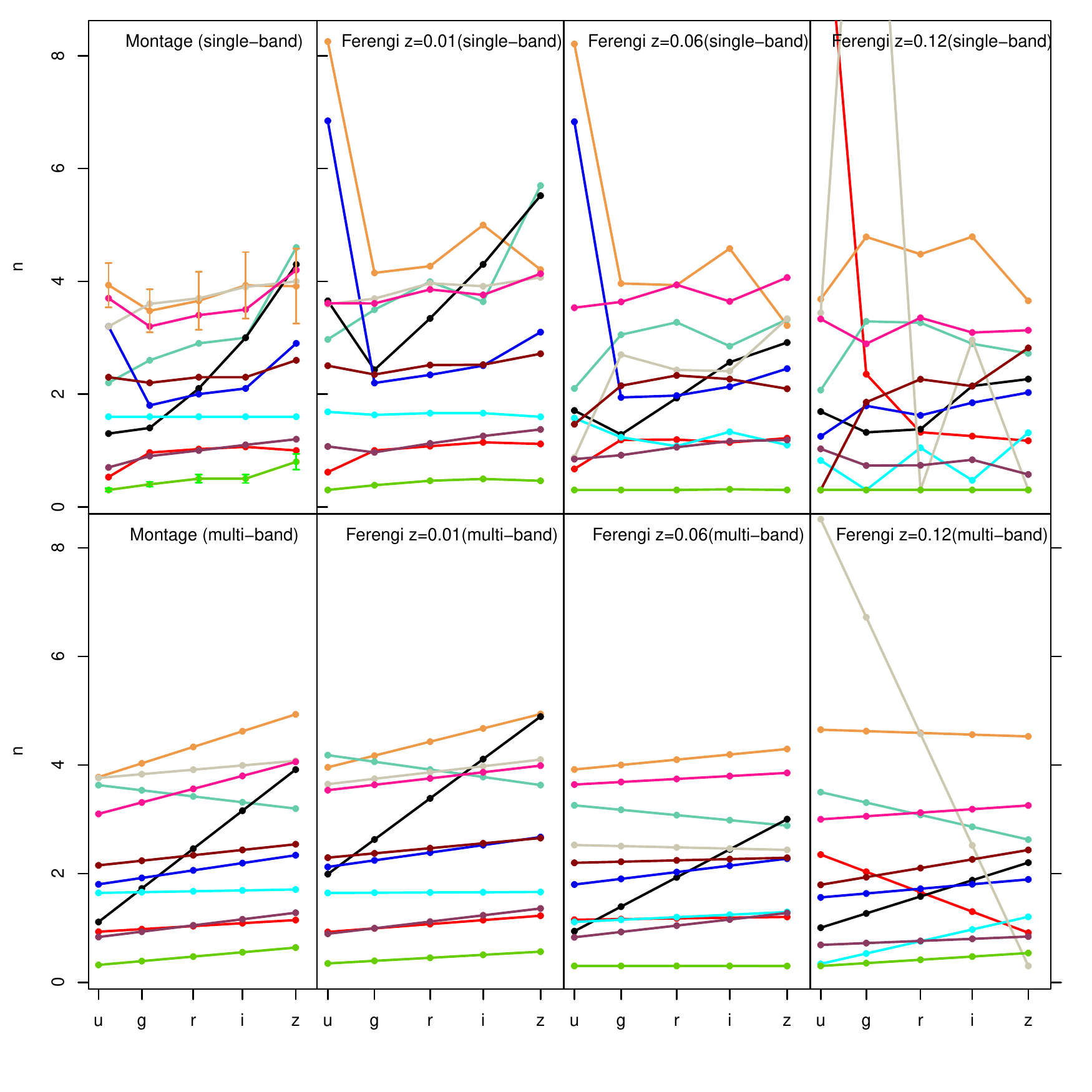}
\caption{Example fitting results for \sersic index.  The figure layout, galaxies and colour coding is the same as Fig.~\ref{fig:reffband}.}
\label{fig:nband}
\end{figure*}

\begin{figure*}
\centering
\includegraphics[height=12cm,width=13.0cm]{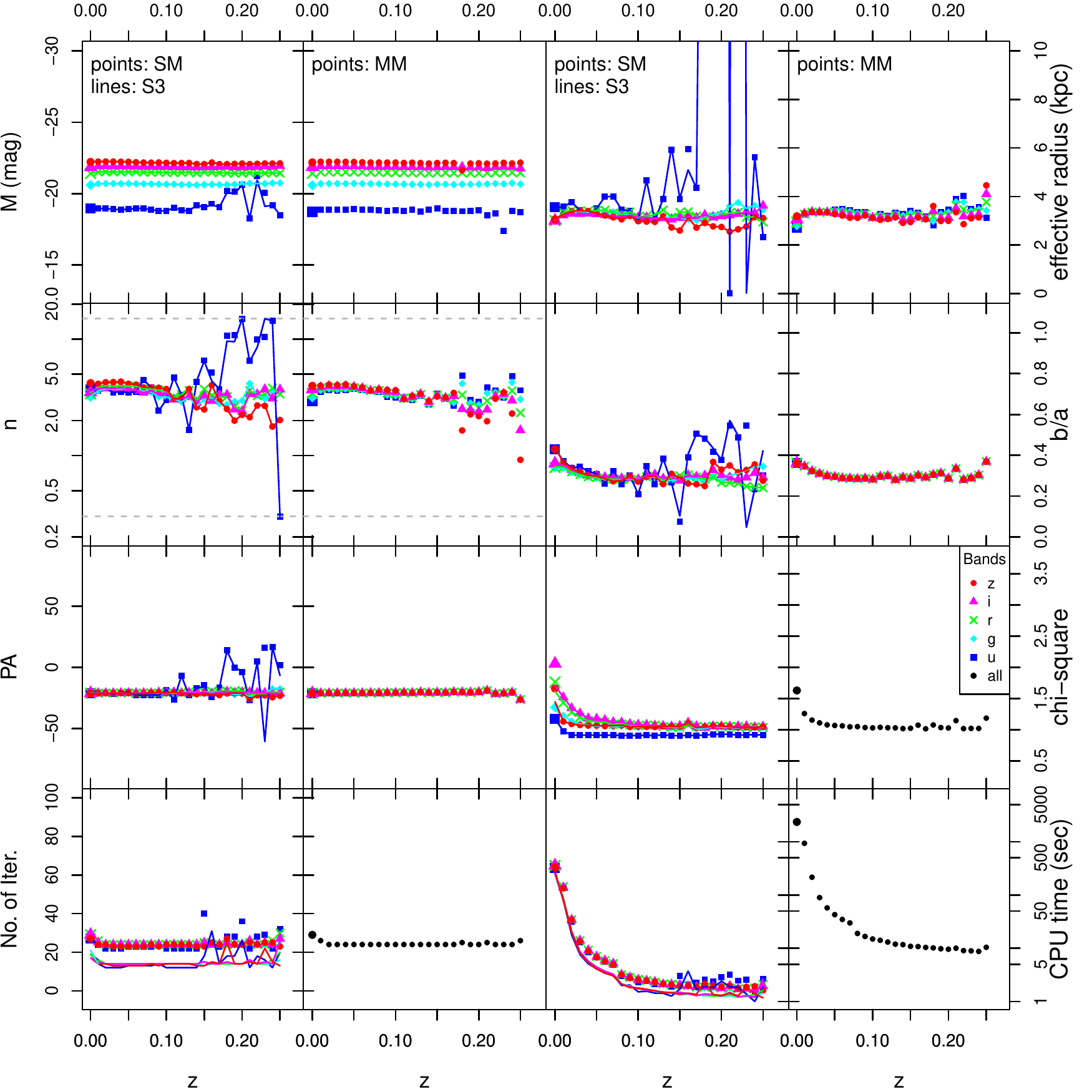}
\includegraphics[height=12cm,width=3.0cm]{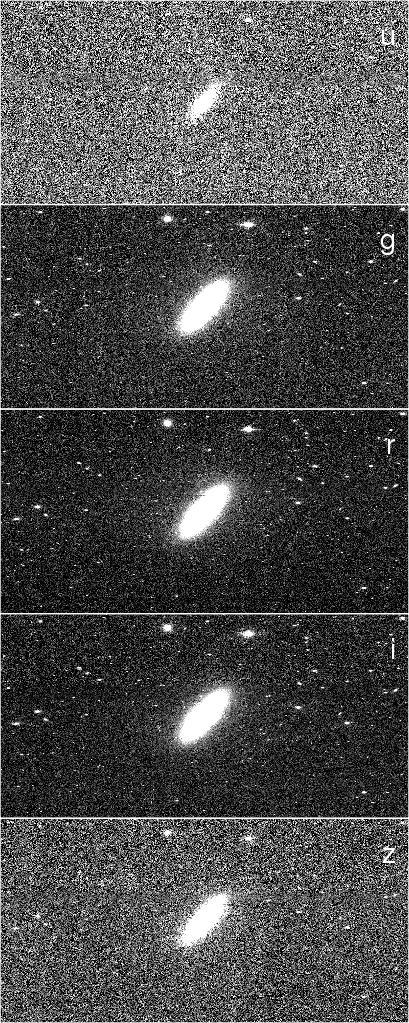}
\caption{Left: A series of plots presenting the variation of recovered parameters as a function of redshift for galaxy NGC4570. First and third columns show the results of single-band fits (points:\galfitm, lines:\galfitthree) while the second and the fourth columns show results of our multi-band fits. The points at redshift zero in each panel give the values for the original galaxy image, while the rest of the points represent the artificially redshifted images. The dashed horizontal lines in the \sersic index panels indicate soft constraint applied on \sersic index. Note that, for example, in case of multi-band fitting the smoothness of the $u$-band magnitudes is a result of the other profile parameters being naturally constrained by the higher signal-to-noise bands, rather than being due to any externally imposed constraints.  Right: The five $u, g, r, i, z$ SDSS images, created with \montage, for the same galaxy.}
\label{fig:gal1}
\end{figure*}

\section{Fitting}
\label{sec:fit}

\subsection{\galfit and \galfitm}
\label{sec:galfitm}
We have adapted \galfit version 3.0.2 for the requirements of this project, with the consent of the original developer, C.~Peng.  To differentiate our modified version from the standard release we refer to it as \galfitm.  For reference, all the work in this paper uses \galfitm version 0.1.2.1.  The code will be publicly released in the near future.  Development is continuing, primarily to improve ease-of-use and incorporate the additional features mentioned below.  However, the general performance of the technique is expected to remain as presented in this paper.

\galfit fits a variety of 2D analytic functions to galaxy images.  Multiple components may be added, for example a detailed model might include a bulge, disc, bar, nucleus and the sky background.  Using the Levenberg-Marquardt (LM) algorithm, \galfit minimises the $\chi^2$ residual between a galaxy image and the PSF-convolved model by modifying the free parameters. The final model is accepted when $\chi^2$ reaches a minimum.  The calculation of $\chi^2$ is weighted by a sigma map, which may either be provided or internally created.  The reader can refer to \citet{tex:PH02,tex:PH10} for a detailed description of \galfit.

The standard version of \galfitthree accepts only one input image with which to constrain the model fit.  It was therefore necessary to make a number of significant modifications to enable the use of multi-band data.  However, most of the original code and its structure is used unchanged, and we intend our modified version to be backward compatible when used with single-band data (see Section \ref{sec:Mvs3} for test of this issue).

Our modified code, \galfitm, can accept an arbitrary number of (pixel-registered) images of the same region of sky at different wavelengths (e.g., we use images in five wavelength bands for this study, while Paper III uses nine bands).  To these images \galfitm fits a single, wavelength-dependent, model.  As for \galfitthree, this model may comprise one or more component functions, each with a number of parameters, e.g. centre position ($x_{\rm c}$, $y_{\rm c}$), magnitude ($m$), effective radius ($r_{\rm e}$), \sersic index ($n$), axial ratio ($b/a$) and position angle ($\theta$) for a single-\sersic function.  To extend these component functions to multi-wavelength, their free parameters are replaced with functions of wavelength. These are chosen to be Chebyshev polynomials, as they possess several convenient properties (see Paper I and III for more details).  In the fitting algorithm, the standard parameters are thus replaced by the coefficients of these polynomials.

For each standard parameter the user can select the maximum polynomial order for which the coefficients are free to vary in the fitting process.  For example, some parameters may be set to have a specific constant value or vary with wavelength in a manner that is fixed, e.g., one might fix $n = 1$ to model an exponential disk component. Other parameters may be allowed to vary as a constant function of wavelength, e.g. one might like to allow the centre $x$ and $y$ coordinates to vary, but require that they are the same in every band.  Still other parameters may be permitted to vary with wavelength as linear, quadratic, or higher-order functions, e.g., one could choose to allow $r_{\rm e}$ to vary quadratically with wavelength, in order to account for colour gradients.  As one allows higher order coefficients to be fit, the function becomes more flexible, such that when the number of free coefficients for a standard parameter is the same as the number of input images, that parameter can effectively vary independently between bands.

The implementation of this technique has been achieved in a general manner, so all of the standard \galfit component functions are available, and all of the standard parameters of those functions are treated identically.   For full details please refer to Paper I.  In this work the polynomials, describing how the model parameters vary between images, are functions of wavelength. However, note that the value corresponding to each band is user-definable.  It is therefore straightforward to use alternatives, e.g. the logarithm of wavelength, as the variable for these polynomials.


With \galfitm, one is faced with a decision of how much freedom to give each parameter to vary with wavelength.  This can be advised by expectations from other studies, physical intuition, and limitations imposed by the signal-to-noise and resolution of the images. A useful approach is to consider the variations seen when the parameters are fit to each image independently. For example, in the top left panel of Fig.~\ref{fig:reffband} we plot the variation of effective radius with wavelength, from single-band fits, for eleven example galaxies.  Six of these galaxies show no trend, one (red line) shows a linear increase, and four of the largest galaxies exhibit a more complex variation of $r_{\rm e}$ with wavelength (see Section \ref{sec:ellipt} for further discussion of this issue). Considering \sersic index, plotted in a similar fashion in the top left panel of Fig.~\ref{fig:nband}, trends of increasing $n$ with wavelength are evident for the majority of objects.


\galfit incorporates constraints, which impose limits on the variation of parameters during the fitting process.  Some of these are hard-coded (such as ensuring that sizes cannot become negative) while others may be optionally specified by the user. Constraints are useful to improve the reliability and efficiency early in the fitting process, by excluding regions of parameter space which are believed to be unphysical or eliminated by other considerations. However, if the minimisation routine repeatedly encounters constraints, this is an indication that a good model fit to the data cannot be achieved. In this case at least some of the resulting parameters will typically lie very close to a constraint boundary.  These parameters for such a fit are likely to be seriously biased, and hence it is sensible to discard them from further analysis. 

The extension to multi-band fitting required the implementation of constraints to be significantly modified in \galfitm. If a proposed parameter step would violate a constraint, \galfitthree typically resolves the conflict by simply setting the offending parameter to the value at the constraint boundary.  For multi-band fits, however, constraints on parameters may be violated in some bands but not others and there is a non-trivial relationship between the standard parameter at a given wavelength and the polynomial coefficients which are the true fit parameters. 

In \galfitm, constraint violations are avoided by simply not changing the offending parameter. For instance, if an individual step would violate a constraint, then the coefficients corresponding to the affected parameter are not changed on that step. The subsequent step may not violate constraints, in which case the parameter will continue to converge toward its optimum value. However, in the standard implementation of GALFITs LM algorithm, if a step successfully reduced the $\chi^2$, then the algorithm is `encouraged' and the next step will be larger. In some cases this can lead to one or two parameters repeatedly attempting to violate constraints and hence becoming `frozen' for the duration of the fit.  To mitigate this, we found that periodically substantially reducing the global step size would often lead to `frozen' parameters being able to take a small step towards their constraint boundary and hence `thawing'.  Tests (see comparison in Figure \ref{fig:gal3vgalm-mag0}) show that this approach works well, and \galfitthree and \galfitm typically return very similar results, even when constraints are important (see Section~\ref{sec:Mvs3} and Paper III). In some cases the parameter continues to migrate to the constraint boundary, so the final fit is treated as unusable.



In order to store the increased outputs in an efficient manner, the FITS file produced by \galfit has been extended for \galfitm. In addition to the original, model and residual images for all bands, this file contains the PSFs and several new tables. These tables contain all the input details and results of the fit, as well as some additional information, including the number of iterations, timings and a record of when constraints were encountered during the fit.



\begin{figure*}
\centering
\includegraphics[height=12cm,width=13.0cm]{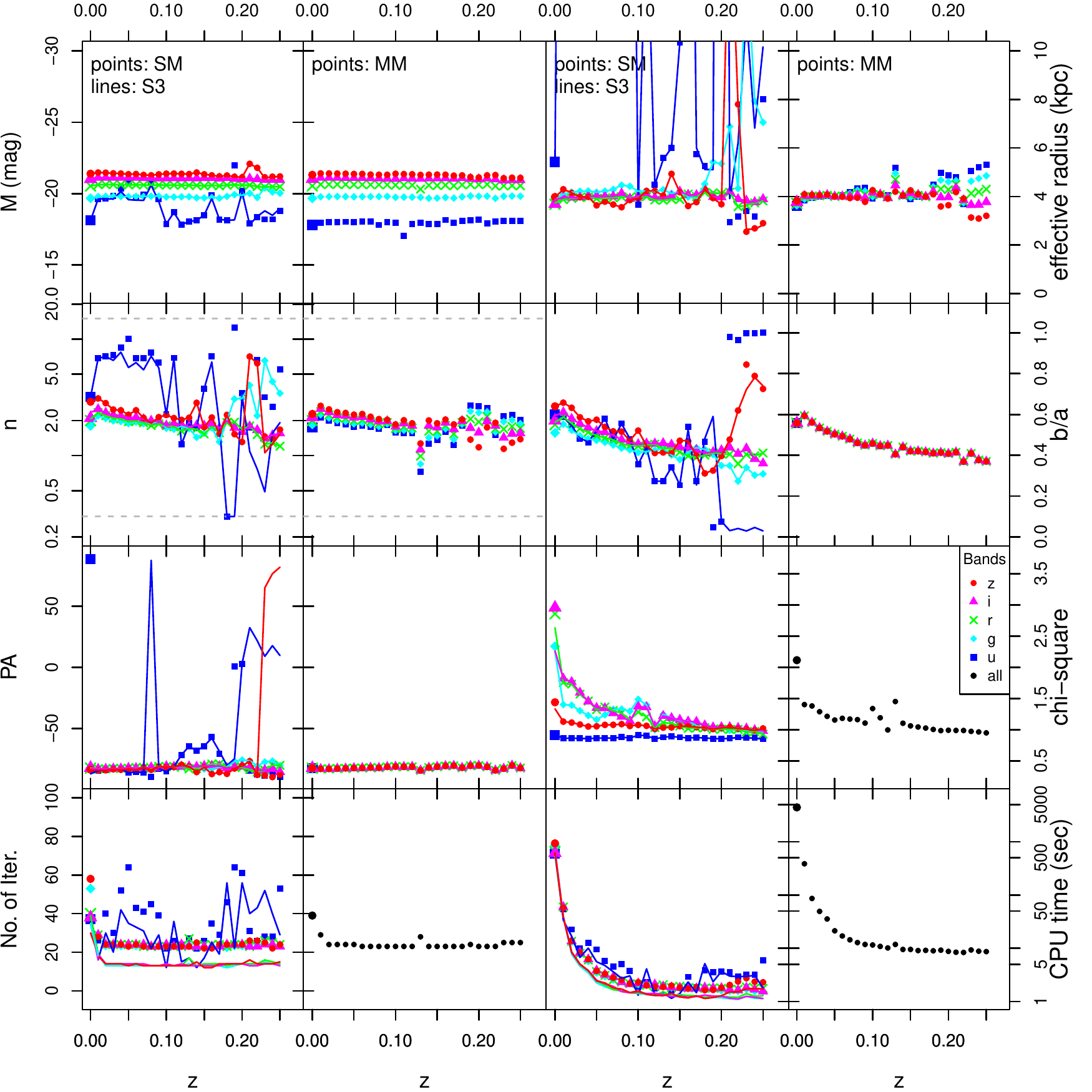}
\includegraphics[height=12cm,width=3.0cm]{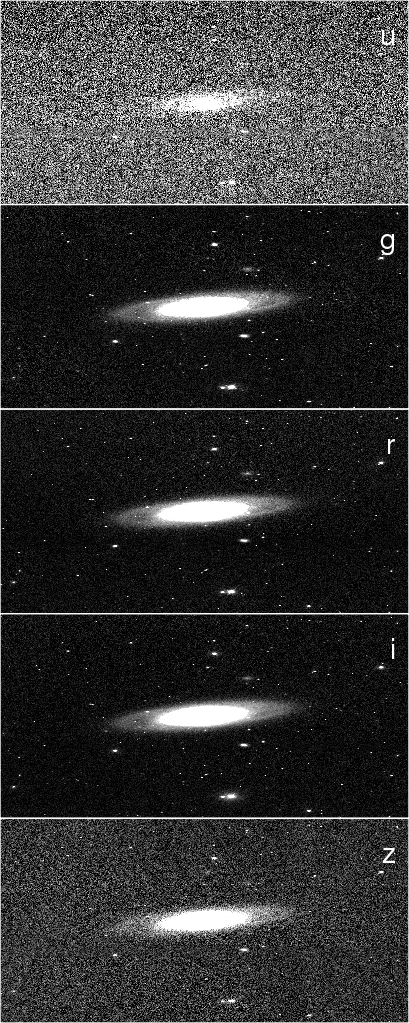}
\caption{A series of plots presenting the variation of recovered parameters as a function of redshift for galaxy NGC4274. The layout is the same as Fig.~\ref{fig:gal1}.}
\label{fig:gal2}
\end{figure*}

\begin{figure*}
\centering
\includegraphics[height=12cm,width=13.0cm]{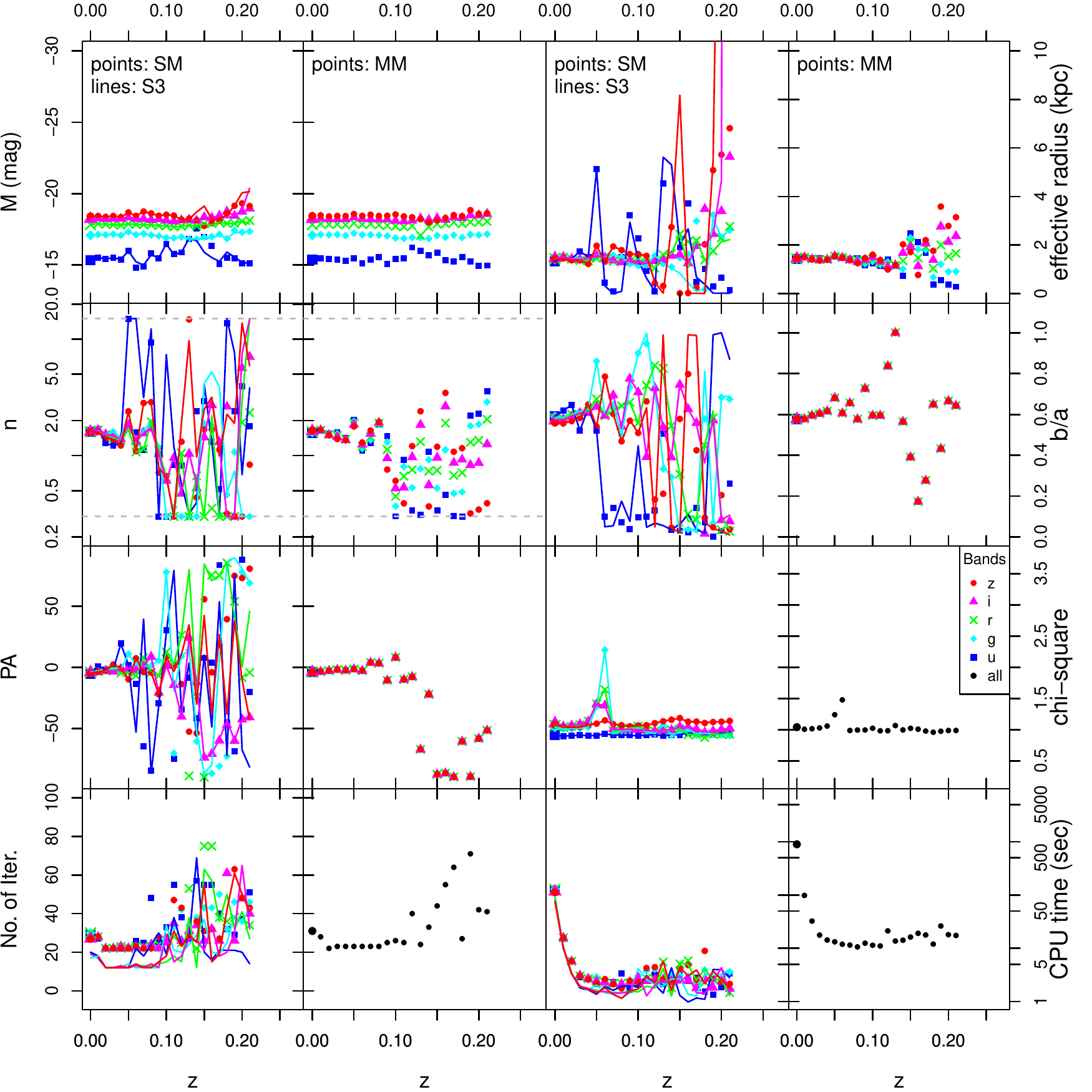}
\includegraphics[height=12cm,width=3.0cm]{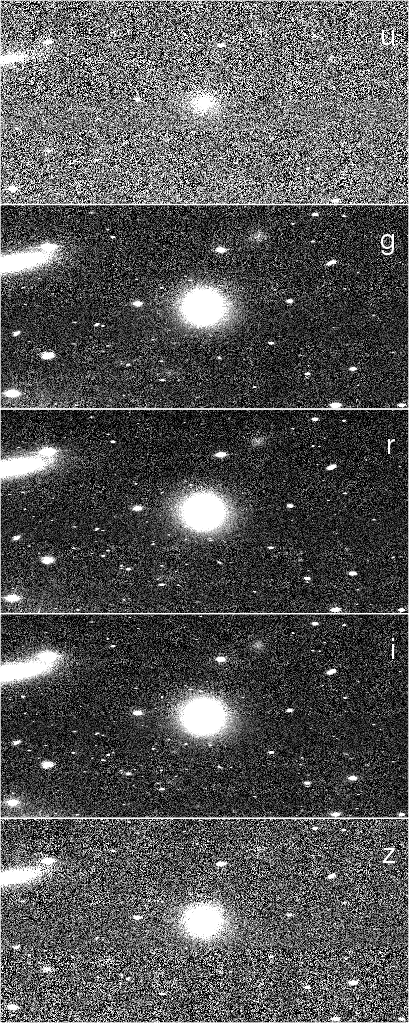}
\caption{A series of plots presenting the variation of recovered parameters as a function of redshift for galaxy NGC4431. The layout is the same as Figs.~\ref{fig:gal1} and \ref{fig:gal2}.  In this example both single- and multi-band fits fail after redshift $\sim 0.07$. In this case we believe that is due to our inability to construct an accurate mask beyond this redshift, i.e. SExtractor no longer deblends two neighbouring objects.  The output model therefore changes shape and magnitude with redshift.}
\label{fig:gal3}
\end{figure*}

\begin{figure*}
\centering
\includegraphics{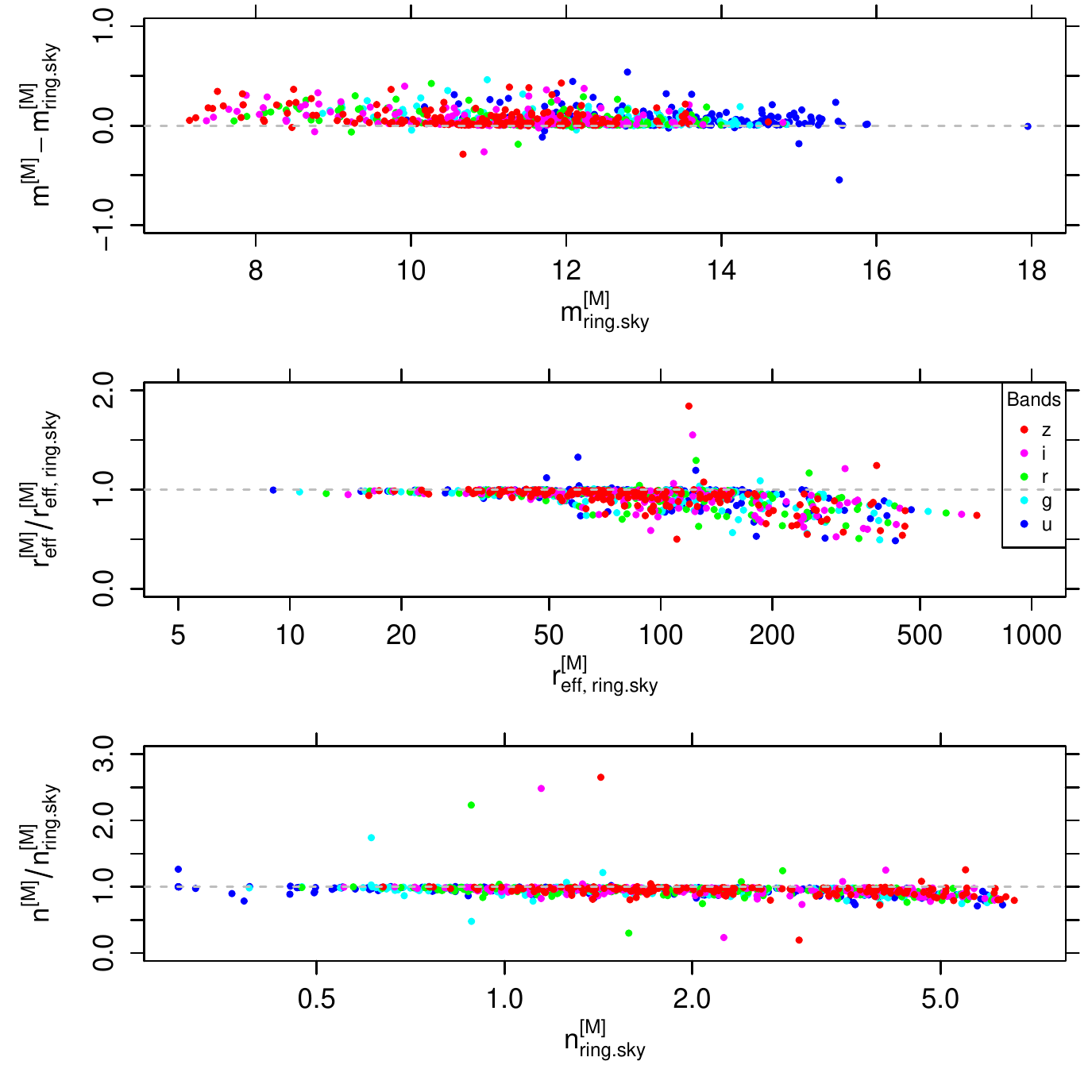}
\caption{Plots showing the difference in parameters from multi-band fitting (MM) when measured using two different sky values, as described in \S \ref{sec:uncert}.  From top to bottom, the panels show difference (or ratio) of magnitude, effective radius and sersic index as a function of the same parameter.}
\label{fig:ellipsky}
\end{figure*}

\subsection{Structural Parameters}
\label{sec:structpar}

We obtain the structural parameters of our galaxies by fitting two-dimensional elliptical single-\sersic models using both \galfitthree and \galfitm. The radial surface brightness profile is given by the \sersic function \citep{tex:S68}:
\begin{equation}
I(r) ~ = ~ I_{\rm e} \exp \left\{ -b_{n}  \left[  \left(  \frac{r}{r_{\rm e}} \right)^{1/n} -1 \right]  \right\} 
\label{eq:sersic}
\end{equation}
where $r_{\rm e}$ is the effective or half-light radius, $I_{\rm e}$ is the intensity at the effective radius, $n$ is the \sersic index and $b_{\rm n}$ is a function of $n$ \citep{tex:GD05}. The term $r_{\rm e}$ describes the physical size of the galaxy and $n$ gives an indication of the concentration of the light distribution.  When $n$ is equal to the values $0.5$, $1$, $4$, the \sersic profile is equivalent to a Gaussian, exponential and \citet{tex:V59} profile, respectively.  Most galaxies are considered to be primarily two component systems, comprising a disk with an exponential ($n=1$) profile, and a bulge, typically well represented by a \sersic function with $n > 1$.  At low resolution, a single \sersic function can do a reasonable job of representing the overall profile of such a two-component system.  In this case, the resulting parameters reflect a combination of both the disk and the bulge.  On the other hand, elliptical galaxies are generally regarded as being single component systems, well described by a \sersic profile (although, this picture too breaks down when considered in detail).



We run three types of fit, each of which is performed on the $u$, $g$, $r$, $i$, $z$ band images for all of our original and redshifted galaxies. For the first (\fita) we fit each band individually with \galfitthree (single-band fitting), for the second (\fitb) we repeat the same fits using \galfitm (again single-band fitting), and finally (\fitc) we fit each galaxy with \galfitm using all the five bands simultaneously (multi-band fitting).

In the multi-band fitting we choose to allow magnitudes to vary completely freely between bands. In practice, this means that we set the wavelength dependence of magnitude to be described by a quartic polynomial, i.e. with as many coefficients as the number of bands.  We allow full freedom as we wish to avoid any potential biases on the recovered magnitudes, and hence colours, which may result from assuming a lower-order polynomial dependence.

For effective radius and \sersic index we elect to allow linear variations with wavelength. In contrast to magnitude, we expect physical variations in these parameters to be smaller and smoother, particularly when compared to their measurement uncertainties and overt the limited optical wavelength range considered in this paper  ($\sim350$--$900$ nm).  This decision is also supported by the outcomes of single-band fitting.  Figures \ref{fig:reffband} and \ref{fig:nband} show the fitting results, effective radius and \sersic index respectively, for eleven example galaxies fitted in our original (left column) and artificially redshifted images (remaining columns).  The single-band results (top row) for most of our galaxies are consistent with constant or linear trends with wavelength for both $n$ and $r_{\rm e}$.  These trends are well-represented by the multi-band fits (bottom row).  Note that $n$, in particular, consistently shows significant trends with wavelength.  Allowing only constant $n$ in the multi-band fits, or imposing structural parameters measured in one band to measure fluxes in the other bands,  would fail to account for this behaviour.  Note that due to low signal-to-noise ratios some recovered parameters in Fig.~\ref{fig:reffband} are out of the plotting area. We choose to not expand the scale of the figure in order to include these individual values but to focus on the reliable values instead.

Finally, we assume that the shapes of our galaxies do not change within the range of wavelength defined by SDSS bands, so we specify the galaxy center, the axis ratio and the position angle to be constant with wavelength.  We have performed an inspection of plots similar to Figs.~\ref{fig:reffband} and \ref{fig:nband} for these parameters, and find no significant evidence to contradict this decision.  These parameters are still free to vary during the fit, just without any variation with wavelength. 

In all the three runs (\fita, \fitb, \fitc) we use the same initial parameters for galaxy center $(x_{\rm c},  y_{\rm c})$, magnitude $(m)$, effective radius ($r_{\rm e}$), \sersic index $(n)$, axis ratio $(b/a)$, position angle ($\theta$) and sky background value (although different values are used for each galaxy image, as described below). 

First, we fit the sample of original galaxy images, produced directly from SDSS data using \montage. The initial magnitude values for the minimisation are taken from the SDSS aperture magnitudes\footnote{http://cas.sdss.org/astro/en/tools/crossid}, the center ($x,y$) initial values where set to the middle of each image\footnote{The \montage images are centred on each galaxy.}, while for the rest of the parameters we use constant initial values for all the galaxies in all the bands ($r_{\rm e}=10$ pixels, $n=1.5$, $b/a=0.5$, $\theta=0$). The initial effective radius value was chosen to be smaller than all the measured effective radii, relying on \galfit's ability to accurately determine the size of a well-resolved galaxy independent of the initial value.  All the parameters are allowed to vary during the fitting process, with the exception of the sky background, which is kept fixed. We additionally apply a soft constraint on \sersic index, requiring $n$ to vary within the range $0.3$ to $15$, and on the center, such that $x$ and $y$ may vary by no more than the square root of their initial values.  These constraints are not important for the high quality original images, but we include them to be consistent with the fits to the artificially redshifted images.

Once we have established accurate fits for the original images, we repeat the same three runs of the fitting process, as described above  (\fita, \fitb, \fitc), on the artificially redshifted images. The output parameter values from the \galfitm multi-band fits (\fitc) to the original images (given in Table \ref{table:first}) are used as initial values for fitting the lowest redshift artificial images.  The output parameters from these fits were then used to derive initial values for all the higher redshift images.  The initial values for \sersic index, axis ratio and position angle were the same at all redshifts as, in principle, the geometry should remain constant with redshift.  Apparent magnitude and effective radius were cosmologically adjusted, such that the initial absolute values of these quantities are constant with redshift.  In order to ensure we do not introduce wavelength trends through the initial parameters, we use the same initial values of effective radius, \sersic index, axis ratio and position angle for all the bands.  These are therefore initially constant with wavelength, even if they are subsequently allowed to vary with wavelength during the fitting process. The common initial value is determined by taking the median of the values for each band.

For each galaxy we visually inspect all recovered parameters for both original and artificial images. In Figs.~\ref{fig:gal1}, \ref{fig:gal2} and \ref{fig:gal3} we present a summary of the results for three example galaxies  together with the five $u, g, r, i, z$ original images.  Similar plots are available for all the 163 nearby galaxies. In these figures, the first and the third columns show the results of the single-band fits (points: \fitb, lines: \fita) while the second and the fourth columns show the results of the multi-bands fits (\fitc).  The first row of panels shows the absolute magnitude ($M$) and effective radius, where the fit was successful, both for original and artificially-redshifted images.  The second row shows the \sersic index and axis ratio, while the third row shows presents the position angle and the minimised chi-square. Finally, the last row shows the number of iterations needed to fit each galaxy and the cpu time taken by the fitting process.  For simplicity, in these figures, absolute magnitude and effective radius are determined assuming distances simply derived from the observed redshift and adopted cosmology.  The values shown in these figures are therefore different to those shown in Section~\ref{sec:ellipt} and Table~\ref{table:ellipt}, where accurate distance estimates are used.

Figure \ref{fig:gal1} presents the recovered structural parameters for the galaxy NGC4570.  The first thing to mention is that, ignoring $u$-band for a moment, the trends with artificial redshift are pleasingly flat for both single- and multi-band fits.  We also note that the results for the single-band approach agree very well between \galfitthree (\fita, lines in columns 1 and 3) and \galfitm (\fitb, points in columns 1 and 3). The results for single- and multi-band (\fitc, points in columns 2 and 4) techniques also generally agree very well.
If we now consider the $u$-band, the lowest signal-to-noise band, we see that this too agrees between all fit types to $z \sim 0.06$. However, for the single-band methods, beyond $z \sim 0.06$, first $n$ begins to show an increasing scatter, then $r_{\rm e}$ and $b/a$, and finally $M$.  Other bands start to show significant increases in scatter for $z \ga 0.11$, particularly for $n$.  In contrast, the multi-band fitting results remain reliable for all bands to higher redshifts.  Significantly increased scatter does not set in for any parameters until $z \ga 0.16$.

The more reliable behaviour of our multi-band method is a result of their structural parameters being constrained by more data, with many of the values required to display a degree of consistency between bands.  Remember that there is no restriction on the variation of magnitude with wavelength, but effective radius and \sersic index may only vary linearly, while axis ratio and position angle are constant with wavelength.  The behaviours seen in Fig.~\ref{fig:gal1} are typical for many of the galaxies in our sample.

We show two more examples, for NGC4274 in Fig.~\ref{fig:gal2} and NGC4431 in Fig.~\ref{fig:gal3}. NGC4274 is another case where the individual $u$-band fits show a large scatter, while the multi-band results remain sensible, and consistent with those obtained for the original image.  Again, our multi-band approach also reduces the scatter in the other bands, enabling more sensible results to be recovered for high artificial redshifts, where the data quality is very poor. NGC4431 is a much less luminous galaxy, for which at $z \ga 0.05$ both single- and multi-band fitting have high scatter and the structure of the galaxy cannot be extracted.  Prior to this, all methods produce reasonable results. For NGC4431, and similar cases where single- and multi-band fitting processes failed to give consistent results with redshift, we investigated the reason by examining the galaxy images, input masks, output models and residuals (via figures similar to Fig.~\ref{fig:ferengi}). In all cases we found that the divergencies in the derived parameters were due to the main galaxy becoming blended with nearby objects or being undetectable in multiple bands.

In general, multi-band fitting improves the stability of the results, particularly for the lower signal-to-noise bands, increasing the distance out to which meaningful multi-wavelength structural information can be recovered for a galaxy of a given luminosity.  However, it cannot help in cases where the galaxy becomes unresolved or undetected in multiple bands (i.e., more than two of the five used in this work). The parameter that typically shows the most improvement is effective radius, followed by \sersic index and then magnitude.  Even though the multi-band method improves the reliability of the parameters, it is fairly common for galaxies in our sample to show systematic trends in some parameters with artificial redshift.  These are usually present to a similar degree with all fitting methods.  We believe that these are mostly a result of the decreasing image quality.  However, some of these trends may be attributable to the inability of the \sersic function to represent the profile of some galaxies, as well as deblending and sky measurement issues.  The observed systematic biases with redshift do not obey a consistent pattern that we have been able to discern, with the exception of \sersic index.  This commonly decreases as the galaxy becomes artificially redshifted to greater distances.  This issue is discussed further in Section~\ref{sec:MvsS}.

 In order to demonstrate the capabilities of the MM approach for more accurately measuring the physical properties of galaxies, we chose to limit some parameters freedom to vary with wavelength, with reference to the trends seen in the results of our single-band fits.  As we will see in the results below, these assumptions allow us to more accurately recover parameters in noisy images.  However, it is important to recognise that, while these constraints increase the stability of the fits, there is also the risk of introducing systematic biases in cases where the true wavelength dependence of the profile does not correspond to that assumed.  Low signal-to-noise images would be most susceptible to such systematics as the parameters would be mostly constrained by the higher signal-to-noise data.

For instance, if the light profiles of galaxies in the (noisy) $u$-band are significantly different to that in $gri$, restricting the effective radius to linear variations with wavelength would impose the structure determined in $gri$ on $u$.  As well as introducing a bias on $r_{{\rm e},u}$, this may also result in a systematic error in magnitude, despite it being allowed to vary freely. On the other hand, if we had tried to fit the low signal-to-noise $u$-band image independently, the much-reduced information regarding the center and profile of the galaxy would result in large statistical uncertainties, which would propagate into uncertainties in the magnitude.

Our method produces measurements for the parameters in all bands.  However, for low signal-to-noise images there will be a partial interpolation or extrapolation from the higher signal-to-noise data.  What signal there is in the low signal-to-noise bands will have some influence on the fit.  Any systematic bias will therefore be less than would have resulted from an extrapolation based only on the high signal-to-noise bands.  Such systematics may be reduced by giving the model more freedom to vary with wavelength, at the cost of increased statistical uncertainties on the resulting parameters in the low signal-to-noise bands.  A further consideration is that with low signal-to-noise data there is a danger of overfitting, for example, the \sersic profile adapting to fit a noise peak rather than the galaxy.  Even when fitting the bands independently, the possibility still remains for significant systematic errors.

A compromise must therefore be made by any user of our MM method.  One can allow more freedom for parameters to vary with wavelength, giving (perhaps) less potential for systematics biases, but resulting in higher statistical uncertainties; or vice versa.  The flexibility of our multi-band fitting approach allows the user to balance systematic and statistical uncertainties by using independent observational results, physical insight and knowledge of their dataset to guide their assumptions for how galaxy parameters vary with wavelength.





\begin{figure*}
\centering
\includegraphics[height=13cm,width=14.0cm]{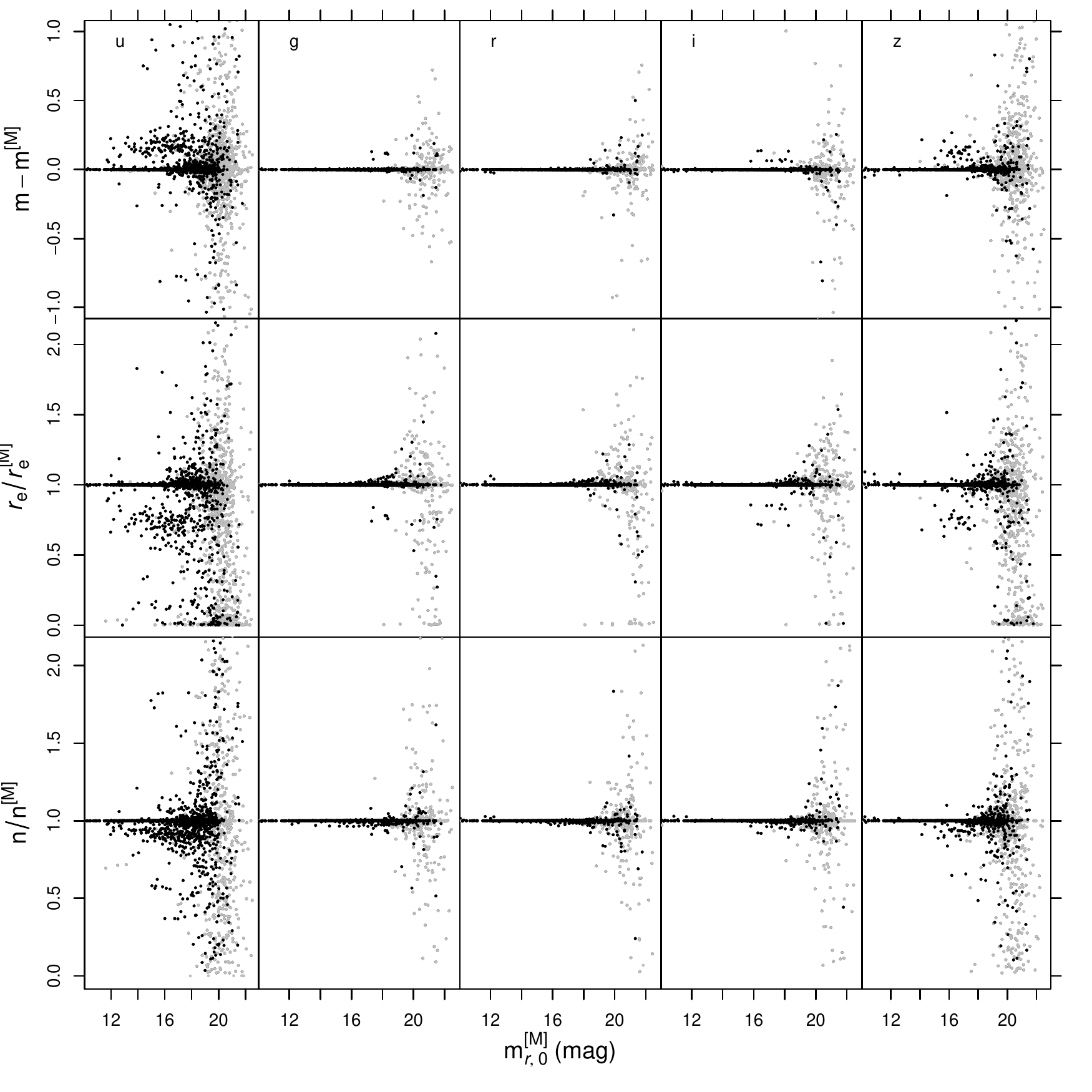}
\caption{Comparison of \galfitthree single-band (\fita) fits versus \galfitm single-band (\fitb) fits. This plot shows the recovered \sersic model parameters as a function of the $r$-band apparent magnitude as cosmologically scaled from the \fitb fit to the lowest-redshift artificial image for each galaxy (using this magnitude, rather than the value measured for each image, avoids scatter on the recovered parameters from entering the horizontal axis of this plot). Upper panels: difference in magnitude; middle panels: ratio of effective radii; lower panels: ratio of \sersic indices. Only parameters for the artificially-redshifted images are included in this figure. The grey points indicate that at least one fit (\fita or \fitb) ended in close proximity to a constraint, and for which the recovered parameters are hence deemed to be unreliable.}
\label{fig:gal3vgalm-mag0}
\end{figure*}

\subsection{Uncertainties on structural parameters}
\label{sec:uncert}
\galfit internally computes statistical errors from the covariance matrix constructed by the Levenberg-Marquardt algorithm.  These errors would be statically correct if the only source of uncertainty were Poisson noise. However, in reality they are substantially underestimated, as they do not take into account the parameter dependencies on (external) sky estimation, PSF accuracy, model mismatch and correlated noise. For example, variations within the acceptable range of sky values may result in systematic parameter uncertainties much greater than the formal \galfit errors \citep{tex:VD12}.

We determine the sky value in the original galaxies using two different methods, in order to obtain a rough indication of how this uncertainty affects the structural parameters.  Figure \ref{fig:ellipsky} shows the parameter differences for the two sky measurements, which are described in Section \ref{sec:reduction}. The purpose of this comparison is not to fully study the parameter dependency on sky, as this has been done by various other studies (e.g., \citealt{tex:HM07}), but rather to provide an indicative uncertainty value for our measurements. The two different sky estimations lead to uncertainties per band ($u$, $g$, $r$, $i$, $z$) for $m$ ($\pm0.13$, $\pm0.09$, $\pm0.1$, $\pm0.11$, $\pm0.12$), $r_{\rm e}$ ($\pm12$\%, $\pm11$\%, $\pm12$\%, $\pm14$\%, $\pm15$\% ) and $n$ ($\pm9$\%, $\pm11$\%, $\pm14$\%, $\pm15$\%, $\pm17$\%). These uncertainties refer only to the original (\montage-derived) images using multi-band fitting and not the redshifted images. They are reproduced at the top of Table \ref{table:first}.
 
As the purpose of this paper is focus on fitting single-\sersic functions, we do not attempt to fit more complex models, or derive the parameters of bulges and disks separately.  This will be the subject of a subsequent paper.  Despite their simplicity, single-\sersic models are often capable of giving a good overall representation of galaxy surface brightness profiles, even in the case of multi-component systems.  In such cases the values for \sersic index and effective radius will reflect the superposition of the components.  For example, an $n \sim 4$ bulge together with an exponential disk of similar brightness, will usually result in an intermediate $n$ for the single-\sersic fit, e.g., $n \sim 2.5$.  On the other hand, elliptical galaxies are often considered to be single-component systems, well represented by a single-\sersic at all radii.

\galfit assumes that the intrinsic profile being fit is perfectly represented by the assumed model.  Mismatches between the real profile and the model, e.g., as in the case of galaxies with a bulge and disk, produce an additional contribution to the uncertainties in model parameters, beyond those estimated by \galfit.  Furthermore, model mismatch may make model parameters more sensitive to systematic biases as a function of signal-to-noise and/or resolution, as mismatches become less significant as image quality decreases.  This issue is discussed further in the following section.

\begin{figure*}
\centering
\includegraphics[height=13cm,width=14.0cm]{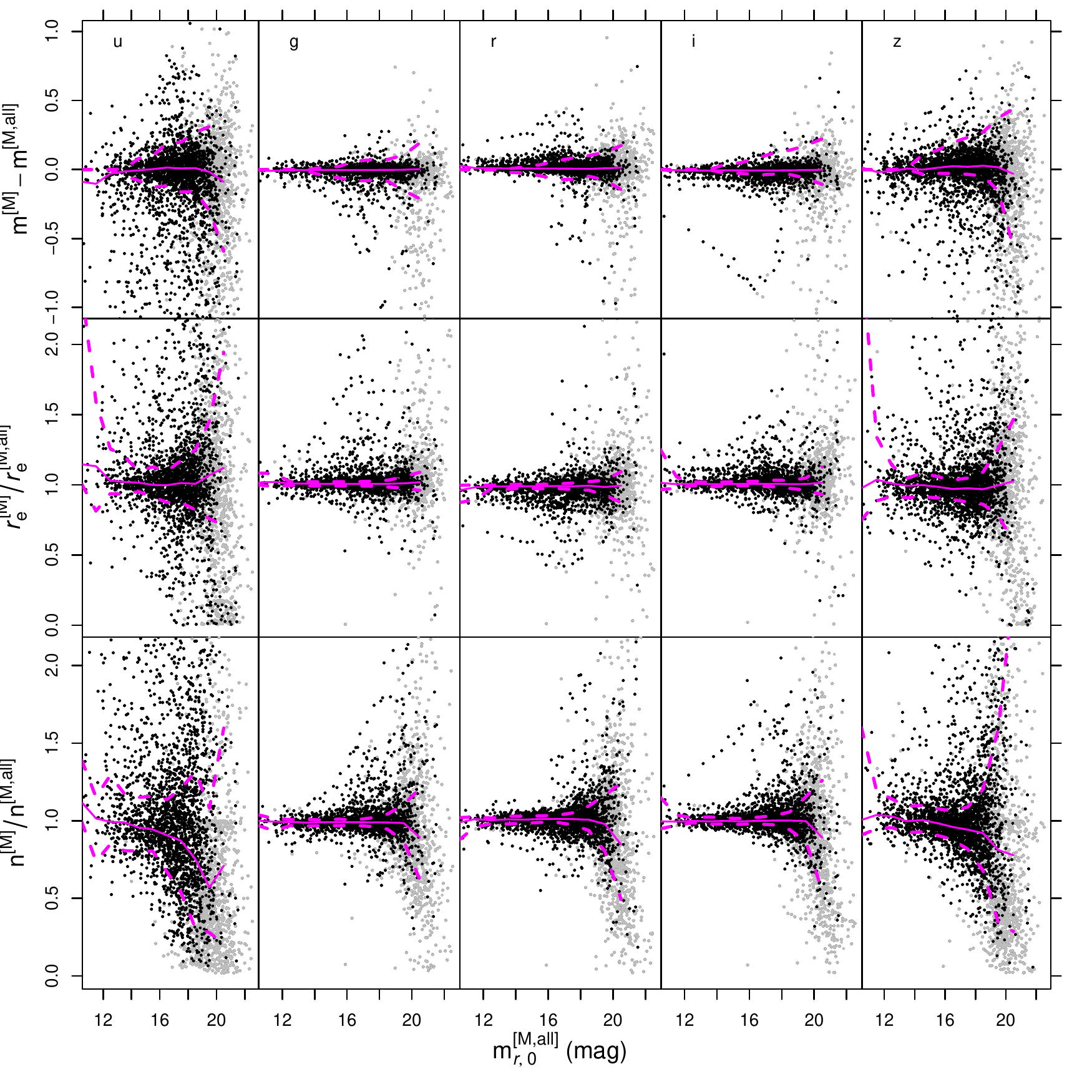}
\caption{Comparison between \galfitm single-band (\fitb) fits versus \galfitm multi-band (\fitc) fits, plotted in a similar manner to Fig.~\ref{fig:gal3vgalm-mag0}. We show the recovered \sersic model parameters as a function of the $r$-band apparent magnitude as cosmologicaly scaled from the \fitc fit to the lowest-redshift artificial image for each galaxy.  Columns show results for $u$, $g$, $r$, $i$, $z$ bands, respectively. Upper panels: difference in magnitude; middle panels: ratio of effective radii; lower panels: ratio of \sersic indices.  Values of $m^{[{\rm M}]} -m^{[{\rm M} ,\rmn{all}]} > 0$, $r_{\rm e}^{[{\rm M}]}/r_{\rm e}^{[{\rm M} ,\rmn{all}]} > 1$ and $n^{[{\rm M}]}/n^{[{\rm M} ,\rmn{all}]} > 0$, indicate that the magnitude is fainter, effective radius larger and \sersic index greater, respectively, when measured in the single-band image compared to that obtained using the multi-band approach. The solid lines indicate the median, while dashed lines are the 16th- and 84th-percentiles of the distribution, all determined within one magnitude bins. 
Only parameters for the artificially-redshifted images are included in this figure. The grey points indicate that at least one fit (\fitb or \fitc) ended in close proximity to a constraint, and for which the recovered parameters are hence deemed to be unreliable. }
\label{fig:galmvgalm-mag0}
\end{figure*}

\section{Results}
\label{sec:results}

\subsection{\galfitm versus \galfitthree}
\label{sec:Mvs3}
Having applied various modifications to the original \galfitthree, we first wish to ascertain that \galfitm continues to return the same (or at least similar) results when applied to the same data with an identical configuration. We therefore compare the single-band \galfitthree (\fita) outputs with those from single-band \galfitm (\fitb) and, investigate the cause of any differences.

In Fig.~\ref{fig:gal3vgalm-mag0}, we compare the derived parameters from the single-band fits for the artificially redshifted sample only. We plot the difference between the \galfitthree parameters ($n$, $r_{\rm e}$, $m$) and  \galfitm parameters ($n^{[{\rm M}]}$, $r_{\rm e}^{[{\rm M}]}$, $m^{[{\rm M}]}$) as a function of the `true' apparent $r$-band magnitude.  This magnitude ($m_{r,0}^{[{\rm M}]}$) is determined by cosmologically transforming the value obtained from the single-band fit to the most local artificially-redshifted image.  Using this magnitude, rather than the value measured for each image, avoids scatter on the recovered parameters from entering the horizontal axis of this plot.  Each column shows a different band, from $u$ on the left, to $z$ on the right. We include all the parameters even when they are in close proximity to a constraint. However, we indicate these unreliable fits by plotting their points in grey.  For simplicity we do not show the parameters derived from the original \montage images.

As can be seen from Fig.~\ref{fig:gal3vgalm-mag0}, for the majority of the objects both methods return the same results, thus we get a straight horizontal line at zero or one. We find that the fraction of galaxies for which the magnitude difference is less that $0.01$~mag is 58 per cent, while this fraction increases to 71 per cent for $\Delta m < 0.05$~mag. Most of the more deviant points we find are consistent within $\Delta m < 0.1$~mag up to $m_r  \sim 17$~mag for $u$- and $z$-bands and $m_r \sim 20$~mag for the rest of the bands. As galaxies get fainter we notice a gradual increase of the scatter, which becomes large after $m_r > 20$. In the second row of panels, for  $r_{\rm e}$, we also notice that there is some fraction of galaxies with ratio very close to zero. This is caused by cases where the recovered \galfitthree effective radius is close, or equal, to the lower-limit constraint at $0.1$~pixels, whereas for \galfitm these values are avoided.

The main reason for differences in the recovered parameters between \galfitthree and \galfitm is due to the handling of constraints between the different codes, although there may also be contributions from a number of more minor modifications (see Section \ref{sec:galfitm} and Paper I). From inspection of Fig.~\ref{fig:gal3vgalm-mag0} we note that much, but not all, of the scatter in these plots is due to objects that have their parameters derived from fits which end up on constraints. Thus, the scatter in this plot, and the significant differences between \galfitm and \galfitthree tend to occur only for galaxies which have poorly-constrained fits in \galfitthree, and which are probably sensitive to the initial conditions, in addition to the details of the fitting routine.

 Finally, note that there is a cloud of points, particularly noticeable in the $u$-band, which is systematically offset to fainter magnitude, smaller size and lower \sersic index.  This contains the majority of objects which disagree between \fita and \fitb, but for which the parameters do not end on a constraint boundary, and corresponds to a variety of objects and redshifts. 
We note that this feature is much more prevalent in low signal-to-noise images, where the $\chi^2$ surface is likely to be complicated, with multiple minima, but we are not yet entirely certain of the reasons for its appearance.  However, all of the fits in the cloud have encountered constraints during the fitting process, and we strongly suspect that differences in the way constraints are handled by \galfitm and \galfitthree are responsible.  We believe that the approach taken by \galfitm is superior to that of \galfitthree, particularly in low signal-to-noise cases.

With noisy data it is quite common for the first few minimisation steps to be large and encounter constraints.  In such cases \galfitthree will typically continue minimising from the constraint boundary, while \galfitm will continue from the previous acceptable value of the offending parameter.  The two approaches may therefore converge to different minima.  However, the one found by GALFITM is more likely to be in a `reasonable' part of parameter space.  Alternatively, in both \galfitm and \galfitthree it is conceivable for the fit parameters to become `frozen' due to repeatedly encountering constraints in a noisy and complicated $\chi^2$-space.  As described earlier, \galfitm attempts to avoid this by periodically taking a very small step, which may lead to systematic differences in its results versus \galfitthree.

 Even though the reason for these occasional systematic differences is somewhat unclear, we have confirmed that \galfitm does indeed produce better results than \galfitthree in such cases, by comparing the final $\chi^2$ values for \fitb and \fita.  These are generally consistent, but in cases where the two methods return very different parameters the \fitb fit usually has a lower $\chi^2$, and is therefore a better representation of the data.

\begin{figure*}
\centering
\includegraphics[height=13cm,width=14.0cm]{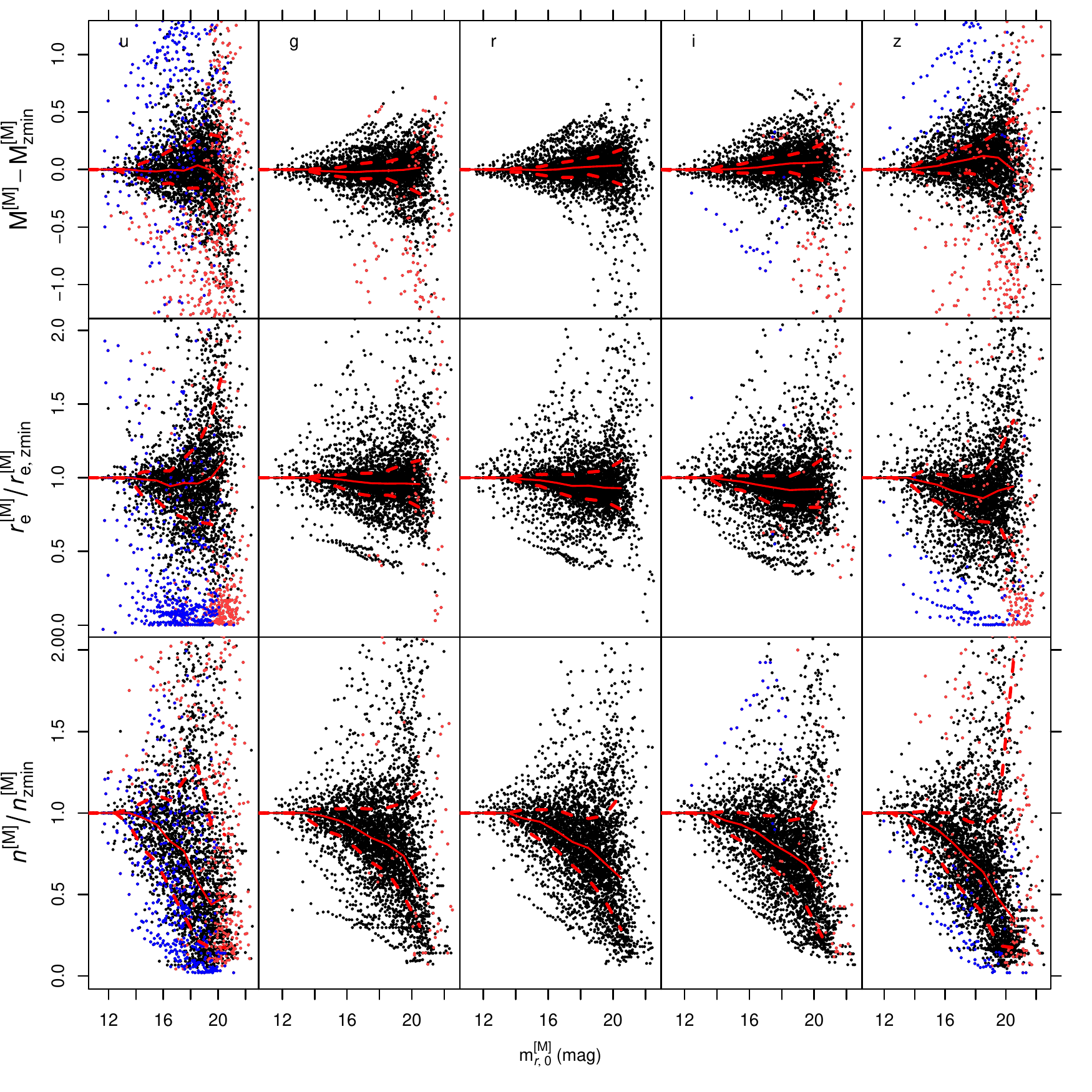}
\caption{Parameter deviations for single-band \galfitm fits (\fitb) relative to the lowest redshift ($z_{\rm min}$) fit, plotted as a function of the $r$-band apparent magnitude, as cosmologically scaled from the \fitb fit to the lowest-redshift artificial image for each galaxy. Values of $M^{[{\rm M}]} -M^{[{\rm M}]}_{\rm zmin} > 0$, $r_{\rm e}^{[{\rm M}]} /r^{[{\rm M}]}_{\rm e,zmin} > 1$ and $n^{[{\rm M}]} /n^{[{\rm M}]}_{\rm zmin} > 1$, indicate that the magnitude is fainter, effective radius larger and \sersic index greater, respectively, when measured in the higher-redshift image compared to that obtained from the $z_{\rm min}$ image. The black points denote the recovered parameters for the 3863 artificially-redshifted galaxies.  The blue points denote the galaxies that  $r_{{\rm e},x,{\rm zmin}^{[{\rm M}]}}/r^{[{\rm M}]}_{{\rm e},r,{\rm zmin}} > 3$ where $x$ represents the band in consideration. The number of `blue' outliers is 464, 0, 25 and 100 for the $u, g, i, z$ bands, respectively. The red points are additional outliers selected with the $r_{{\rm e},x}^{[{\rm M}]}/r^{[{\rm M}]}_{{\rm e},r} > 3$ or $r_{{\rm e},x}^{[{\rm M}]}/r^{[{\rm M}]}_{{\rm e},r} <0.33$ criteria (see text for more details).  The number of `red' outliers not included in the `blue' sample are 596, 144, 99, 323. The solid red lines indicate the median, while dashed red lines are the 16th- and 84th-percentiles of the distribution, all determined within one magnitude bins.  The magnitudes and effective radii obtained from the redshifted images have been cosmologically converted to their absolute values at the true redshift of each galaxy for the purpose of these comparisons.}
\label{fig:indiv-zmin}
\end{figure*} 

\begin{figure*}
\centering
\includegraphics[height=13cm,width=14.0cm]{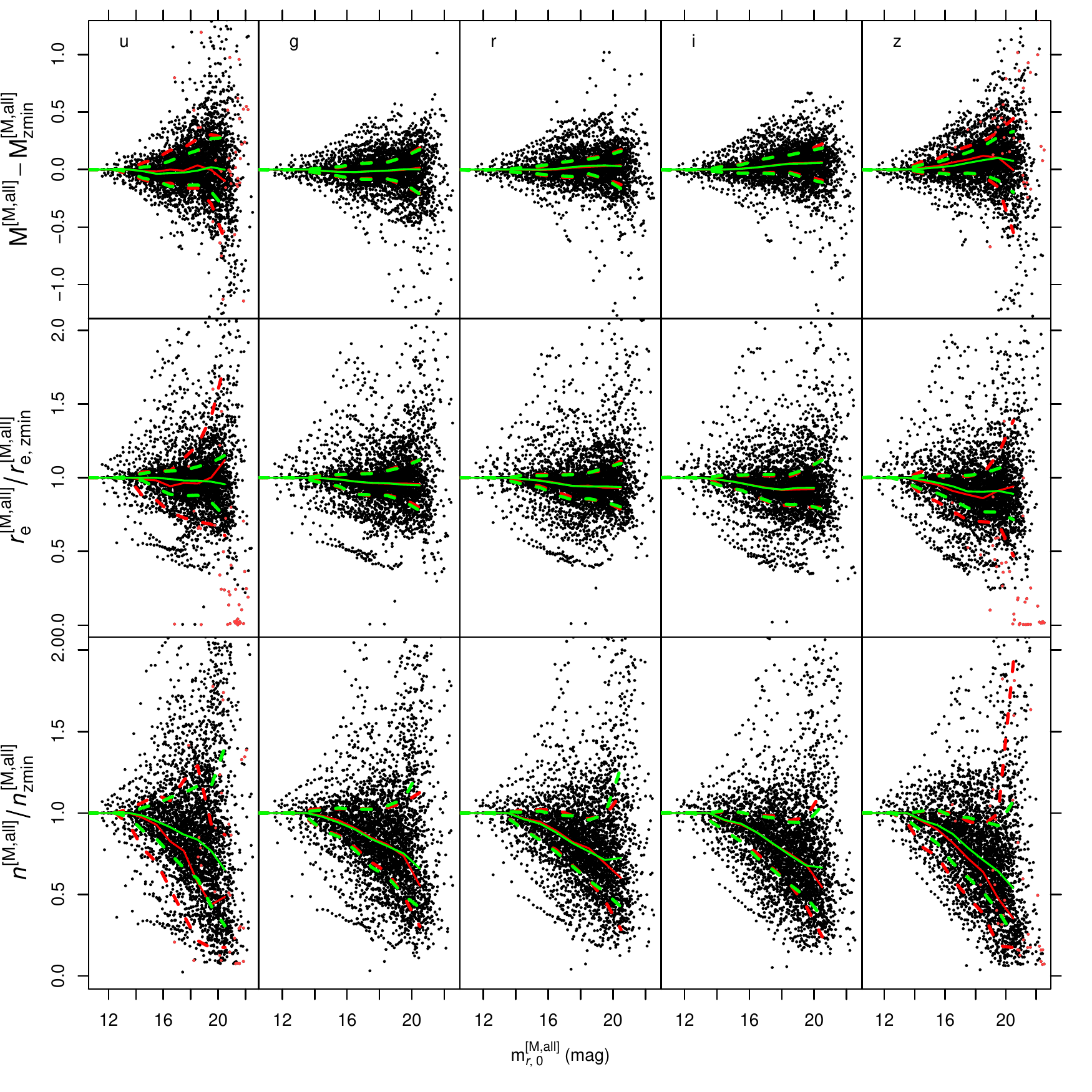}
\caption{In a similar manner to Fig.~\ref{fig:indiv-zmin}, this figure shows parameter deviations for multi-band \galfitm fits (\fitc) relative to the lowest redshift ($z_{\rm min}$) fit, plotted as a function of the $r$-band apparent magnitude, as cosmologically scaled from the \fitc fit to the lowest-redshift artificial image for each galaxy. The black points denote the recovered parameters for the 3863 artificially-redshifted galaxies. The red points have been selected in the same way as Fig.~\ref{fig:indiv-zmin}, but using \fitc data in this case. The number of the red outliers is 43, 0, 0 and 43 for $u, g, i, z$ bands respectively. We do not find any blue outliers using \fitc data. The solid green lines indicate the median, while dashed green lines are the 16th- and 84th-percentiles of the distribution, all determined within one magnitude bins.  To more easily compare to the \fitb fits, we include the lines from Fig.~\ref{fig:indiv-zmin} in red.  For consistency, we calculate the green lines using exactly the same galaxies as were used in determining the red lines. The magnitudes and effective radii obtained from the redshifted images have been cosmologically converted to their absolute values at the true redshift of each galaxy for the purpose of these comparisons.}
\label{fig:combzmin}
\end{figure*}

\begin{figure*}
\centering
\includegraphics[height=13cm,width=14.0cm]{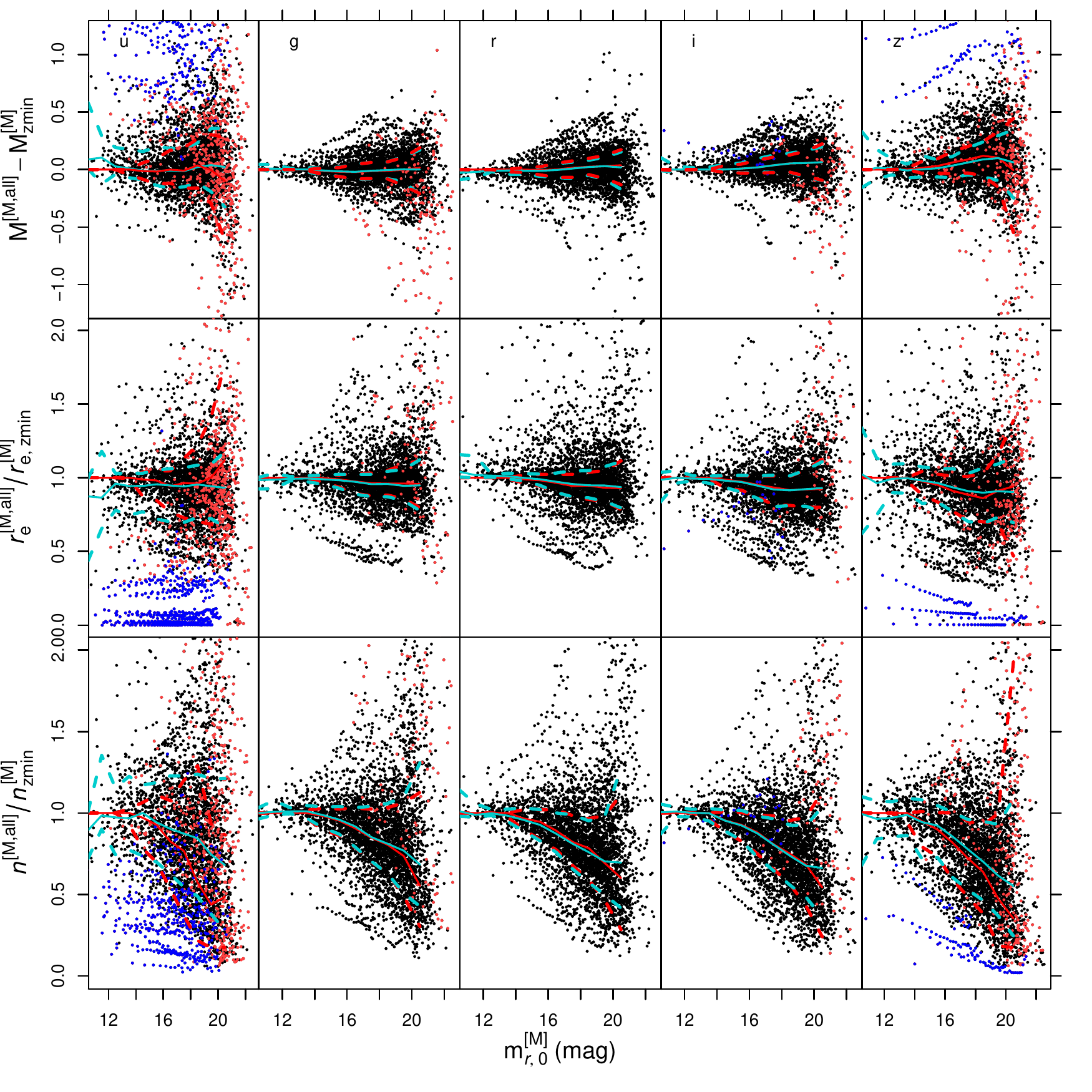}
\caption{ In a similar manner to Fig.~\ref{fig:indiv-zmin}, this figure shows parameter deviations for multi-band \galfitm fits (\fitc) relative to the lowest redshift ($z_{\rm min}$) \fitb fit, plotted as a function of the $r$-band apparent magnitude, as cosmologically scaled from the \fitc fit to the lowest-redshift artificial image for each galaxy. The black points denote the recovered parameters for the 3863 artificially-redshifted galaxies.  The red and blue points have been selected in the manner way as in Fig~\ref{fig:indiv-zmin}, based only on the \fitb results, since this method is responsible for their outlying values. The solid cyan lines indicate the median, while dashed cyan lines are the 16th- and 84th-percentiles of the distribution, all determined within one magnitude bins.  To more easily compare to the \fitb fits, we include the lines from Fig.~\ref{fig:indiv-zmin} in red.  For consistency, we calculate the green lines using exactly the same galaxies as were used in determining the red lines.  The magnitudes and effective radii obtained from the redshifted images have been cosmologically converted to their absolute values at the true redshift of each galaxy for the purpose of these comparisons.}
\label{fig:mixzmin}
\end{figure*}

\subsection{Multi-band versus single-band}
\label{sec:MvsS}
Having checked that applying \galfitm to single-band data (\fitb) is sufficiently consistent with \galfitthree (\fita), we now move on to compare the performance of single- versus multi-band methods.  In particular, we wish to investigate how multi-band fitting modifies the uncertainties on the measured parameters, and check whether it introduces any systematic deviations compared with the single-band results. In Fig.~\ref{fig:galmvgalm-mag0}, we compare the parameters from single-band fits using \galfitm (\fitb; $n^{[{\rm M}]}$, $r_{\rm e}^{[{\rm M}]}$, $m^{[{\rm M}]}$) with those from multi-band fits (\fitc; $n^{[{\rm M},\rmn{all}]},r_{\rm e}^{[{\rm M},\rmn{all}]},m^{[{\rm M},\rmn{all}]}$, where `all' indicates that all five $u$, $g$, $r$, $i$, $z$ bands have been used in the fit).  This figure plots the difference in the magnitudes, and the ratios of effective radius and \sersic index, as a function of the apparent magnitude, in an equivalent manner to Fig.~\ref{fig:gal3vgalm-mag0}.

As expected, both algorithms are consistent in extracting parameters in the $g$, $r$ and $i$ bands, while in the shallower bands ($z$ and especially $u$) we see a weaker correspondence between the two methods. There are no discernible systematic differences between the \fitb and \fitc parameters except for the \sersic index values in the $u$ and $z$ bands. The systematic differences in the $u$- and $z$-band \sersic index values will be investigated with the subsequent figures, and discussed further below.

Although most galaxies behave similarly and constitute the main body of points, there are occasional exceptions.  For example, an outlying `line' of points can be seen in the magnitude plots above ($r$-band) and below ($g$ and $i$-band) the main sequence of points. This line is caused by one galaxy, NGC 4374, an elliptical galaxy known to host a strong active galactic nucleus \citep{tex:CP06}.  Visually inspecting our NGC 4374 redshifted images reveals a prominent, unresolved nucleus. This extra central flux appears to produce inconsistent results for \fitb and \fitc; fits using both methods have issues.  This serves to highlight that in cases where the assumed profile is a bad match to reality, no method will produce satisfactory results.  A more appropriate way to fit this galaxy would be to add a central PSF component in the model.


The scatter in this plot is a simple function of object magnitude; the fainter an object, the greater the scatter between the \fitb and \fitc parameters. We quantify this scatter in a robust manner, as half of the range between the 16th- and 84th-percentiles (which is equivalent to the standard deviation for a Gaussian distribution), in order to avoid being overly influenced by a relatively small number of severe outliers. We also exclude all fits that ended up in close proximity to any of the constraints in either the single- or multi-band fits.  At $m_{r,0}^{[{\rm M, all}]}=18$ the typical scatter per band ($u$, $g$, $r$, $i$, $z$) for $m$ is $(\pm0.18,\pm0.02,\pm0.02,\pm0.02,\pm0.10)$, for  $r_{\rm e}$ is ($\pm33$\%, $\pm3$\%, $\pm4$\%, $\pm3$\%, $\pm16$\% ) and for $n$ is ($\pm29$\%, $\pm4$\%, $\pm7$\%, $\pm7$\%, $\pm33$\%).

The scatter is worst in the case of the \sersic index, mainly because this parameter is most sensitive to changes in the details in the fitting.  Magnitudes are integrated quantities, and are therefore generally the easiest galaxy properties to extract from imaging.  Determining the effective radius may be thought of as finding a radius for which the integrated flux within and without are equal.  The behaviour of the profile within each of those two regions does not matter, and there is no ambiguity in the definition for any arbitrary monotonic profile. However, determining \sersic index requires further information, particularly regarding the behaviour of the profile at its peak and in its tail.  Higher \sersic indices imply a greater shift of flux from around the half-light radius and into both the centre and outskirts of the profile, but these pieces of information need not be consistent for an arbitrary monotonic profile. The \sersic index is therefore more difficult to measure accurately and is more susceptible to changes in signal-to-noise and resolution.  Furthermore, as we have seen in Figs.~\ref{fig:reffband} and \ref{fig:nband}, \sersic index is typically a stronger function of wavelength than effective radius.  It may be expected to show more significant changes between single- and multi-band methods.

The origin of the scatter in Fig.~\ref{fig:galmvgalm-mag0} is obviously a result of the single- and multi-band fitting methods measuring different values. However, from this figure it is not clear if one or both methods introduce the scatter, i.e, whether one of the methods is superior. To further investigate the advantages of multi-band versus single-band fitting, we test the stability of the structural parameters as the galaxy becomes fainter and less resolved, as a result of the artificial redshifting. We have already shown examples, in Figs.~\ref{fig:nband}, \ref{fig:gal1} and \ref{fig:gal2}, for individual galaxies where the increased stability of multi-band fitting is apparent. Now, in Figs.~\ref{fig:indiv-zmin}, \ref{fig:combzmin} and \ref{fig:mixzmin} we demonstrate that this improvement is true for the whole sample of artificially-redshifted galaxies.

In these figures, we show the deviations of the fit parameters relative to the values obtained from the image with the lowest artificial-redshift ($z_{\rmn{min}}$).  We choose to compare to this image, rather than the original, as the extreme change in resolution means that some of the nearest galaxies show offsets between parameters from the original and $z_{\rmn{min}}$ images.  We wish to avoid including this additional scatter in these plots, and instead choose to focus on changes over redshift ranges more typical for galaxy surveys (i.e. $z > 0.01$).
Obviously, the apparent magnitude and angular size of galaxies will change as we measure them in successively more distant artificially-redshifted images.  To avoid this known cosmological trend from complicating our interpretation, in these figures we use absolute magnitude and physical size, in kpc, determined using the same cosmology used by \ferengi to create the artificial images.

Firstly, in Fig.~\ref{fig:indiv-zmin} we examine the ability of the single-band method (\fitb) to recover the redshifted parameters. The first row of panels shows the offsets in absolute magnitude measured in all the artificially-redshifted images, relative to the value obtained for the lowest artificial redshift ($z_{\rm min}$) for each galaxy.  Similarly, in the second row we show the ratio of the effective radius relative to that for the $z_{\rm min}$ image, and finally, the third row shows the same for \sersic index. The data in each panel is plotted against $r$-band apparent magnitude ($m_{r,0}^{[{\rm M}]}$), cosmologically scaled from the value for the \fitb fit to the $z_{\rm min}$ image for each galaxy, as also used in Fig.~\ref{fig:gal3vgalm-mag0}. The columns show the parameter deviations for each of the $u$, $g$, $r$, $i$, $z$ bands.  We overplot the median and the 16th- and 84th-percentiles as a function of magnitude to guide the eye and help identify systematic trends, while reducing the influence of severe outliers. In determining these lines, we exclude all fits that ended up in close proximity to any of the constraints in either the single- or multi-band fits (these points are not shown).  For reference, the fraction of galaxies for which the single-band fit failed or ended on constraints is 20\%, 6\%, 4\%, 5\%, 13\% for $u$, $g$, $r$, $i$, $z$, respectively. Coloured outlying points, selected as described below, are also excluded when determining the trend lines, such that they represent the behaviour of the main bulk of fits.

We investigated the origin of the outlying points in these plots, most obvious in the $r_{\rm e}$, $u$-band panel in Fig.~\ref{fig:indiv-zmin}.  The `lines' of points offset from the main distribution are primarily due to apparent inconsistencies in the \emph{lowest}-redshift ($z_{\rm min}$) fits.  In these cases the effective radius is unreasonably large in the $z_{\rm min}$ fit, but becomes more reasonable at higher redshifts, leading to a systematic offset in $r_{\rm e}^{[{\rm M}]} / r^{[{\rm M}]}_{\rm e,zmin}$.  To illustrate this we identify cases where the $z_{\rm min}$ fits have an effective radius, in the band under consideration, that is more than three times larger than the effective radius in the $r$-band.  These galaxies are denoted by blue points.  Variation in effective radius by more than a factor of three over optical wavelengths is very unlikely to be physical, e.g. see Fig.~\ref{fig:reffband}. The excessively large effective radii in these $z_{\rm min}$ fits tend to be accompanied by high \sersic indices and overly-bright magnitudes.  This behaviour is discussed further below.

Where the $z_{\rm min}$ fits are consistent between bands, further outliers may also be identified by considering cases where instead the fits at higher redshift become inconsistent.  To illustrate this, we denote with red points those fits where the effective radius in a specific band ($u,g,i,z$) differs from the equivalent $r$-band value by more than a factor of three.  These points do not display a consistent systematic behaviour in \sersic index and magnitude.  

We select both types of outlier using the \fitb results in Fig.~\ref{fig:indiv-zmin} and \ref{fig:mixzmin} and \fitc results in Fig.~\ref{fig:combzmin}.  In this way we identify cases where either the \fitb or \fitc method fails to fit a reasonable model, under the assumption that the $r$-band fit results are robust.

In Fig.~\ref{fig:combzmin} we show a similar plot to Fig.~\ref{fig:indiv-zmin}, but now for the results of our multi-band (\fitc) method.  The median and (robustly-determined) scatter are shown by green lines. The multi-band approach dramatically reduces, by a factor of 4-40, the fraction of fits which fail or produce results in close proximity to constraints.  These fractions are 5\%, 0.1\%, 0.1\%, 0.1\% and 2\% for $u$, $g$, $r$, $i$, $z$, respectively. We also find dramatically fewer `red' outliers, and no `blue' outliers at all.  To allow a fair comparison, we therefore determine the green lines only using those galaxies with both successful \fitb and \fitc fits and excluding the galaxies that are outliers both in \fitb and \fitc results. To facilitate the comparison we include in Fig.~\ref{fig:combzmin} the \fitb median and scatter lines from Fig.~\ref{fig:indiv-zmin} in red.

For the \fitc results, shown by the green lines in Fig.~\ref{fig:combzmin}, at $m_{r,0}^{[{\rm M,all}]}=18$ the typical scatter per band ($u$, $g$, $r$, $i$, $z$) for $m$ is $(\pm0.09,\pm0.07,\pm0.07,\pm0.07,\pm0.13)$, for  $r_{\rm e}$ is ($\pm9$\%, $\pm7$\%, $\pm10$\%, $\pm10$\%, $\pm14$\% ) and for $n$ is ($\pm18$\%, $\pm17$\%, $\pm17$\%, $\pm16$\%, $\pm22$\%).
 
In contrast, for the single-band \fitb results, shown by the red lines in Figs. ~\ref{fig:indiv-zmin} and \ref{fig:combzmin}, at $m_{r,0}^{[{\rm M,all}]}=18$ the typical scatter per band ($u$, $g$, $r$, $i$, $z$) for $m$ is $(\pm0.14,\pm0.07,\pm0.09,\pm0.07,\pm0.18)$, for  $r_{\rm e}$ is ($\pm16$\%, $\pm8$\%, $\pm10$\%, $\pm10$\%, $\pm21$\% ) and for $n$ is ($\pm26$\%, $\pm18$\%, $\pm20$\%, $\pm17$\%, $\pm33$\%). Recall that both the blue and red outliers were excluded before we determined these scatters. 

In the upper row of panels in Fig.~\ref{fig:combzmin}, we can compare the recovery of magnitude for \fitb and \fitc fits as the galaxies become more distant and hence smaller and fainter.  Both \fitb and \fitc results show similar systematic trends.  These are absent in the bluer bands, but in the red there appears to be a small bias toward recovering fainter absolute magnitudes for objects viewed at larger distances (and hence with fainter apparent magnitudes).  The systematic trends are discussed further below.  There is a hint that the bias in the $z$-band is slightly smaller for \fitc.  The two methods show similar scatter in $g$, $r$ and $i$, but for the lower signal-to-noise bands, $z$ and especially $u$, the \fitc results show significantly smaller scatter.

In the middle row, we compare the results for effective radius. Again, we see that in the higher signal-to-noise bands the methods return almost identical results.  However, the \fitc method significantly reduces the scatter in $z$, and makes a dramatic improvement in $u$-band. Part of the reason for this reduction in scatter is a result of allowing only limited variations in this parameter with wavelength. By design, the $r_{\rm e}$ in the different bands can only vary as a linear function of wavelength.  Therefore the $u$-band size is constrained using information from all the bands.  Recall that, in Section \ref{sec:structpar}, we chose the linear variation with the aim of improving the fitting results, by applying the physical assumption that these systems should only show smooth variations with wavelength, observing that single-band (\fitb) parameters are consistent with linear variation within the uncertainties (see top left panels of Figs.~\ref{fig:reffband} and \ref{fig:nband}).

Note that there are outlying lines of points below the main body of points in all of the effective radius panels in both Figs.~\ref{fig:indiv-zmin} and \ref{fig:combzmin}. These points have not be selected as outliers (at least in $g$, $r$, $i$) by either the `blue' or `red' criteria, because their effective radii in each band are reasonably consistent with the $r$-band values. We have found that these points are the result of three galaxies (NGC 3631, NGC 4321 and NGC 4725), which share the characteristic that fits to their lower artificially-redshifted images measure substantially higher effective radius than is measured on either the original (\montage) images, or the higher-redshift (\ferengi) images.  All methods (\fita, \fitb and \fitc) result in similar behaviour for these galaxies.  These are all approximately face-on spirals with small, prominent bulges and extended faint disks.  Not unsurprisingly, it may be that this distinctively two-component morphology is more susceptible to systematic variations in \sersic fit parameters at intermediate resolution.  Preliminary results from bulge-disk fits confirm this impression.

Finally, the bottom row of Fig.~\ref{fig:combzmin}, presents the results for \sersic index.  The comparison between \fitb and \fitc is very similar to that for effective radius. \sersic indices recovered by our multi-band (\fitc) method, with $n$ varying linearly with wavelength, show reduced scatter in $z$ and, in particular, $u$ bands.  The systematic trends seen for $r_{\rm e}$ and $n$ versus $m_{r,0}^{[{\rm M,all}]}$ are also significantly reduced in $z$ and $u$ when using the \fitc method.

Overall, the comparison of \fitb versus \fitc fitting shows that for the high signal-to-noise bands ($g$, $r$ and $i$) the improvement in extracting parameters is minor, while for low signal-to-noise bands ($u$ and $z$) there is a clear reduction of both the systematic and statistical uncertainties for fainter galaxies.  Another important advantage of \fitc versus \fitb is that the multi-band approach increases the number of galaxies that are successfully fitted. The improvement is small for samples of large bright galaxies but can be very significant when we move to fainter samples. This advantage of multi-band fitting is more evident in Paper III, were we deal with thousands of faint galaxies, and show that this approach has the ability to obtain meaningful measurements even in extreme cases where the galaxy is not visible in some of the bands.


In Figs.~\ref{fig:indiv-zmin} and \ref{fig:combzmin} we notice a systematic effect whereby both \fitb and \fitc methods return systematically fainter magnitudes, smaller effective radii and lower \sersic indices when going to fainter apparent magnitudes (i.e., for galaxies artificially-redshifted to greater distances).  A significant source of systematic uncertainty in both figures originates from the sky determination.  In principle, for the artificially-redshifted images we know the fundamental sky level, as it has been explicitly simulated by \ferengi.  However, in practice there are faint, unresolved, and hence unmasked, objects in the images, in addition to the galaxy of interest.  In order to account for the contribution of these objects we need to measure the apparent sky value.  At higher artificial redshifts, neighbouring objects become fainter (due to surface brightness dimming), cover a larger area of the image (due to the worsening angular resolution) and are less likely to be detected by \sex and hence masked.  The contribution of such sources to the sky therefore varies, and so a sky value must be independently determined for each artificially-redshifted image.

During the \ferengi artificial-redshifting process, the angular size of the original image, and hence the number of pixels that sample it, are reduced.  At $z=0.25$ the artificial images have decreased to about $5$ per cent of the size (and so $0.3$ per cent of the area) of the first redshifted image (i.e. with $z=0.01$).  Furthermore, the PSF convolution acts to increase the size of the galaxy relative to the image.  For high artificial redshifts, the resulting small images do not provide enough free sky pixels for a secure sky determination and may lead to an overestimation of the sky value. This would result in systematic trends in $m$, $r_{\rm e}$ and $n$ in the directions seen. However, we suspect  that this is not the only cause of the observed trends for three reasons. Firstly, the trends, particularly in the case of \sersic index, can be seen to set in at relatively bright magnitudes, and hence low redshifts, where all images still contain a large fraction of sky pixels.  Secondly, for single-band fitting, the strength of the trend increases for bands with lower signal-to-noise, suggesting that the systematics are dependent on noise, and not just resolution. Thirdly, the trends are significantly improved by multi-band fitting, even though the same sky is used for both methods.

A plausible alternative cause of these trends is systematic mismatches between the true galaxy profiles and the \sersic model.  Many of our galaxies are spirals, and are hence better described by two-components, a bulge and a disk, than they are by a single-\sersic function. Even elliptical galaxies are not always well-represented by \sersic profiles, particularly in their centres, as discussed further in Section~\ref{sec:ellipt}. One consequence of the bulges in spiral galaxies is that they display excess light in their centres, when compared to a single-\sersic profile fit at larger radii. This excess will be resolved, and be apparent with high signal-to-noise, for large galaxies at low redshift.  It would therefore be expected that a fit to the full image would attempt to fit this feature, resulting in a relatively large \sersic index.  For example, for an exponential disk with a bright, well resolved $n \sim 4$ bulge we might obtain a single-\sersic profile with $n \sim 3$.  At higher redshift, the same galaxies would be less resolved and the noise relatively higher, lowering the significance of the central bulge.  The fit will therefore be more constrained than previously by flux at larger radii.  A lower \sersic index for the single-component fit would be a natural outcome, perhaps $n \sim 2$ in our above example. Due to the covariance between $n$, $r_{\rm e}$ and $m$, shifts to lower $n$ would be accompanied by smaller  $r_{\rm e}$ and fainter $m$ as is seen \citep{tex:GM09,tex:YW11}. This is an important issue that should be considered carefully by any study investigating the evolution of galaxy structural parameters with redshift.

With increasing redshift, neighbouring objects often become blended with the main galaxy.  This effect is present in both real higher-redshift data, and in the artificially-redshifted images.
As mentioned in Section \ref{sec:ferengi}, a downside of the redshifting process is that \ferengi treats both galaxies and stars in the same way.  Together with the high frequency of stars in the original images, this means that blending with similarly bright neighbours  probably occurs more often in our images than it would in real data.  Masking is not always effective, due to the difficult task of identifying and separating these blended neighbours.  In such cases, the recovered parameters, for both in single- and multi-band fits, may be distorted. This highlights how important it is to carefully mask and fit neighbouring objects, particularly when fitting faint galaxies.  It is also important to recognise that inseparable blends will always introduce a contamination when attempting to measure galaxy structural parameters in low resolution data.  These effects will introduce additional scatter in the recovered parameters, primarily at faint apparent magnitudes.

Masking issues should affect both \fitb and \fitc methods equally, while systematic trends are present for both, but somewhat reduced for \fitc.  The differences in the scatter between these two methods compares the performance of the two methods. From this we have shown that our multi-band technique reduces the statistical uncertainties associated with lower signal-to-noise and lower physical resolution images and enables the structural parameters of galaxies to be more accurately measured, when compared with fitting each band independently.

Now, after studying the performance of the \fitb and \fitc methods separately in Figs.~\ref{fig:indiv-zmin} and \ref{fig:combzmin}, we can better understand the behaviour seen in Fig.~\ref{fig:galmvgalm-mag0}.  First we note that the scatter in Fig.~\ref{fig:galmvgalm-mag0}, particularly for the $g$, $r$ and $i$ bands, is significantly smaller than that in Figs.~\ref{fig:indiv-zmin} and \ref{fig:combzmin}.  This must be because galaxies experience similar systematic trends with both \fitb and \fitc; Fig.~\ref{fig:galmvgalm-mag0} is just showing residual differences between the two methods.  The same argument explains the systematic decline of the \sersic index ratio in Fig.~\ref{fig:galmvgalm-mag0}.  Both methods measure smaller values of $n$ as the galaxies become more distant, but the systematic effect is greater for \fitb than \fitc (Fig.~\ref{fig:combzmin}; red line versus green line, respectively). This difference results in the moderately trend in the purple line in Fig.~\ref{fig:galmvgalm-mag0}.

We have shown that \fitc is more stable and robust to redshift effects, in terms of the fraction of successful fits, reduced systematic biases and smaller statistical scatter.  However, we have not fully tested whether the \fitc results are more accurate, i.e. whether its results are closer to the \emph{true} values than \fitb. Recall that the assumption of a particular dependence of $n$ and $r_{\rm e}$ on wavelength (linear in this case) may introduce systematic biases if it is overly restrictive.  To investigate this we assume that the lowest redshift ($z_{\rm min}$) single-band (\fitb) fit accurately represents the structure of each galaxy (or at least is less affected by systematic effects than \fitc).  We will test the accuracy of the \fitc fits by looking for any systematic bias between the \fitc results and the $z_{\rmn{min}}$ \fitb results. 
 
In Fig.~\ref{fig:mixzmin} we present a hybrid of Figs.~\ref{fig:indiv-zmin} and \ref{fig:combzmin}, showing the parameter differences between our multi-band (\fitc) fits to each artificially-redshifted image and the single-band  (\fitb) fits to the lowest redshift image. In a similar manner to the previous figures, we determine the median and scatter lines using only those images with successful fits using both \fitb and \fitc methods, and excluding the `blue' and `red' outliers, as defined in Fig.~\ref{fig:indiv-zmin}. We plot the median and scatter in blue. In order to aid the comparison we include the red lines from Fig.~\ref{fig:indiv-zmin}.  In the upper row, we can compare the recovered magnitudes, in the middle row we compare the recovered effective radius and in the lower panels the \sersic indexes, as the galaxies become more distant. As seen before, there is little change in behaviour for $g$, $r$ and $i$, but there are clear differences for the lower signal-to-noise bands, $z$ and especially $u$. 

For distant, faint galaxies the scatter of the \fitc results is similar, or slightly better, than \fitb, and the biases are significantly improved.  For such galaxies, the benefits of the improved precision and robustness of the multi-band approach outweigh any systematic inaccuracies.  For nearby, bright galaxies there remains a significant scatter, which is associated with differences between \fitb and \fitc fits that persist even for reasonable quality imaging.  At high signal-to-noise, one might suspect that \fitb results are more accurate and \fitc results biased.  However, despite the scatter, there is little evidence for a systematic difference between the two methods at the bright end.  Furthermore, by comparing Figs. ~\ref{fig:indiv-zmin} and \ref{fig:combzmin}, it is clear that the vast majority of outliers in Fig.~\ref{fig:mixzmin} are due to inconsistencies in the \fitb fits (blue and red points), even for good quality imaging.  Also recall that the number of fits which encounter constraints, and are hence unusable, is many times greater for \fitb. We incline to trust the \fitc results over the \fitb results as the latter are more likely to return fits with errors. Therefore, we conclude that the benefits of improved robustness in the multi-band approach largely outweigh concerns about systematic biases.  Nevertheless, we advise that future studies should consider the scientific consequences of systematic biases resulting from assuming a restrictive wavelength dependence for $n$ and $r_{\rm e}$, and if necessary increase the flexibility of this dependence.  This is extremely easy to do in \galfitm, by increasing the order of the relevant polynomial when setting up the fit.

\begin{figure*}
\centering
\includegraphics[width=14.0cm]{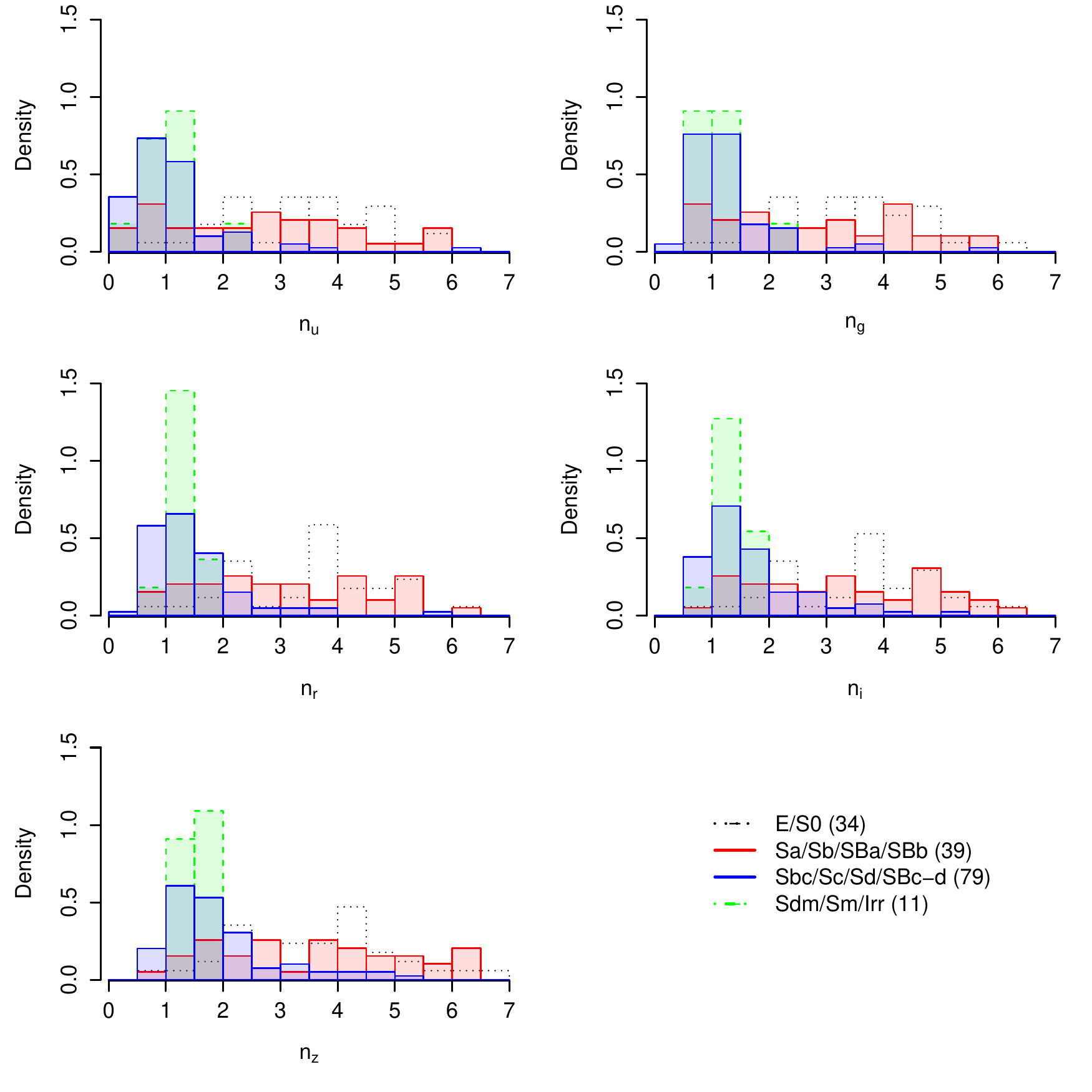}
\caption{Distribution of \sersic indices obtained from \fitc fits to the original imaging for our sample of 163 galaxies. Each histogram shows the distribution for a specific range of morphologies.  Each histogram has been normalised so the total area (integral) is equal to one.  Each panel plots the results for a different band.}
\label{fig:morph-n}
\end{figure*}

\begin{figure*}
\centering
\includegraphics[height=14cm,width=13.0cm]{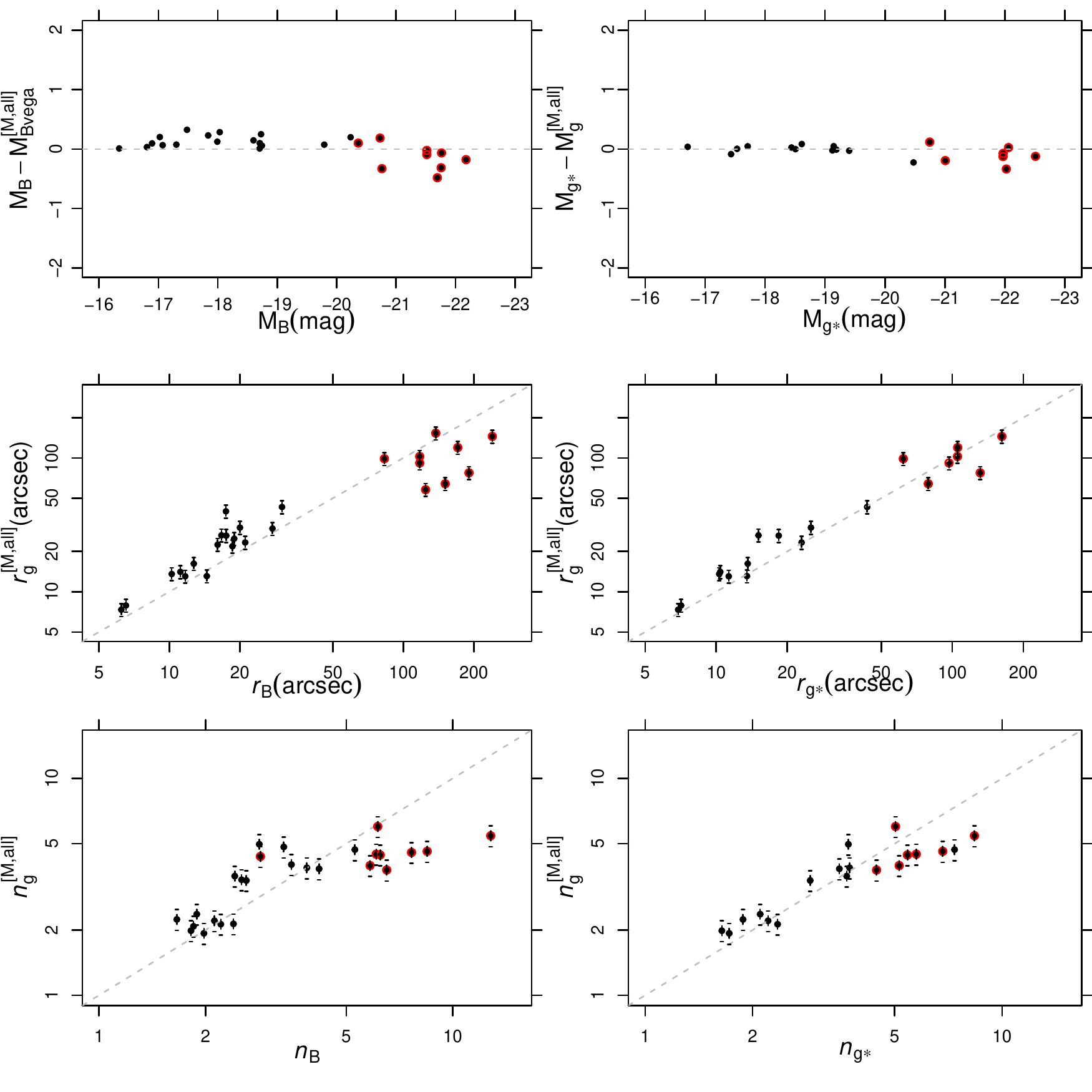}
\caption{Comparison between structural parameters derived in this work and those obtained by C93 for 26 common elliptical galaxies (left column) and by C10 for 19 common galaxies (right column). The red points highlight galaxies with $r_{e,{\rm B}}>50$ arcsec. Error bars in $r_{g}$ and $n_{g}$ show the estimated uncertainties of 11 per cent in each.}
\label{fig:ellip}
\end{figure*}

\subsection{Structural parameters from the original \montage images}

In Fig.~\ref{fig:morph-n} we plot the distribution of \sersic indices for the original images, using $n$ values obtained from our multi-band (\fitc) single-\sersic fits. The black dotted lines show the distribution of elliptical and lenticular galaxies. The red lines show early-type spirals, while blue lines show late-type spirals, both including barred galaxies. Finally, green dashed lines show pure disk galaxies and irregulars. 
This plot illustrates how \sersic index changes with both morphological type and wavelength. For the E/S0 sample we can see that the majority of early-type galaxies have \sersic indices between 3 and 5, and show only minor variations with wavelength.  It is interesting to note a hint of a bimodality, with $n$ peaked at values around both 4 and 2.   We have investigated this and find that there is a sign that this is a result of ellipticals inhabiting the high-$n$ peak with S0s tending to have lower-$n$. However, low number statistics prevent us from pursuing this further.

Early-type spirals show a wide range of $n$ values from $\sim 0.5$ to $\sim 6.5$.  The morphological classification of spiral galaxies is usually based up on both the tightness of their spirals and their bulge-to-disk ratio.  These are generally considered to be correlated, such that earlier-type spirals are associated with both tighter-winding arms and higher bulge-to-disk ratios.  We would therefore expect that early-type spiral galaxies would have more prominent bulges, compared to late-type spirals, and hence higher \sersic indices.  However, as we can see from Fig.~\ref{fig:morph-n}, early-type spirals have equal large and small $n$ values and this distribution shows no obvious changes with wavelength.  It appears that visually classified early-type spirals can possess a wide range of $n$, and hence bulge-to-disk ratio.

Next we consider late-type spiral galaxies. We see that late spirals (Sbc -- Sd) are concentrated around $n=1$, with a small tail to higher $n$, as expected if they are typically dominated by an exponential disk.  Bulgeless disks (Sdm/Sm) and irregular galaxies are even more concentrated around $n \sim 1$. Furthermore, for both samples of late-type galaxies, as we move from blue to red wavelengths we see the distribution shift consistently to higher \sersic indices, from $\la 1$ in $u$-band to $\sim 1.5$ in $z$-band).

That effect can be attributed to a combination of two phenomena. First, the small bulge or pseudobulge that these galaxies contain will become more prominent at longer wavelengths, due to it typically having a redder colour than the disk.  Second, at bluer wavelengths the light emission is more attenuated by dust, which may act to flatten the profile at shorter wavelengths \citep{tex:PP13}.

\subsubsection{Elliptical sample properties}
\label{sec:ellipt}

All 26 elliptical galaxies\footnote{as classified in \citet{tex:CC93} and/or the NASA/IPAC Extragalactic Database (NED)} in our sample also have \sersic profile parameters  measured by the classic work of \citet[hereafter C93]{tex:CC93}.  More recently, \citet[hereafter C10]{tex:CC10} have also measured these parameters for 19 galaxies in our sample using imaging from SDSS.    In this section we investigate the reliability of our method by comparing our derived parameters with those measured by C93 and C10.  For this purpose, we use only the structural parameters obtained from multi-band fits to the original \montage images.  In order to perform a sensible comparison with C93, our $g$-band AB magnitudes have been converted to $B$-band values and to the Vega zeropoint system. We adopt the transformations provided by \citet{tex:BR07}:
\begin{equation}
M_{B}  = M_{g} + 0.2354 + 0.3915 [ (M_{g}-M_{r})-0.6102 ] \ ,
\label{eq:gtoB}
\end{equation}
\begin{equation} 
M_{B,\rm Vega} = M_{B,\rm AB} + 0.09\ .
\end{equation}
In Table \ref{table:ellipt} we list the absolute magnitude from our multi-band fit, after converting to $B$-band Vega magnitudes,  the galaxy distance, the galactic extinction in $g$-band from NED \citep{tex:SF11}, C93's $B$-band absolute magnitude ($M_{B}^{*}$), effective radius ($r_{\rmn{e},B}^{*}$) and their equivalent profile \sersic index ($n_{\rmn{eq},B}^{*}$).  As C93's magnitudes and effective radii are presented in different format to that produced by \galfit, we have derived $M_{B}^{*}$ and $r_{\rmn{e},B}^{*}$ from equation (9) of C93 using $b_{n}  = 1.999n - 0.3271$, $c_{n} = (2.5b_{n})/\ln(10)$ \citep{tex:GD05} and the distances found in Table \ref{table:ellipt}. The parameters measured by C10 can be easily extracted directly from their paper and compared to our without any conversions, and so are not included in Table \ref{table:ellipt}. All the absolute magnitude values presented in this section have been calculated assuming the distances found in Table \ref{table:ellipt}.


In Fig.~\ref{fig:ellip} we show the comparison between the parameters derived in this work and those from both C93 and C10.  C93 were the first to introduce \sersic index as the third free parameter of the \sersic function, until then it was common to keep it fixed at a value of 4. In the left panels of Fig.~\ref{fig:ellip} we compare our multi-band results those of C93. We see that the two sets of derived properties are in reasonable agreement.  However, there is a systematic trend, such that large galaxies, with $r_{{\rm e},B} \ga 50$ arcsec, are measured to have smaller sizes and fainter magnitudes using our method, and vice versa for smaller galaxies.
In the right panels of Fig.~\ref{fig:ellip} we compare with C10.  These measurements show much better agreement with ours, independent of the galaxy size.  However, in both lower panels of Fig.~\ref{fig:ellip} we note that where both C93 or C10 find $n_{\rmn{eq},B}^{*} \ga 5$ our indices remain around $n \sim 5$.

That C93 measure slightly different values compared to both this study and C10 is not unexpected.  Both this study and C10 are based on recent imaging data from SDSS and apply modern fitting techniques, whereas C93 was one of the first papers to apply this fitting method, using lower quality images.  They are therefore more likely to suffer from systematic uncertainties, especially for the most extended galaxies in their sample.  Measuring the effective radius and \sersic index of large elliptical galaxies is a not as easy as it might seem. Their low surface brightness haloes can be very extended and are difficult to determine \citep{tex:KF09}. As \citet{tex:HM07} have shown, a galaxy with $n=4$ is much harder to fit than a galaxy with $n=1$, primarily due to uncertainties in the sky estimate.  Small overestimations of the background effectively clip the outer faint halo and result in an underestimation of effective radius, magnitude and \sersic index. 

Furthermore, large elliptical galaxies are known to show light profiles that diverge from exact \sersic functions. It has been shown that bright ($M_{B}<-20.5$) elliptical galaxies have cores in their central regions, which deviate from a \sersic model in the sense that the surface brightness is lower than would be expected from extrapolating the outer \sersic profile inwards (e.g., \citealt{tex:FB94,tex:G11}).
Usually, detailed fitting of these objects either excludes the inner part of the galaxy, uses an additional function to model the central light distribution (e.g. \citealt{tex:LA95,tex:FC06a}) or adopts a core-\sersic function \citep{tex:TE04}. For instance, both studies that we compare with (C93 and C10) have excluded the inner few arcsec of each galaxy light profile from the fitting process.  The decision of how much of the inner region should be excluded from the fit typically requires an individual visual inspection of the one dimensional surface brightness distribution of each galaxy. This process is beyond of the aim of this paper.  Furthermore, we prefer to keep the overall fitting process homogeneous and automated, such that our results can inform the interpretation of studies involving much larger samples. The core-\sersic function could possibly provide better fits to the ellipticals in our original images but would be difficult to constrain in case of our artificially-redshifted sample. An added complication is that, at present, the core-\sersic function is not implemented in \galfit.  

\begin{figure}
\centering
\includegraphics[width=0.5\textwidth]{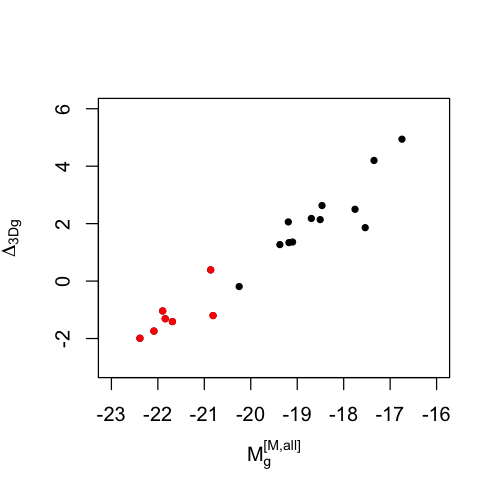}
\caption{The $\Delta_{3Dg}$ parameter quantifies the light deficit (core galaxies, negative values) or light excess (cuspy or power law galaxies, positive values) that appear in the central region of early type galaxies. We plot here values of  $\Delta_{3Dg}$, obtained from \citet{tex:GF11} for 19 of our 26 elliptical galaxies, versus absolute $g$-band magnitude. The galaxies shown in red points are the same as in Fig.~\ref{fig:ellip}, i.e., those with $r_{e,{\rm B}}>50$ arcsec.}
\label{fig:corecusp}
\end{figure}

To further investigate the characteristics of these galaxies, and specifically the discrepancies we see compared to C93 and C10 at high-$n$, in Fig.~\ref{fig:corecusp} we plot the $\Delta_{3Dg}$ parameter for 19 of our elliptical galaxies, as derived by \citet{tex:GF11}. This parameter identifies whether galaxies have a flux deficit or excess in their inner regions. Negative $\Delta_{3Dg}$ values are indicative of a core structure in the center of a galaxy (see \citet{tex:GF11} for more details).  We highlight in red the same large galaxies as in Fig.~\ref{fig:ellip}.  It is clear that the galaxies that show the largest deviation from C93, in left panel in Fig.~\ref{fig:ellip}, display negative values in Fig.~\ref{fig:corecusp}.  Therefore, the galaxies with central flux deficits are those that show different $n$ and $r_{\rm e}$, between the different studies. Similar investigation regarding galaxy size discrepancies can be found in C10 their figure 11.  Central deficits and excesses in elliptical galaxies only affect the inner few percent of $r_{\rm e}$, so these deviations from the \sersic function are only resolved for the larger galaxies, which typically show central deficits.   In the case of smaller galaxies, the excess region is not resolved in the SDSS imaging and thus does not significantly impact upon our \sersic function fits.  Finally, note that both C93 and C10 use one-dimensional fitting methods, in contrast to our two-dimensional approach, so the difference between our \sersic values may partly result from this \citep{tex:VD12}.


At higher redshifts, e.g. in our artificially-redshifted images, we would expect that small deviations from \sersic profiles in the central regions would rapidly become less important.  We have seen in Section~\ref{sec:MvsS} that the `central excesses' associated with bulges in disk galaxies result in a trend such that the recovered \sersic index decreases with increasing redshift.  Ellipticals with a central deficit should show the opposite behaviour.  Indeed, upon examination we found that all seven galaxies with negative $\Delta_{3Dg}$ values show an increase in their measured \sersic index with redshift or stay flat.  This contrasts with the remaining ellipticals -- which mostly show flat, or occasionally decreasing, trends -- and spirals, which often display systematic decreases in $n$ with redshift (see Section~\ref{sec:MvsS}).  This supports the picture that when we resolve, and attempt to fit, the centres of cored elliptical galaxies we recover smaller \sersic indices, while as galaxies become less resolved with increasing redshift, the \sersic function is less affected by the central behaviour and the fitting process returns higher $n$ values.

\section{Summary}  
\label{sec:sum}

In this paper we have presented a modified version of \galfitthree, named \galfitm, which can constrain wavelength-dependent models of galaxy surface-brightness profiles using multi-band imaging data. We assess the performance of this new feature by applying it to fit elliptical single-\sersic profile models to SDSS $ugriz$ imaging for 4026 galaxies (163 at low redshift and 3863 artificially redshifted to greater distances).

We show that \galfitm produces very similar results to \galfitthree  when used to fit single-band imaging. After we have secured that we have not substantially changed the basic behaviour of the code, we compare the recovered parameters of \galfitm when fitting single-band images individually to those obtained by fitting multiple bands simultaneously. We find that these two methods perform similarly for obtaining structural parameters in the $g$, $r$ and $i$ bands, but that our multi-band approach shows greater differences in the lower signal-to-noise $z$ and $u$ bands.

In order to demonstrate that the multi-band method produces more robust results, we study the stability of the recovered structural parameters as the galaxies become more distant, and hence the images become noisier and less well resolved.  We show that the \galfitm multi-band technique results in significant reductions in both the statistical and systematic uncertainties on structural parameter measurements.  Additionally, the multi-band technique significantly increases the fraction of galaxies that return successful fitting results. These improvements are most noticeable for low quality images, but in all the tests we performed multi-band fitting is found to improve upon, or at least do no worse than, single-band results in terms of accuracy and robustness.

In datasets with a variety of signal-to-noise levels, our approach involves a degree of interpolation or extrapolation of parameters constrained in the high signal-to-noise images to help determine parameter values for the lower signal-to-noise images.  There is thus a balance to be struck between reducing statistical uncertainties and avoiding systematic errors.  Our method provides the flexibility to adapt the assumptions for the wavelength dependence of the model, depending on the galaxy population being studied, the nature of the dataset and the aim of the analysis.

Using the multi-band fitting method we measure and present single-\sersic structural parameters for the 163 galaxies in our sample, as measured on the full-quality SDSS images.   We compare the distributions of \sersic index for subsamples with different visual morphologies, finding an anticipated trend for spiral galaxies to show higher \sersic indices in redder bands. Focussing on elliptical galaxies, we compare our multi-band measurements with structural parameters obtained by two independent studies.  Overall, we find good agreement between our multi-band fitting results and single-band fitting results from the literature, and are able to offer explanations for the small discrepancies that exist.

We conclude that using multi-band fitting improves over single-band fitting for the extraction of structural parameters, particularly for datasets containing multi-band images with a mixture of signal-to-noise and resolution.  Another advantage of this method, which has not been explored in this paper, but will be utilised in forthcoming work, is that it naturally produces robust colour gradients. The multi-wavelength approach produces a more physically consistent model, which we anticipate will enable us to make more meaningful inferences than would otherwise be possible.

\section*{Acknowledgments}
This publication was made possible by NPRP grant \# 08-643-1-112 from the Qatar National Research Fund (a member of Qatar Foundation). The statements made herein are solely the responsibility of the authors.  BH and MV are supported by this NPRP grant.  SPB gratefully acknowledges an STFC Advanced Fellowship.  We would like to thank Carnegie Mellon University in Qatar and The University of Nottingham for their hospitality.  We would like to thank the referee for the constructive comments that improved the paper.

Funding for the SDSS and SDSS-II has been provided by the Alfred P. Sloan Foundation, the Participating Institutions, the National Science Foundation, the U.S. Department of Energy, the National Aeronautics and Space Administration, the Japanese Monbukagakusho, the Max Planck Society, and the Higher Education Funding Council for England. The SDSS Web Site is http://www.sdss.org/. The SDSS is managed by the Astrophysical Research Consortium for the Participating Institutions. The Participating Institutions are the American Museum of Natural History, Astrophysical Institute Potsdam, University of Basel, University of Cambridge, Case Western Reserve University, University of Chicago, Drexel University, Fermilab, the Institute for Advanced Study, the Japan Participation Group, Johns Hopkins University, the Joint Institute for Nuclear Astrophysics, the Kavli Institute for Particle Astrophysics and Cosmology, the Korean Scientist Group, the Chinese Academy of Sciences (LAMOST), Los Alamos National Laboratory, the Max-Planck-Institute for Astronomy (MPIA), the Max-Planck-Institute for Astrophysics (MPA), New Mexico State University, Ohio State University, University of Pittsburgh, University of Portsmouth, Princeton University, the United States Naval Observatory, and the University of Washington. 

NED is operated by the Jet Propulsion Laboratory, California Institute of Technology, under contract with the National Aeronautics and Space Administration.

\bibliography{references}

\onecolumn

\setlength{\textwidth}{6.8in}
\LTcapwidth=\textwidth
\begin{landscape}
\scriptsize 
\begin{center}
\begin{longtable}{llddddddddddddddd}
\caption{Galaxy Sample and the fit results as derived from the multi-band fitting; Column(1): Galaxy name; Column(2): Hubble type from NED;  Column(3-7): apparent magnitudes in $u$, $g$, $r$, $i$, $z$ passbands;  Column(8-12): effective radius;   Column(13-17): \sersic index.  See Section \ref{sec:fit} for more details. 
\label{table:first} }\\

\hline  \\[-2.0ex]
\multicolumn{1}{l}{Name} & 
\multicolumn{1}{l}{Type } & 
\multicolumn{1}{c}{$m_u$} & 
\multicolumn{1}{c}{$m_g$} & 
\multicolumn{1}{c}{$m_r$} & 
\multicolumn{1}{c}{$m_i$} & 
\multicolumn{1}{c}{$m_z$} & 
  \multicolumn{1}{c}{$r_{\rmn{e},u}$} &
  \multicolumn{1}{c}{$r_{\rmn{e},g}$} &
  \multicolumn{1}{c}{$r_{\rmn{e},r}$} &
  \multicolumn{1}{c}{$r_{\rmn{e},i}$} &
  \multicolumn{1}{c}{$r_{\rmn{e},z}$} &
  \multicolumn{1}{c}{$n_{u}$ } &
  \multicolumn{1}{c}{$n_{g}$} &
  \multicolumn{1}{c}{$n_{r}$}  &
    \multicolumn{1}{c}{$n_{i}$}  &
  \multicolumn{1}{c}{$n_{z}$}  \\[0.5ex]

\multicolumn{1}{l}{    } & 
\multicolumn{1}{l}{    } & 
\multicolumn{1}{c}{(mag)} & 
\multicolumn{1}{c}{(mag)} & 
\multicolumn{1}{c}{(mag)} & 
\multicolumn{1}{c}{(mag)} & 
\multicolumn{1}{c}{(mag)} & 
\multicolumn{1}{c}{(\arcsec)} & 
\multicolumn{1}{c}{(\arcsec)} &
\multicolumn{1}{c}{(\arcsec)} &
\multicolumn{1}{c}{(\arcsec)} &
\multicolumn{1}{c}{(\arcsec)} & 
\multicolumn{1}{c}{ } &
\multicolumn{1}{c}{  } &
\multicolumn{1}{c}{ } &
\multicolumn{1}{c}{     } &
\multicolumn{1}{c}{     }  \\[0.5ex]

\multicolumn{1}{l}{    } & 
\multicolumn{1}{l}{    } & 
\multicolumn{1}{c}{ ($\pm0.13$)  } & 
\multicolumn{1}{c}{ ($\pm0.09$)  } & 
\multicolumn{1}{c}{ ($\pm0.1$)  } & 
\multicolumn{1}{c}{ ($\pm0.11$)  } & 
\multicolumn{1}{c}{ ($\pm0.12$)  } & 
\multicolumn{1}{c}{ ($\pm12$\%) } & 
\multicolumn{1}{c}{ ($\pm11$\%)  } &
\multicolumn{1}{c}{ ($\pm12$\%)  } &
\multicolumn{1}{c}{ ($\pm14$\%)  } &
\multicolumn{1}{c}{ ($\pm15$\%)  } & 
\multicolumn{1}{c}{ ($\pm10$\%)  } &
\multicolumn{1}{c}{ ($\pm11$\%)  } &
\multicolumn{1}{c}{ ($\pm14$\%)  } &
\multicolumn{1}{c}{ ($\pm15$\%)  } &
\multicolumn{1}{c}{ ($\pm17$\%)  }  \\[0.5ex]

\hline\hline  \\[-1.8ex]
\endfirsthead

\multicolumn{10}{c}{ \tablename\ \thetable{} -- continued from previous page} \\[1mm]
\hline  \\[-2.0ex]
\multicolumn{1}{l}{Name} & 
\multicolumn{1}{l}{Type } & 
\multicolumn{1}{c}{$m_u$} & 
\multicolumn{1}{c}{$m_g$} & 
\multicolumn{1}{c}{$m_r$} & 
\multicolumn{1}{c}{$m_i$} & 
\multicolumn{1}{c}{$m_z$} & 
  \multicolumn{1}{c}{$r_{\rmn{e},u}$} &
  \multicolumn{1}{c}{$r_{\rmn{e},g}$} &
  \multicolumn{1}{c}{$r_{\rmn{e},r}$} &
  \multicolumn{1}{c}{$r_{\rmn{e},i}$} &
  \multicolumn{1}{c}{$r_{\rmn{e},z}$} &
  \multicolumn{1}{c}{$n_{u}$ } &
  \multicolumn{1}{c}{$n_{g}$} &
  \multicolumn{1}{c}{$n_{r}$}  &
    \multicolumn{1}{c}{$n_{i}$}  &
  \multicolumn{1}{c}{$n_{z}$}  \\[0.5ex]

\multicolumn{1}{l}{    } & 
\multicolumn{1}{l}{    } & 
\multicolumn{1}{c}{(mag)} & 
\multicolumn{1}{c}{(mag)} & 
\multicolumn{1}{c}{(mag)} & 
\multicolumn{1}{c}{(mag)} & 
\multicolumn{1}{c}{(mag)} & 
\multicolumn{1}{c}{(\arcsec)} & 
\multicolumn{1}{c}{(\arcsec)} &
\multicolumn{1}{c}{(\arcsec)} &
\multicolumn{1}{c}{(\arcsec)} &
\multicolumn{1}{c}{(\arcsec)} & 
\multicolumn{1}{c}{ } &
\multicolumn{1}{c}{  } &
\multicolumn{1}{c}{ } &
\multicolumn{1}{c}{     } &
\multicolumn{1}{c}{     }  \\[0.5ex]

\multicolumn{1}{l}{    } & 
\multicolumn{1}{l}{    } & 
\multicolumn{1}{c}{ ($\pm0.13$)  } & 
\multicolumn{1}{c}{ ($\pm0.09$)  } & 
\multicolumn{1}{c}{ ($\pm0.1$)  } & 
\multicolumn{1}{c}{ ($\pm0.11$)  } & 
\multicolumn{1}{c}{ ($\pm0.12$)  } & 
\multicolumn{1}{c}{ ($\pm12$\%) } & 
\multicolumn{1}{c}{ ($\pm11$\%)  } &
\multicolumn{1}{c}{ ($\pm12$\%)  } &
\multicolumn{1}{c}{ ($\pm14$\%)  } &
\multicolumn{1}{c}{ ($\pm15$\%)  } & 
\multicolumn{1}{c}{ ($\pm10$\%)  } &
\multicolumn{1}{c}{ ($\pm11$\%)  } &
\multicolumn{1}{c}{ ($\pm14$\%)  } &
\multicolumn{1}{c}{ ($\pm15$\%)  } &
\multicolumn{1}{c}{ ($\pm17$\%)  }  \\[0.5ex]

\hline \hline
\endhead

\hline
\endfoot

\hline
\endlastfoot
IC 0724          &      Sa                      &        14.68  &  12.86  &  12.05  &  11.67  &  11.41  &   33.30  &   31.78  &   29.98  &   28.25  &   26.38  &  4.41  &  4.48  &  4.56  &  4.64  &  4.73    \\
IC 1067          &      SB(s)b               &        14.65  &  13.02  &  12.31  &  11.93  &  11.68  &   27.70  &   27.45  &   27.14  &   26.85  &   26.53  &  2.63  &  2.87  &  3.16  &  3.44  &  3.74    \\
IC 1125          &      S/Irr                    &        15.04  &  13.88  &  13.29  &  12.99  &  12.75  &   18.15  &   17.77  &   17.31  &   16.88  &   16.41  &  1.04  &  1.14  &  1.26  &  1.37  &  1.49    \\
IC 1158          &      SAB(r)c              &       14.98  &  13.31  &  12.51  &  12.11  &  11.90  &   31.41  &   39.98  &   50.17  &   59.89  &   70.42  &  0.30  &  0.87  &  1.55  &  2.20  &  2.90    \\
IC 3468          &     E1                        &       15.20  &  13.63  &  12.95  &  12.63  &  12.40  &   19.63  &   19.73  &   19.85  &   19.97  &   20.10  &  1.83  &  1.86  &  1.90  &  1.93  &  1.97    \\
IC 3540          &     S0(2)                   &       15.87  &  14.37  &  13.75  &  13.42  &  13.21  &    8.55  &    8.63  &    8.73  &    8.83  &    8.93  &  0.84  &  0.85  &  0.86  &  0.88  &  0.89    \\
IC 3653          &     E3                        &       15.90  &  14.12  &  13.32  &  12.92  &  12.64  &    7.16  &    7.23  &    7.31  &    7.39  &    7.47  &  1.87  &  1.95  &  2.03  &  2.11  &  2.20    \\
NGC 0428     &     SAB(s)m             &       12.83  &  11.81  &  11.44  &  11.24  &  11.18  &   46.53  &   44.68  &   42.48  &   40.38  &   38.11  &  1.00  &  1.06  &  1.12  &  1.18  &  1.24   \\
NGC 0450     &     SAB(s)cd            &       13.51  &  12.42  &  12.05  &  11.89  &  11.84  &   45.35  &   43.92  &   42.23  &   40.61  &   38.85  &  1.17  &  1.35  &  1.57  &  1.77  &  1.99   \\
NGC 0701     &     SB(rs)c                &       13.77  &  12.52  &  11.86  &  11.51  &  11.26  &   33.52  &   33.10  &   32.61  &   32.13  &   31.62  &  0.83  &  0.95  &  1.08  &  1.22  &  1.36   \\
NGC 0853     &     Sm pec                &      14.04  &  13.03  &  12.57  &  12.34  &  12.15  &   16.98  &   18.17  &   19.58  &   20.93  &   22.39  &  0.92  &  0.97  &  1.02  &  1.07  &  1.12    \\
NGC 0941     &     SAB(rs)c              &      14.00  &  12.89  &  12.46  &  12.24  &  12.15  &   26.77  &   26.37  &   25.89  &   25.44  &   24.94  &  1.02  &  1.06  &  1.11  &  1.16  &  1.21    \\
NGC 1042     &     SAB(rs)cd           &       12.93  &  11.59  &  11.04  &  10.74  &  10.66  &   66.48  &   65.36  &   64.02  &   62.75  &   61.37  &  0.45  &  0.60  &  0.77  &  0.94  &  1.12    \\
NGC 1068     &     (R)SA(rs)b          &       10.74  &   9.42  &   8.77  &   8.44  &   8.16  &   39.93  &   38.73  &   37.30  &   35.93  &   34.45  &  3.87  &  3.70  &  3.50  &  3.31  &  3.10   \\
NGC 1084     &     SA(s)c                 &       12.86  &  11.26  &  10.69  &  10.39  &  10.21  &   35.97  &   35.79  &   35.58  &   35.38  &   35.16  &  0.99  &  0.97  &  0.94  &  0.91  &  0.88    \\
NGC 1087     &     SAB(rs)c             &       12.45  &  11.37  &  10.86  &  10.57  &  10.39  &   42.40  &   43.22  &   44.20  &   45.13  &   46.14  &  0.60  &  0.71  &  0.85  &  0.98  &  1.12    \\
NGC 1299     &     SB(rs)b                &      14.43  &  13.25  &  12.70  &  12.40  &  12.17  &   14.62  &   14.72  &   14.84  &   14.95  &   15.08  &  0.83  &  0.98  &  1.17  &  1.35  &  1.54    \\
NGC 2541     &     SA(s)cd                &      13.19  &  12.08  &  11.64  &  11.45  &  11.38  &   80.03  &   77.20  &   73.83  &   70.62  &   67.14  &  1.26  &  1.33  &  1.40  &  1.48  &  1.55    \\
NGC 2543     &     SB(s)b                 &       14.47  &  12.88  &  12.05  &  11.61  &  11.29  &   29.73  &   32.09  &   34.91  &   37.59  &   40.50  &  0.33  &  1.13  &  2.09  &  3.00  &  3.99    \\
NGC 2639     &     (R)SA(r)a            &       13.95  &  12.23  &  11.40  &  11.01  &  10.71  &   19.51  &   19.03  &   18.46  &   17.91  &   17.32  &  2.22  &  2.21  &  2.20  &  2.20  &  2.19    \\
NGC 2684     &     S                           &      14.72  &  13.49  &  12.93  &  12.61  &  12.40  &   14.48  &   14.73  &   15.02  &   15.30  &   15.60  &  0.54  &  0.64  &  0.75  &  0.86  &  0.98    \\
NGC 2701     &     SAB(rs)c              &      13.85  &  12.60  &  12.05  &  11.75  &  11.57  &   26.73  &   27.56  &   28.56  &   29.50  &   30.53  &  0.30  &  0.61  &  0.99  &  1.34  &  1.73    \\
NGC 2712     &     SB(r)b                  &      14.01  &  12.54  &  11.82  &  11.43  &  11.19  &   32.11  &   32.29  &   32.50  &   32.71  &   32.93  &  0.30  &  0.84  &  1.49  &  2.11  &  2.78    \\
NGC 2742     &     SA(s)c                 &       13.34  &  12.00  &  11.38  &  11.01  &  10.80  &   44.86  &   44.55  &   44.18  &   43.83  &   43.45  &  0.39  &  0.47  &  0.57  &  0.66  &  0.77    \\
NGC 2775     &     SA(r)ab               &       12.15  &  10.34  &   9.51  &   9.03  &   8.75  &   56.50  &   60.15  &   64.49  &   68.64  &   73.13  &  3.29  &  3.47  &  3.68  &  3.89  &  4.11    \\
NGC 2776     &      SAB(rs)c             &      13.14  &  11.93  &  11.42  &  11.15  &  10.99  &   32.25  &   31.40  &   30.40  &   29.43  &   28.39  &  1.16  &  1.36  &  1.60  &  1.83  &  2.08    \\
NGC 2841     &     SA(r)b                   &     11.34  &   9.63  &   8.87  &   8.44  &   8.15  &  102.69  &  100.40  &   97.69  &   95.09  &   92.28  &  2.44  &  2.46  &  2.47  &  2.48  &  2.49    \\
NGC 2967     &      SA(s)c                 &      13.38  &  12.05  &  11.43  &  11.10  &  10.90  &   28.96  &   28.39  &   27.71  &   27.06  &   26.36  &  1.07  &  1.15  &  1.25  &  1.35  &  1.46    \\
NGC 3023     &      SAB(s)c pec       &      14.15  &  13.13  &  12.64  &  12.38  &  12.22  &   24.77  &   25.65  &   26.70  &   27.70  &   28.78  &  1.51  &  1.63  &  1.78  &  1.91  &  2.06    \\
NGC 3055     &      SAB(s)c               &     13.72  &  12.55  &  12.02  &  11.75  &  11.54  &   27.96  &   27.99  &   28.04  &   28.08  &   28.12  &  0.92  &  1.03  &  1.16  &  1.28  &  1.41    \\
NGC 3246     &      SABdm                &     14.74  &  13.42  &  12.97  &  12.72  &  12.59  &   30.98  &   30.38  &   29.67  &   28.99  &   28.25  &  1.01  &  1.14  &  1.30  &  1.45  &  1.61    \\
NGC 3259     &      SAB(rs)bc            &     14.46  &  13.21  &  12.66  &  12.38  &  12.20  &   19.57  &   18.45  &   17.12  &   15.85  &   14.47  &  1.15  &  1.20  &  1.26  &  1.32  &  1.39    \\
NGC 3310     &     SABbc pec            &    11.93  &  11.19  &  10.88  &  10.77  &  10.62  &   12.89  &   13.09  &   13.32  &   13.54  &   13.79  &  1.24  &  1.36  &  1.50  &  1.63  &  1.77    \\
NGC 3359     &      SB(rs)c                 &     12.25  &  11.10  &  10.72  &  10.49  &  10.38  &   76.25  &   73.75  &   70.78  &   67.95  &   64.88  &  0.91  &  1.00  &  1.11  &  1.21  &  1.32    \\
NGC 3423     &      SA(s)cd                &     12.58  &  11.33  &  10.91  &  10.67  &  10.47  &   55.51  &   54.93  &   54.24  &   53.57  &   52.86  &  0.59  &  0.79  &  1.02  &  1.24  &  1.48    \\
NGC 3430     &      SAB(rs)c              &     13.30  &  12.03  &  11.47  &  11.19  &  10.99  &   42.65  &   40.86  &   38.74  &   36.71  &   34.51  &  0.76  &  0.87  &  0.99  &  1.11  &  1.25    \\
NGC 3486     &     SAB(r)c                 &     12.01  &  10.82  &  10.28  &   9.98  &   9.83  &   64.89  &   64.45  &   63.92  &   63.42  &   62.87  &  1.66  &  2.18  &  2.79  &  3.38  &  4.01    \\
NGC 3488     &      SB(s)c                  &     14.61  &  13.34  &  12.83  &  12.53  &  12.35  &   24.29  &   24.03  &   23.71  &   23.41  &   23.09  &  0.55  &  0.69  &  0.85  &  1.00  &  1.17  \\
NGC 3521     &     SAB(rs)bc             &     10.83  &   9.29  &   8.55  &   8.16  &   7.89  &  118.97  &  113.77  &  107.58  &  101.68  &   95.29  &  2.77  &  2.78  &  2.80  &  2.82  &  2.83   \\
NGC 3583     &      SB(s)b                  &     13.48  &  12.01  &  11.29  &  10.88  &  10.63  &   25.14  &   24.62  &   24.00  &   23.41  &   22.77  &  1.27  &  1.48  &  1.73  &  1.96  &  2.22   \\
NGC 3589     &      Sd                          &    15.19  &  14.15  &  13.79  &  13.60  &  13.52  &   25.91  &   25.76  &   25.58  &   25.41  &   25.22  &  0.46  &  0.54  &  0.62  &  0.71  &  0.80  \\
NGC 3593     &     SA(s)0                    &    12.91  &  11.26  &  10.44  &  10.03  &   9.79  &   65.41  &   60.51  &   54.68  &   49.12  &   43.09  &  2.67  &  2.68  &  2.69  &  2.70  &  2.70  \\
NGC 3631     &      SA(s)c                  &     12.63  &  10.94  &  10.43  &  10.14  &   9.97  &   66.40  &   66.75  &   67.16  &   67.56  &   67.98  &  1.16  &  1.43  &  1.75  &  2.06  &  2.40  \\
NGC 3642     &      SA(r)bc                 &    13.14  &  11.82  &  11.26  &  10.95  &  10.77  &   36.49  &   35.61  &   34.56  &   33.55  &   32.47  &  2.76  &  3.00  &  3.29  &  3.56  &  3.86  \\
NGC 3756     &      SAB(rs)bc            &     13.20  &  11.83  &  11.23  &  10.90  &  10.70  &   53.34  &   52.31  &   51.08  &   49.90  &   48.63  &  0.60  &  0.63  &  0.67  &  0.71  &  0.76   \\
NGC 3888     &      SAB(rs)c               &    13.77  &  12.56  &  11.97  &  11.65  &  11.42  &   19.51  &   19.08  &   18.57  &   18.09  &   17.57  &  1.05  &  1.20  &  1.38  &  1.55  &  1.74  \\
NGC 3893     &     SAB(rs)c                &    12.05  &  10.86  &  10.36  &  10.08  &   9.91  &   42.18  &   41.69  &   41.10  &   40.54  &   39.93  &  1.48  &  1.61  &  1.75  &  1.89  &  2.05    \\
NGC 3898     &      SA(s)ab                &    12.82  &  11.07  &  10.31  &   9.91  &   9.67  &   39.59  &   38.51  &   37.22  &   35.99  &   34.65  &  4.66  &  4.73  &  4.82  &  4.91  &  5.00  \\
NGC 3938     &     SA(s)c                     &   11.95  &  10.70  &  10.17  &   9.85  &   9.70  &   62.93  &   61.82  &   60.50  &   59.23  &   57.87  &  0.91  &  1.13  &  1.39  &  1.65  &  1.92    \\
NGC 3982     &      SAB(r)b                 &    13.09  &  11.96  &  11.42  &  11.15  &  10.92  &   18.99  &   18.88  &   18.75  &   18.63  &   18.50  &  0.63  &  0.78  &  0.95  &  1.11  &  1.29    \\
NGC 3992     &      SB(rs)bc                &   12.16  &  10.43  &   9.62  &   9.15  &   8.88  &   96.56  &  103.69  &  112.18  &  120.28  &  129.04  &  0.96  &  1.55  &  2.25  &  2.92  &  3.64   \\
NGC 4030     &      SA(s)bc                 &    12.00  &  10.72  &  10.08  &   9.73  &   9.49  &   40.11  &   38.44  &   36.44  &   34.53  &   32.47  &  1.78  &  1.86  &  1.95  &  2.03  &  2.12    \\
NGC 4041     &      SA(rs)bc                &   12.78  &  11.59  &  11.01  &  10.74  &  10.55  &   25.94  &   24.16  &   22.03  &   20.01  &   17.81  &  2.07  &  2.06  &  2.06  &  2.06  &  2.05    \\
NGC 4102     &     SAB(s)b                  &   13.14  &  11.50  &  10.66  &  10.22  &   9.87  &   32.87  &   32.21  &   31.42  &   30.66  &   29.85  &  3.03  &  3.55  &  4.17  &  4.77  &  5.41    \\
NGC 4108     &      (R)SAc                   &   14.00  &  12.88  &  12.38  &  12.13  &  11.96  &   14.98  &   14.34  &   13.56  &   12.82  &   12.02  &  1.09  &  1.21  &  1.35  &  1.49  &  1.64    \\
NGC 4108B  &      SAB(s)dpec         &     14.88  &  14.16  &  13.82  &  13.68  &  13.59  &   18.59  &   19.06  &   19.63  &   20.17  &   20.76  &  0.98  &  1.16  &  1.38  &  1.58  &  1.81    \\
NGC 4116     &      SB(rs)dm              &    13.31  &  12.23  &  11.83  &  11.60  &  11.50  &   58.88  &   57.75  &   56.39  &   55.10  &   53.70  &  1.18  &  1.27  &  1.37  &  1.48  &  1.59    \\
NGC 4123     &      SB(r)c                    &    13.17  &  11.78  &  11.15  &  10.80  &  10.58  &   65.68  &   68.77  &   72.44  &   75.94  &   79.74  &  0.46  &  1.02  &  1.70  &  2.34  &  3.03    \\     
NGC 4168     &      E2                          &    13.54  &  11.74  &  10.98  &  10.57  &  10.34  &   31.45  &   31.07  &   30.62  &   30.18  &   29.72  &  2.90  &  2.92  &  2.93  &  2.95  &  2.97    \\
NGC 4210     &      SB(r)b                   &    14.73  &  12.87  &  12.06  &  11.62  &  11.41  &   18.10  &   26.35  &   36.17  &   45.53  &   55.67  &  0.30  &  1.34  &  2.58  &  3.76  &  5.04    \\
NGC 4215     &      SA(r)0+                &     14.34  &  12.50  &  11.68  &  11.23  &  10.91  &   15.43  &   17.42  &   19.78  &   22.04  &   24.48  &  2.78  &  3.27  &  3.85  &  4.41  &  5.02    \\   
NGC 4254     &      SA(s)c                   &    11.46  &  10.21  &   9.61  &   9.27  &   9.10  &   63.18  &   61.49  &   59.47  &   57.55  &   55.46  &  1.21  &  1.40  &  1.61  &  1.82  &  2.05    \\  
NGC 4255     &      SB(r)0                   &    14.83  &  13.04  &  12.23  &  11.78  &  11.47  &   10.04  &   10.44  &   10.91  &   11.36  &   11.85  &  3.40  &  3.69  &  4.02  &  4.34  &  4.69    \\  
NGC 4261     &      E2-3                      &    12.63  &  10.74  &   9.94  &   9.50  &   9.24  &   46.71  &   45.52  &   44.10  &   42.74  &   41.27  &  4.01  &  4.00  &  3.99  &  3.98  &  3.96    \\      
NGC 4268     &      SB0/a                    &    15.02  &  13.14  &  12.31  &  11.85  &  11.52  &   11.79  &   12.75  &   13.89  &   14.98  &   16.15  &  2.52  &  2.80  &  3.14  &  3.46  &  3.81    \\   
NGC 4270     &      S0                          &    14.31  &  12.54  &  11.76  &  11.32  &  11.02  &   16.87  &   18.08  &   19.52  &   20.90  &   22.39  &  2.04  &  2.29  &  2.59  &  2.88  &  3.19    \\  
NGC 4273     &      SB(s)c                   &    13.27  &  12.14  &  11.60  &  11.33  &  11.13  &   26.02  &   25.75  &   25.42  &   25.11  &   24.77  &  1.18  &  1.31  &  1.45  &  1.59  &  1.75     \\ 
NGC 4274     &     (R)SB(r)ab             &   12.81  &  10.94  &  10.08  &   9.60  &   9.30  &   56.52  &   57.22  &   58.05  &   58.85  &   59.71  &  1.74  &  1.86  &  2.00  &  2.13  &  2.28      \\  
NGC 4281     &      S0+                        &    13.47  &  11.61  &  10.75  &  10.29  &   9.94  &   29.75  &   31.40  &   33.36  &   35.24  &   37.26  &  2.76  &  3.17  &  3.67  &  4.15  &  4.66     \\     
NGC 4303     &      SAB(rs)bc             &    11.67  &   9.96  &   9.17  &   8.70  &   8.45  &   49.58  &   79.96  &  116.14  &  150.66  &  188.02  &  0.30  &  1.75  &  3.47  &  5.11  &  6.89    \\   
NGC 4321     &      SAB(s)bc              &    11.39  &   9.91  &   9.18  &   8.75  &   8.53  &   95.02  &  101.43  &  109.06  &  116.34  &  124.22  &  0.40  &  0.94  &  1.58  &  2.20  &  2.86    \\  
NGC 4339     &      E0                          &    13.76  &  11.91  &  11.12  &  10.71  &  10.41  &   25.40  &   26.26  &   27.28  &   28.25  &   29.31  &  3.55  &  3.70  &  3.88  &  4.05  &  4.24     \\ 
NGC 4342     &     S0-                          &    14.71  &  12.76  &  11.94  &  11.50  &  11.14  &    6.06  &    6.22  &    6.42  &    6.61  &    6.81  &  3.04  &  3.25  &  3.49  &  3.72  &  3.97           \\  
NGC 4352     &      SA0                       &    14.69  &  12.98  &  12.23  &  11.84  &  11.58  &   22.60  &   23.04  &   23.57  &   24.07  &   24.62  &  2.91  &  3.04  &  3.18  &  3.32  &  3.47       \\   
NGC 4360     &      E                            &    14.87  &  12.98  &  12.13  &  11.68  &  11.37  &   17.72  &   18.47  &   19.36  &   20.20  &   21.12  &  3.86  &  4.05  &  4.27  &  4.48  &  4.72        \\  
NGC 4365     &      E3                          &    11.94  &  10.03  &   9.21  &   8.77  &   8.48  &   73.81  &   74.72  &   75.80  &   76.83  &   77.95  &  3.95  &  4.03  &  4.14  &  4.23  &  4.34           \\ 
NGC 4370     &     Sa                           &    14.87  &  13.20  &  12.32  &  11.86  &  11.50  &   21.78  &   21.04  &   20.16  &   19.32  &   18.40  &  0.88  &  1.09  &  1.35  &  1.59  &  1.85        \\      
NGC 4374     &      E1                          &    11.58  &   9.60  &   8.72  &   8.18  &   7.85  &   58.77  &   66.93  &   76.65  &   85.93  &   95.97  &  3.86  &  4.16  &  4.51  &  4.85  &  5.21           \\    
NGC 4378     &      (R)SA(s)a             &    13.61  &  11.76  &  11.00  &  10.63  &  10.44  &   38.03  &   35.09  &   31.58  &   28.24  &   24.62  &  4.57  &  4.51  &  4.45  &  4.38  &  4.31         \\     
NGC 4387     &     E5                           &    14.51  &  12.68  &  11.90  &  11.48  &  11.20  &   13.24  &   13.31  &   13.38  &   13.46  &   13.53  &  2.12  &  2.15  &  2.19  &  2.23  &  2.26        \\    
NGC 4415     &     S0/a                        &    14.98  &  13.31  &  12.62  &  12.25  &  12.03  &   19.57  &   19.46  &   19.34  &   19.21  &   19.08  &  1.86  &  1.86  &  1.86  &  1.86  &  1.86          \\      
NGC 4431     &     SA(r)0                     &   15.22  &  13.53  &  12.79  &  12.40  &  12.18  &   22.18  &   22.30  &   22.43  &   22.57  &   22.71  &  1.57  &  1.58  &  1.60  &  1.62  &  1.64       \\     
NGC 4434     &      E0/S0(0)                &   14.32  &  12.55  &  11.79  &  11.38  &  11.11  &   12.01  &   12.15  &   12.31  &   12.46  &   12.62  &  3.48  &  3.61  &  3.76  &  3.90  &  4.05         \\   
NGC 4436     &      dE6/dS0                &   15.34  &  13.70  &  12.99  &  12.62  &  12.40  &   21.01  &   21.15  &   21.31  &   21.47  &   21.64  &  2.00  &  2.02  &  2.04  &  2.07  &  2.10       \\  
NGC 4450     &      SA(s)ab                 &   12.20  &  10.40  &   9.58  &   9.13  &   8.85  &   68.97  &   73.79  &   79.52  &   85.00  &   90.93  &  2.66  &  2.99  &  3.38  &  3.75  &  4.16       \\ 
NGC 4452     &     S0(9)                       &   14.37  &  12.61  &  11.87  &  11.48  &  11.20  &   27.09  &   27.04  &   26.99  &   26.93  &   26.88  &  1.43  &  1.38  &  1.33  &  1.28  &  1.23         \\  
NGC 4458     &     E0/E1                      &   14.20  &  12.44  &  11.69  &  11.26  &  11.01  &   22.19  &   23.14  &   24.28  &   25.37  &   26.55  &  4.02  &  4.20  &  4.41  &  4.61  &  4.83          \\   
NGC 4459     &      SA(r)0                    &   12.76  &  10.81  &   9.87  &   9.34  &   8.99  &   31.22  &   36.83  &   43.51  &   49.88  &   56.78  &  2.95  &  3.49  &  4.13  &  4.74  &  5.41           \\  
NGC 4464     &      E3                           &   14.85  &  13.04  &  12.24  &  11.80  &  11.50  &    7.36  &    7.72  &    8.15  &    8.57  &    9.01  &  3.08  &  3.30  &  3.57  &  3.83  &  4.10             \\ 
NGC 4472     &     E2/S0                      &   10.61  &   8.76  &   7.98  &   7.54  &   7.30  &  127.21  &  121.94  &  115.67  &  109.68  &  103.20  &  4.11  &  4.07  &  4.01  &  3.95  &  3.90        \\ 
NGC 4473     &      E5                           &   12.48  &  10.65  &   9.87  &   9.48  &   9.22  &   38.17  &   37.03  &   35.67  &   34.37  &   32.96  &  4.30  &  4.29  &  4.27  &  4.26  &  4.25    \\  
NGC 4474     &      S0 pec                    &  13.73  &  12.00  &  11.25  &  10.85  &  10.61  &   26.91  &   26.34  &   25.67  &   25.03  &   24.33  &  3.67  &  3.70  &  3.73  &  3.76  &  3.80    \\
NGC 4476     &      SA(r)0                     &  14.26  &  12.71  &  12.02  &  11.68  &  11.47  &   17.13  &   16.40  &   15.53  &   14.70  &   13.80  &  3.02  &  2.96  &  2.90  &  2.83  &  2.76    \\
NGC 4478     &      E2                            &  17.94  &  11.90  &  11.14  &  10.71  &  10.44  &   12.83  &   12.84  &   12.85  &   12.85  &   12.86  &  1.88  &  1.89  &  1.90  &  1.91  &  1.93    \\
NGC 4480     &      SAB(s)c                  &  14.16  &  12.87  &  12.31  &  11.98  &  11.79  &   29.37  &   28.59  &   27.65  &   26.76  &   25.79  &  0.46  &  0.65  &  0.88  &  1.10  &  1.33    \\
NGC 4486     &      E0/E pec               &    10.87  &   9.11  &   8.33  &   7.90  &   7.62  &   98.06  &   94.62  &   90.52  &   86.62  &   82.39  &  3.37  &  3.32  &  3.27  &  3.21  &  3.16    \\
NGC 4496A   &     SB(rs)m                  &   13.09  &  11.94  &  11.49  &  11.26  &  11.13  &   49.32  &   48.81  &   48.21  &   47.63  &   47.00  &  0.77  &  0.86  &  0.97  &  1.08  &  1.19    \\
NGC 4517A   &      SB(rs)dm               &  14.08  &  12.81  &  12.37  &  12.14  &  12.13  &   61.74  &   61.10  &   60.34  &   59.62  &   58.84  &  0.78  &  0.89  &  1.02  &  1.14  &  1.28    \\
NGC 4545     &      SB(s)cd                  &   14.17  &  12.93  &  12.44  &  12.16  &  11.99  &   28.75  &   28.18  &   27.50  &   26.85  &   26.14  &  0.66  &  0.74  &  0.84  &  0.94  &  1.04   \\
NGC 4550     &      SB0                         &  13.84  &  12.13  &  11.37  &  10.99  &  10.75  &   20.43  &   19.90  &   19.27  &   18.67  &   18.01  &  1.75  &  1.76  &  1.78  &  1.79  &  1.80    \\
NGC 4551     &      E                              &  14.33  &  12.48  &  11.67  &  11.25  &  10.95  &   15.61  &   15.73  &   15.88  &   16.02  &   16.17  &  2.24  &  2.29  &  2.34  &  2.40  &  2.45    \\
NGC 4552     &      E                              &  12.16  &  10.27  &   9.49  &   9.10  &   8.84  &   47.53  &   44.72  &   41.38  &   38.19  &   34.73  &  4.74  &  4.52  &  4.27  &  4.03  &  3.76   \\
NGC 4564     &      E6                            &  13.39  &  11.56  &  10.77  &  10.36  &  10.09  &   24.72  &   24.38  &   23.96  &   23.57  &   23.14  &  3.27  &  3.34  &  3.41  &  3.49  &  3.57    \\
NGC 4570     &      S0(7)                       &  13.22  &  11.36  &  10.52  &  10.09  &   9.75  &   23.10  &   23.99  &   25.06  &   26.07  &   27.16  &  2.94  &  3.17  &  3.44  &  3.69  &  3.97    \\
NGC 4600     &      S0(6)                       &  14.64  &  13.04  &  12.32  &  11.92  &  11.70  &   19.40  &   19.42  &   19.45  &   19.48  &   19.51  &  1.50  &  1.51  &  1.53  &  1.54  &  1.55   \\
NGC 4621     &      E5                            &  11.92  &  10.06  &   9.28  &   8.85  &   8.66  &   76.34  &   72.64  &   68.24  &   64.04  &   59.49  &  5.20  &  5.17  &  5.13  &  5.09  &  5.05    \\
NGC 4623     &      SB0+                       &  14.49  &  12.79  &  12.04  &  11.64  &  11.36  &   25.73  &   25.90  &   26.10  &   26.30  &   26.51  &  2.11  &  2.19  &  2.29  &  2.38  &  2.49    \\
NGC 4636     &      E/S0-1                     &  11.53  &   9.69  &   8.90  &   8.43  &   8.15  &  103.24  &  105.57  &  108.34  &  110.99  &  113.85  &  3.52  &  3.65  &  3.81  &  3.97  &  4.13    \\
NGC 4638     &      S0-                           &  13.57  &  11.76  &  11.00  &  10.61  &  10.33  &   17.75  &   17.47  &   17.14  &   16.82  &   16.47  &  3.08  &  3.11  &  3.14  &  3.18  &  3.22    \\
NGC 4649     &      E2                            &  11.28  &   9.32  &   8.49  &   8.04  &   7.78  &   76.47  &   74.56  &   72.28  &   70.11  &   67.76  &  3.35  &  3.35  &  3.35  &  3.35  &  3.35    \\
NGC 4653     &      SAB(rs)c d              &  13.88  &  12.58  &  12.07  &  11.80  &  11.66  &   38.94  &   38.08  &   37.06  &   36.08  &   35.03  &  1.04  &  1.18  &  1.34  &  1.49  &  1.66    \\
NGC 4660     &      E5                            &  13.51  &  11.71  &  10.93  &  10.56  &  10.28  &   13.59  &   13.26  &   12.87  &   12.49  &   12.09  &  3.75  &  3.65  &  3.54  &  3.44  &  3.33    \\
NGC 4668     &      SB(s)d                     &  14.47  &  13.39  &  12.96  &  12.74  &  12.58  &   19.03  &   19.23  &   19.47  &   19.69  &   19.94  &  0.54  &  0.61  &  0.70  &  0.79  &  0.88    \\
NGC 4698     &      SA(s)ab                   & 12.53  &  10.83  &  10.07  &   9.66  &   9.41  &   56.19  &   55.05  &   53.69  &   52.39  &   50.98  &  4.07  &  4.03  &  3.99  &  3.95  &  3.91    \\
NGC 4725     &      SAB(r)ab                 & 11.38  &   9.55  &   8.72  &   8.28  &   8.05  &  149.69  &  162.67  &  178.13  &  192.89  &  208.86  &  2.85  &  3.44  &  4.14  &  4.80  &  5.52    \\
NGC 4736     &      (R)SA(r)ab  	           & 10.00  &   8.46  &   7.77  &   7.41  &   7.20  &   55.26  &   53.61  &   51.65  &   49.77  &   47.74  &  3.85  &  3.82  &  3.80  &  3.77  &  3.74    \\
NGC 4845     &      SA(s)ab     	           & 13.51  &  11.69  &  10.85  &  10.39  &  10.09  &   61.98  &   60.83  &   59.47  &   58.17  &   56.76  &  0.62  &  0.77  &  0.95  &  1.12  &  1.30   \\
NGC 4900     &      SB(rs)c       	           & 12.81  &  11.70  &  11.19  &  10.87  &  10.66  &   32.25  &   32.91  &   33.70  &   34.45  &   35.27  &  0.31  &  0.38  &  0.47  &  0.55  &  0.64   \\
NGC 4904     &      SB(s)cd    	           & 13.66  &  12.40  &  11.82  &  11.52  &  11.30  &   27.58  &   27.52  &   27.45  &   27.39  &   27.32  &  0.90  &  1.03  &  1.20  &  1.35  &  1.52   \\
NGC 5055     &      SA(rs)bc   	           & 10.36  &   8.98  &   8.25  &   7.82  &   7.57  &  140.70  &  141.08  &  141.54  &  141.98  &  142.45  &  1.83  &  2.02  &  2.25  &  2.47  &  2.71   \\
NGC 5147     &      SB(s)dm    	           & 13.35  &  12.33  &  11.90  &  11.68  &  11.54  &   25.14  &   24.81  &   24.43  &   24.06  &   23.66  &  0.46  &  0.58  &  0.74  &  0.88  &  1.04   \\
NGC 5218     &      SB(s)bpec               & 14.57  &  12.90  &  12.15  &  11.76  &  11.51  &   22.29  &   21.19  &   19.87  &   18.62  &   17.27  &  0.96  &  1.07  &  1.21  &  1.34  &  1.48   \\
NGC 5300     &      SAB(r)c     	           & 13.51  &  12.15  &  11.62  &  11.30  &  11.19  &   51.01  &   50.36  &   49.60  &   48.87  &   48.07  &  0.52  &  0.63  &  0.75  &  0.86  &  0.99   \\
NGC 5334     &      SB(rs)c        		  & 13.51  &  12.14  &  11.61  &  11.31  &  11.22  &   57.49  &   56.51  &   55.34  &   54.23  &   53.02  &  0.56  &  0.64  &  0.75  &  0.85  &  0.96   \\
NGC 5364     &      SA(rs)bcpec            & 12.33  &  10.86  &  10.25  &   9.90  &   9.77  &   90.73  &   88.88  &   86.68  &   84.59  &   82.32  &  1.23  &  1.39  &  1.57  &  1.75  &  1.95    \\
NGC 5376     &      SAB(r)b                    & 14.19  &  12.64  &  11.91  &  11.49  &  11.21  &   23.37  &   23.54  &   23.74  &   23.93  &   24.14  &  0.61  &  0.84  &  1.13  &  1.40  &  1.69    \\
NGC 5430     &      SB(s)b          	           & 14.00  &  12.52  &  11.76  &  11.34  &  11.02  &   24.49  &   24.72  &   24.98  &   25.23  &   25.50  &  1.38  &  1.85  &  2.42  &  2.96  &  3.54    \\
NGC 5480     &      SA(s)c               	  & 13.87  &  12.64  &  12.04  &  11.72  &  11.49  &   22.52  &   23.50  &   24.66  &   25.77  &   26.97  &  0.73  &  0.90  &  1.10  &  1.29  &  1.50    \\
NGC 5584     &      SAB(rs)cd  	           & 13.29  &  12.07  &  11.57  &  11.31  &  11.17  &   53.73  &   53.94  &   54.20  &   54.45  &   54.72  &  0.49  &  0.61  &  0.75  &  0.89  &  1.03    \\
NGC 5624     &      S              		  & 15.10  &  14.06  &  13.62  &  13.40  &  13.21  &   13.95  &   14.68  &   15.56  &   16.40  &   17.30  &  0.86  &  0.86  &  0.86  &  0.86  &  0.87    \\
NGC 5660     &      SAB(rs)c 		  & 13.33  &  12.19  &  11.72  &  11.47  &  11.30  &   31.23  &   31.13  &   31.01  &   30.90  &   30.77  &  0.79  &  1.00  &  1.24  &  1.47  &  1.72    \\
NGC 5667     &      Scd pec  		  & 14.25  &  13.22  &  12.84  &  12.63  &  12.48  &   23.42  &   23.60  &   23.82  &   24.02  &   24.24  &  0.66  &  0.80  &  0.98  &  1.14  &  1.32    \\
NGC 5668     &      SA(s)d     		  & 13.02  &  11.92  &  11.50  &  11.29  &  11.17  &   37.47  &   37.51  &   37.56  &   37.61  &   37.66  &  1.16  &  1.30  &  1.47  &  1.64  &  1.81    \\
NGC 5693     &      SB(rs)d   		  & 14.91  &  13.60  &  13.08  &  12.82  &  12.69  &   23.04  &   22.75  &   22.42  &   22.09  &   21.74  &  0.86  &  0.93  &  1.00  &  1.07  &  1.15    \\
NGC 5713     &      SAB(rs)bcpec 	  & 12.99  &  11.66  &  11.02  &  10.71  &  10.50  &   28.00  &   27.88  &   27.72  &   27.58  &   27.42  &  1.40  &  1.42  &  1.44  &  1.46  &  1.49    \\
NGC 5768     &      SA(rs)c             	  & 14.29  &  12.99  &  12.40  &  12.06  &  11.85  &   20.71  &   20.33  &   19.88  &   19.45  &   18.98  &  0.71  &  0.87  &  1.06  &  1.25  &  1.44    \\
NGC 5774     &      SAB(rs)d           	  & 13.73  &  12.48  &  12.04  &  11.82  &  11.66  &   53.27  &   52.56  &   51.72  &   50.91  &   50.04  &  1.92  &  1.96  &  2.02  &  2.07  &  2.13    \\
NGC 5806     &      SAB(s)b           	  & 13.54  &  11.88  &  11.07  &  10.63  &  10.31  &   41.73  &   43.49  &   45.58  &   47.58  &   49.75  &  1.68  &  2.08  &  2.56  &  3.02  &  3.51    \\
NGC 5850     &      SB(r)b              	  & 13.33  &  11.44  &  10.69  &  10.32  &  10.11  &   82.57  &   76.33  &   68.89  &   61.80  &   54.13  &  4.64  &  4.71  &  4.79  &  4.88  &  4.97    \\
NGC 5879     &      SA(rs)bc                  & 13.17  &  11.86  &  11.29  &  11.00  &  10.82  &   38.17  &   35.31  &   31.90  &   28.64  &   25.12  &  2.33  &  2.37  &  2.42  &  2.47  &  2.52   \\
NGC 5937     &      (R)SAB(rs)b 	           & 14.25  &  12.90  &  12.14  &  11.79  &  11.46  &   20.27  &   19.83  &   19.30  &   18.79  &   18.25  &  1.22  &  1.31  &  1.42  &  1.52  &  1.63    \\
NGC 6070     &      SA(s)cd  		  & 13.93  &  12.09  &  11.19  &  10.68  &  10.38  &   42.48  &   52.20  &   63.78  &   74.83  &   86.79  &  0.30  &  1.06  &  1.98  &  2.84  &  3.79    \\
NGC 6155     &      S              		  & 14.23  &  13.00  &  12.43  &  12.14  &  11.92  &   17.85  &   17.92  &   18.00  &   18.08  &   18.17  &  0.73  &  0.81  &  0.91  &  1.00  &  1.10    \\
NGC 6314     &      SA(s)a    		  & 15.24  &  13.53  &  12.71  &  12.31  &  12.04  &   19.83  &   18.65  &   17.25  &   15.91  &   14.47  &  5.44  &  5.14  &  4.79  &  4.45  &  4.08    \\
NGC 7437     &      SAB(rs)d                  & 14.57  &  13.30  &  12.78  &  12.49  &  12.33  &   28.41  &   29.06  &   29.83  &   30.57  &   31.37  &  0.51  &  0.63  &  0.78  &  0.92  &  1.07    \\
UGCA 021     &      SB(s)d                      &14.17  &  13.07  &  12.63  &  12.38  &  12.26  &   30.91  &   31.22  &   31.58  &   31.93  &   32.30  &  1.03  &  1.13  &  1.26  &  1.37  &  1.50   \\
UGCA 219     &     Scp                            & 15.26  &  14.85  &  14.82  &  14.83  &  14.64  &    3.56  &    4.11  &    4.77  &    5.40  &    6.08  &  3.31  &  3.43  &  3.58  &  3.72  &  3.88    \\   
UGC 02081   &      SAB(s)cd                &  15.55  &  14.22  &  13.72  &  13.47  &  13.35  &   30.33  &   30.61  &   30.94  &   31.25  &   31.59  &  0.84  &  0.95  &  1.09  &  1.22  &  1.36    \\
UGC 04393   &     SBc       		          &  14.61  &  13.45  &  13.05  &  12.82  &  12.67  &   23.44  &   23.60  &   23.79  &   23.98  &   24.18  &  2.03  &  1.99  &  1.96  &  1.92  &  1.89    \\
UGC 06162   &     Sd                             &  14.37  &  13.28  &  12.86  &  12.60  &  12.53  &   40.89  &   40.59  &   40.24  &   39.91  &   39.55  &  0.60  &  0.69  &  0.81  &  0.92  &  1.04    \\ 
UGC 06309   &     SB                             & 14.82  &  13.54  &  12.93  &  12.61  &  12.37  &   21.81  &   21.84  &   21.88  &   21.91  &   21.95  &  0.38  &  0.42  &  0.48  &  0.52  &  0.58    \\  
UGC 06518   &     S                   	            &15.57  &  14.49  &  14.00  &  13.77  &  13.60  &   13.33  &   13.13  &   12.90  &   12.67  &   12.43  &  0.68  &  0.74  &  0.81  &  0.87  &  0.95    \\
UGC 06903   &     SB(s)cd                    & 14.58  &  13.21  &  12.65  &  12.35  &  12.22  &   41.18  &   41.82  &   42.57  &   43.30  &   44.08  &  0.30  &  0.63  &  1.02  &  1.39  &  1.80    \\
UGC 07700   &     SB(s)dm                   & 15.06  &  14.03  &  13.71  &  13.60  &  13.66  &   37.63  &   35.02  &   31.91  &   28.95  &   25.73  &  1.33  &  1.38  &  1.44  &  1.50  &  1.56    \\
UGC 08041   &     SB(s)d                      & 13.94  &  12.71  &  12.24  &  11.98  &  11.90  &   50.74  &   50.52  &   50.26  &   50.01  &   49.75  &  0.58  &  0.69  &  0.81  &  0.93  &  1.06    \\
UGC 08084   &     SB(s)dm              	  & 15.36  &  14.22  &  13.86  &  13.65  &  13.59  &   35.63  &   34.30  &   32.72  &   31.22  &   29.58  &  2.03  &  1.97  &  1.91  &  1.85  &  1.78    \\
UGC 08237   &     (R)SBb    		  & 14.71  &  13.33  &  12.64  &  12.31  &  12.04  &    8.16  &    8.34  &    8.54  &    8.74  &    8.95  &  4.32  &  4.68  &  5.11  &  5.52  &  5.96    \\
UGC 08658   &     SAB(rs)c 		  & 14.73  &  13.33  &  12.83  &  12.57  &  12.46  &   37.27  &   36.11  &   34.72  &   33.40  &   31.97  &  1.10  &  1.19  &  1.30  &  1.40  &  1.51    \\
UGC 09215   &     SB(s)d                      & 13.95  &  12.95  &  12.55  &  12.32  &  12.18  &   30.34  &   31.10  &   32.00  &   32.86  &   33.79  &  1.47  &  1.59  &  1.74  &  1.88  &  2.03    \\
UGC 09741   &     Sc                              & 15.30  &  14.22  &  13.71  &  13.42  &  13.22  &   10.54  &   11.14  &   11.85  &   12.53  &   13.27  &  0.59  &  0.70  &  0.83  &  0.95  &  1.09    \\
UGC 09837   &     SAB(s)c                    & 14.82  &  13.70  &  13.29  &  13.10  &  13.04  &   27.72  &   27.19  &   26.55  &   25.94  &   25.28  &  0.96  &  1.20  &  1.50  &  1.78  &  2.08    \\
UGC 10721   &     Scd                            & 14.76  &  13.60  &  13.07  &  12.79  &  12.56  &   15.99  &   15.68  &   15.31  &   14.96  &   14.58  &  0.63  &  0.79  &  0.98  &  1.17  &  1.36    \\ 
UGC 12709   &     SAB(s)m                   & 15.71  &  14.43  &  14.00  &  13.79  &  13.75  &   37.04  &   36.43  &   35.71  &   35.03  &   34.28  &  0.76  &  0.86  &  0.99  &  1.11  &  1.24    \\
\end{longtable}   
\end{center}
\normalsize
\end{landscape}

\twocolumn

 \begin{table*}
\caption{Subsample of elliptical galaxies. The galaxy type and the galactic correction ($A_{g}$) are from NED (and in brackets from C93).  The absolute magnitude $M_{B}^{*}$, the effective radius ($r_{\rmn{e},B}^{*}$) and the \sersic index ($n_{\rmn{eq},B}^{*}$) have been derived from one-dimensional fitting in C93. The $M_{B}$ is our $g$-band magnitude after conversion to the $B$-band and Vega system. The $r_{\rmn{e},g}$ and the $n_{g}$ are our $g$-band effective radius and \sersic index. The distances have come from \citet{tex:BJ09} unless otherwise specified: $^\rmn{a}$ \citet{tex:TD01}, $^\rmn{b}$ \citet{tex:JB04}. In the case of NGC4360 a mean Virgo distance has been adopted. \label{lastpage}}
  \smallskip
  \centering
  \begin{tabular}{llldddddddd}
  \hline
  \noalign{\smallskip}
Galaxy & VCC  & Type  & \multicolumn{1}{c}{$M_{B}^{*}$}    & \multicolumn{1}{c}{$r_{\rmn{e},B}^{*}$}  &     \multicolumn{1}{c}{$n_{\rmn{eq},B}^{*}$}    &      \multicolumn{1}{c}{distance}       &   \multicolumn{1}{c}{$A_{g}$}  &   \multicolumn{1}{c}{$M_{B}$} &   \multicolumn{1}{c}{$r_{\rmn{e},g}$} &     \multicolumn{1}{c}{$n_{g}$} \\
              &  &            &          \multicolumn{1}{c}{(mag)}            &  \multicolumn{1}{c}{(arcsec)}           &                                  &        \multicolumn{1}{c}{(Mpc)}         &     \multicolumn{1}{c}{(mag)}    &        \multicolumn{1}{c}{(mag)}   & \multicolumn{1}{c}{(arcsec)} &      \\ 
\noalign{\smallskip}
\hline
\hline
\noalign{\smallskip}       
IC3468            &     VCC1422     &        E1                       &   -16.89  &   21.08  &  2.21  &  15.4  &  0.105  &  -16.99                         &    19.73     &  1.86     \\
IC3653            &     VCC1871     &        E3                       &   -16.34  &    6.23  &  1.82  &  15.5  &  0.103  &  -16.35                          &     7.23      &  1.95     \\
NGC4168       &     VCC49          &        E2                       &   -20.23  &   17.44  &  2.53  &  30.9^\rmn{a}  &  0.069  &  -20.43          &     31.07     & 2.29    \\ 
NGC4261       &     VCC345        &        E2-3                   &   -21.76  &  124.53  &  7.65  &  31.6^\rmn{a}   &  0.060   &  -21.44       &     45.52     &  4.00    \\   
NGC4339       &     VCC648        &        E0   (S0)            &   -18.60  &   27.57  &  3.50  &  16.4^\rmn{a}   &  0.084   &  -18.75         &    26.26      &  3.70    \\
NGC4360       &     VCC722        &        E                         &   -17.48  &   18.92  &  3.33  &  16.5  &  0.072  &  -17.80                          &     18.47     &  4.05    \\ 
NGC4365       &     VCC731        &        E3                       &   -21.52  &  117.51  &  6.08  &  23.1  &  0.070   &  -21.42                       &     74.72     &   4.03    \\ 
NGC4374       &     VCC763        &        E1                       &   -21.69  &  190.85  &  8.47  &  18.5  &  0.133  &  -21.21                        &    66.93      &   4.16    \\ 
NGC4387       &     VCC828        &        E5                       &   -17.99  &   10.23  &  2.12  &  18.0  &  0.109    &  -18.12                       &    13.31      &   2.15   \\ 
NGC4415       &     VCC929        &        S0/a   (dE1)       &   -17.02  &   18.59  &  1.85  &  14.9^\rmn{b}  &  0.071  &  -17.23          &     19.46     &   1.86    \\   
NGC4434       &     VCC1025      &        E0/S0(0)  (E0)   &  -18.74  &   14.45  &  4.19  &  22.5  &  0.073  &  -18.80                         &      12.15    &    3.61  \\
NGC4436       &     VCC1036      &        dE6/dS0  (dS0) &  -16.81  &   16.07  &  2.40  &  15.4^\rmn{b}  &  0.091   &  -16.84         &    21.15      &    2.02  \\ 
NGC4458       &     VCC1146      &        E0/E1                  &  -18.03  &   20.03  &  2.84  &  16.3  &  0.079  &  -18.32                         &   23.14       &    4.20   \\ 
NGC4464       &     VCC1178      &        E3                        &  -17.07  &    6.52  &  2.61  &  13.4  &  0.072  &  -17.14                          &   7.72       &     3.30   \\ 
NGC4472       &     VCC1226      &       E2/S0  (E2)          & -22.18  &  240.12  &  6.24  &  16.7  &  0.074  &  -22.00                        &  121.94        &  4.07      \\
NGC4473       &     VCC1231      &        E5                        &  -19.79  &   30.29  &  5.29  &  15.2  &  0.094  &  -19.86                         &    37.03      &    4.29    \\
NGC4478       &     VCC1279      &       E2                         &  -18.71  &   11.69  &  1.98  &  17.1  &  0.081  &  -18.80                         &   12.84       &   1.89     \\
NGC4486       &     VCC1316      &       E0/E pec              &  -21.76  &  170.91  &  6.51  &  16.7  &  0.074  &  -21.70                       &    94.62      &   3.32     \\      
NGC4551       &     VCC1630      &       E                            & -17.84  &   12.71  &  1.89  &  16.2  &  0.127  &  -18.07                         &   15.73       &    2.29     \\
NGC4552       &     VCC1632      &       E   (S0(0))            &  -20.76  &  151.06  &  12.81  &  16.0  &  0.135   &  -20.43                    &    44.72      &    4.52     \\
NGC4564       &     VCC1664      &       E6                          & -18.73  &   17.49  &  2.42  &  15.9  &  0.115  &  -18.98                         &    24.38      &   3.34      \\
NGC4621       &     VCC1903      &        E5                         & -20.36  &   82.98  &  6.14  &  14.9  &  0.110   &  -20.46                        &    72.64      &   5.17       \\
NGC4623       &     VCC1913      &        SB0+ ( E7)          & -17.30  &   16.68  &  1.66  &  13.2  &  0.073  &  -17.38                         &    25.90      &     2.19    \\     
NGC4636       &     VCC1939      &       E/S0-1  (E1)         & -20.73  &  137.61  &  2.86  &  14.7^\rmn{a}  &  0.092  &  -20.91         &    105.57      &  3.65       \\    
NGC4649       &     VCC1978      &       E2  (S02)              & -21.52  &  117.49  &  5.84  &  16.5  &  0.088  &  -21.49                        &     74.56     &   3.35       \\ 
NGC4660       &     VCC2000      &        E5                         & -18.70  &   11.14  &  3.87  &  15.0  &  0.108    &  -18.71                       &   13.26       &    3.65      \\
\noalign{\smallskip}
\hline
\end{tabular}
\label{table:ellipt} 
\end{table*}

\end{document}